\documentclass[a4paper,11pt]{article}
\pdfoutput=1 

\usepackage{jheppub} 

\usepackage[T1]{fontenc} 

\usepackage{pgfplots}
\pgfplotsset{compat=newest}
\usepgfplotslibrary{fillbetween}

\usepackage{amssymb,amsmath}
\usepackage{bbm}
\usepackage{mathtools}
\usepackage{graphicx}
\usepackage{epsfig}
\usepackage{epstopdf}
\usepackage{setspace}
\usepackage{dsfont}
\usepackage{latexsym}
\usepackage{tikz}
\usepackage{psfrag}
\usepackage{booktabs}
\usepackage{bbm}
\usepackage[toc]{appendix}
\usepackage{color}
\usepackage{datetime}
\usepackage{tensor}
\usepackage{lscape}
\usepackage{lipsum}
\usepackage{caption}
\usepackage{subcaption}

\usepackage{tikz}
\usetikzlibrary{arrows,shapes}
\usetikzlibrary{trees}
\usetikzlibrary{patterns}
\usetikzlibrary{matrix,arrows} 				
\usetikzlibrary{positioning}				
\usetikzlibrary{calc,through}				
\usetikzlibrary{decorations.pathreplacing}  
\usepackage{pgffor}							
\usetikzlibrary{decorations.pathmorphing}	
\usetikzlibrary{decorations.markings}
\tikzset{
    vector/.style={decorate, decoration={snake}, draw},
	provector/.style={decorate, decoration={snake,amplitude=2.5pt}, draw},
	antivector/.style={decorate, decoration={snake,amplitude=-2.5pt}, draw},
        smallvector/.style={decorate, decoration={snake,amplitude=1.5pt,post length=0.5mm}, draw},
    fermion/.style={draw=black, postaction={decorate},
        decoration={markings,mark=at position .55 with {\arrow[draw=black]{>}}}},
    fermionbar/.style={draw=black, postaction={decorate},
        decoration={markings,mark=at position .55 with {\arrow[draw=black]{<}}}},
    fermionnoarrow/.style={draw=black},
    gluon/.style={decorate, draw=black,
        decoration={coil,amplitude=4pt, segment length=5pt}},
    scalar/.style={dashed,draw=black, postaction={decorate},
        decoration={markings,mark=at position .55 with {\arrow[draw=black]{>}}}},
    scalarbar/.style={dashed,draw=black, postaction={decorate},
        decoration={markings,mark=at position .55 with {\arrow[draw=black]{<}}}},
    scalarnoarrow/.style={dashed,draw=black},
    electron/.style={draw=black, postaction={decorate},
        decoration={markings,mark=at position .55 with {\arrow[draw=black]{>}}}},
    bigvector/.style={decorate, decoration={snake,amplitude=4pt}, draw},
    arrow/.style={draw=black, postaction={decorate},
        decoration={markings,mark=at position 1 with {\arrow[draw=black]{>}}}},
}

\tikzstyle{block} = [draw, rectangle, 
    minimum height=3em, minimum width=6earticlem]

\definecolor{darkblue}{rgb}{0.2, 0, 0.8}

\numberwithin{equation}{section}

\newcommand{\str}{\text{str}}

\usepackage{comment}

\newcommand{\reef}[1]{(\ref{#1})}

\newcommand{\be}{\begin{equation}}
\newcommand{\ee}{\end{equation}}

\def\be{\begin{equation}}
\def\ee{\end{equation}}
\def\bea{\begin{Eqarray}}
\def\eea{\end{Eqarray}}
\def\ba{\begin{array}}
\def\ea{\end{array}}
\def\bd{\begin{displaymath}}
\def\ed{\end{displaymath}}

\def\tr{{\rm tr}}

\def\a{\alpha}

\def\d{\delta}
  
\def\g{\gamma}

\def\G{\Gamma}

\def\>{\rangle} 
\def\<{\langle} 
\def\Dsl{D \hskip-.6em \raise1pt\hbox{$ / $ } }
\def\to{\rightarrow}

\newcommand{\lra}{\leftrightarrow}

\newcommand{\gap}{\text{gap}}

\title{Corners and Islands in the S-matrix Bootstrap of the Open Superstring}

\author{Justin Berman,}
\author{Henriette Elvang}

\affiliation{
Leinweber Center for Theoretical Physics, Randall Laboratory of Physics,\\  
The University of Michigan, Ann Arbor, MI 48109-1040, USA
}

\emailAdd{jdhb@umich.edu, elvang@umich.edu}

\abstract{
We bootstrap the Veneziano superstring amplitude in 10 dimensions from the bottom-up. Starting with the most general maximally supersymme\-tric Yang-Mills EFT, we input information about the lowest-lying massive states,
which we assume contribute via tree-level exchanges to the 4-point amplitude. 
We show the following: (1) if there is only a single state at the lowest mass, it must be a scalar. 
(2) Assuming a string-inspired gap between the mass of this scalar and any other massive states, the allowed region of Wilson coefficients has a new sharp corner where the Veneziano amplitude is located. 
(3) Upon fixing the next massive state to be a vector,
the EFT bounds have a one-parameter family of corners; these would correspond to models with linear Regge trajectories of varying slopes, one of which is the open superstring. 
(4) When the ratio between the massive scalar coupling and the $\tr F^4$ coefficient is fixed to its string value, the spin and mass of the second massive state is determined by the bootstrap and the  Veneziano amplitude is isolated on a small island in parameter space. 
Finally, we compare with other recent bootstraps approaches, both the pion model and imposing Regge-inspired maximal spin constraints.  
}

\vspace{2cm}

\begin{document} 
 \begin{flushright}
{\tt LCTP-24-10} \\
\end{flushright}
\maketitle
\flushbottom



\raggedbottom
\section{Introduction and Summary of Results}
\label{s:intro}
The effective field theory (EFT) S-matrix bootstrap makes use of fundamental physical assumptions, such as unitarity, analyticity, and  Froissart-like bounds, to constrain the allowed ranges for Wilson coefficients of  
higher-dimensional operators. The so-called ``dual'' formulation of the S-matrix bootstrap, which explicitly rules out coupling parameter space, has been applied to EFTs with a broad range of massless states, including permutation symmetric scalars, pions, photons, gluons, and gravitons \cite{Caron-Huot:2020cmc,Arkani-Hamed:2020blm,Chiang:2021ziz,Albert:2022oes,Caron-Huot:2022jli,Fernandez:2022kzi,Albert:2023jtd,Alberte:2020bdz,Henriksson:2021ymi,Chowdhury:2021ynh,Caron-Huot:2022ugt,deRham:2022sdl,Ma:2023vgc,CarrilloGonzalez:2023cbf,Berman:2023jys,Chiang:2023quf}. More recent implementations of this bootstrap technique study how additional physical assumptions about the massive spectrum can  limit the allowed parameter space and yield new interesting features \cite{Albert:2023bml,Haring:2023zwu}. In particular, \cite{Albert:2023bml} showed that for large-$N$ massless pion scattering, a new corner appeared in the allowed parameter space when information about the spin of the lowest-lying massive particles in the spectrum was included in the bootstrap. 

Motivated by these recent results, we derive bounds on maximally supersymmetric Yang-Mills (SYM) EFTs using the S-matrix bootstrap 
combined with basic assumptions about the lowest massive states in the spectrum of the UV theory. We work at large rank of the gauge group so that multi-trace operators are suppressed and we assume weak coupling to suppress massless loops and ensure that the low-energy expansion is polynomial in the Mandelstam variables.
Using maximal supersymmetry and locality, the general ansatz for the low-energy 4-point scattering amplitudes is\footnote{Polarization-dependent overall factors are accounted for in Section \ref{s:disprep}.}
\begin{equation}
\label{ansatzIntro}
A(s,u)  
= -\frac{s}{u} 
+ s^2 
\Big[
 a_{0,0} 
 + a_{1,0} (s+u)
 + a_{2,0} (s^2+u^2)
 + a_{2,1} su
 + \ldots
\Big]
\, .
\end{equation}
The $a_{k,q}$ are the Wilson coefficients of the maximally supersymmetrized versions of the local single-trace 
 operators $\tr (D^{2k}F^4)$, e.g.~
\be\label{opcoeff}
  a_{0,0} 
  ~\lra~
  \tr F^4\,,
  ~~~~~~
  a_{1,0} 
  ~\lra~
  \tr D^2 F^4 \,,
  ~~~~~~
  a_{2,0}\,,
  a_{2,1}
  ~\lra~
  \tr D^4F^4 \,,
  ~~~\text{etc}.
\ee
There are two independent maximally supersymmetric $\tr D^4F^4$ operators, hence two coefficients are listed. 
From the low-energy perspective, without regard for 
the UV origin of the EFT, the Wilson coefficients 
$a_{k,q}$ in \reef{ansatzIntro} could be any real numbers in units of some UV cutoff. The EFT S-matrix bootstrap allows us to compute
upper and lower bounds on ratios of $a_{k,q}$.

One of the possible UV completions of the EFTs considered here is the open superstring, whose 4-point tree-level scattering process is given by the
 Veneziano amplitude \cite{Veneziano:1968yb},
\be
   A^\text{str}(s,u)
   = -(\alpha's)^2 \frac{\G(-\a's)\G(-\a'u)}{\G(1-\a'(s+u))} \, .
   \label{scalarVeneziano}
\ee   
Its low-energy $\alpha'$-expansion, 
\be
   A^\text{str}(s,u) = -\frac{s}{u}
+ s^2 \bigg(
 \zeta_2 \alpha'^2 
+ \zeta_3 \alpha'^3 (s+u)
+ \zeta_4 \alpha'^4 
\big(s^2 +  u^2\big)
+ \frac{1}{4}\zeta_4 \alpha'^4
s u 
+ \ldots
\bigg)\,,
\label{scalarVenezianoExp}
\ee
corresponds to a specific choice of the coefficients $a_{k,q}$ in \reef{ansatzIntro}. 
One of the goals of this paper is to bootstrap the Veneziano amplitude using as little physical input as possible.

In \cite{Berman:2023jys}, we determined universal two-sided bounds on the Wilson coefficients $a_{k,q}$ of 4-dimensional $\mathcal{N}=4$ SYM EFT assuming the existence of a mass gap, but with no constraints imposed the UV spectrum; hence the notion  of `universal bounds'. 
To compute the bounds, the $a_{k,q}$ are made dimensionless by scaling out powers of the mass gap and the bounds are derived for ratios of couplings 
\be 
  \label{abardef}
  \bar{a}_{k,q} \equiv 
  \frac{a_{k,q}}{a_{0,0}}\,.
\ee 
The universal bounds determine an allowed region in the space of effective couplings  $\bar{a}_{k,q}$. 
The Veneziano amplitude was found in the interior of the allowed region; it was not at any special place such as near a corner or cusp in the boundary. 
However, when the EFT bootstrap was combined with the additional constraint that the amplitude obeyed the string monodromy relations,\footnote{The string monodromy relations \cite{Plahte:1970wy,Stieberger:2009hq,Bjerrum-Bohr:2009ulz,Bjerrum-Bohr:2010mia,Bjerrum-Bohr:2010pnr} are a set of linear relations arising from the disk amplitude via contour deformations of the integration over the vertex operator insertion points.} it was found \cite{Berman:2023jys,Chiang:2023quf} that the two-sided bounds narrowed in on the $a_{k,q}$ values of the Veneziano amplitude  \reef{scalarVenezianoExp}. This was evidence for the earlier conjecture \cite{Huang:2020nqy} that the open string is the unique amplitude compatible with the combined constraints of the EFT bootstrap and the string monodromy relations. 

Isolating the low-energy expansion of the Veneziano amplitude with the monodromy relations has an interesting geometric interpretation,\footnote{We found in \cite{Berman:2023jys} that the general coupling space (i.e.~without monodromy relations imposed) has fewer independent Wilson coefficients than the general EFT-expansion \reef{ansatzIntro} suggests, and we proposed a partially resummed low-energy expansion of the 4-point amplitude. We do not use the partially resummed form in this paper.}  
but it is not  
satisfactory to bootstrap the string amplitude by assuming one of its salient  worldsheet properties.  
Instead, it would be much more desirable to find evidence of string theory (here specifically  the Veneziano amplitude) from a more particle-based approach.

In this paper, we pursue a bottom-up EFT S-matrix bootstrap of the Veneziano amplitude in $D=10$ dimensions. The new ``ingredient'' in the bootstrap is information about the lowest massive states. In particular, the open string   has a spin 0 state at mass-squared $1/\alpha'$, a spin 1 state at $2/\alpha'$, a spin 2 state at $3/\alpha'$, etc, as its leading Regge trajectory. 
We find that even just assuming that the lowest-lying massive state is a scalar and that there is a suitable gap to the next (otherwise unspecified) state restricts the allowed parameter space so that the string is now found  very near a corner in the resulting bounds. Moreover, we show that with input about only the lowest-mass state's spin and coupling to the massless states relative to $a_{0,0}$,
the Veneziano amplitude is isolated on islands that shrink in size as more constraints from higher-derivative operators are included.\footnote{The Veneziano amplitude is not the unique UV completion of a maximally supersymmetric YM EFT; other options include the Coulomb branch amplitudes. In that case, the massive states couple quadratically and therefore have to appear in loops. In contrast, we assume that the lowest mass states are exchanged at tree-level.} 

More generally, we also examine the effects of basic low-spectrum input on the bounds in the EFT S-matrix bootstrap.


\subsection*{Setup and Summary of Results}

We compute bounds on the ratios of Wilson coefficients in 
\reef{abardef}. This is done by using analyticity and the Froissart bound to derive dispersive representations for each $a_{k,q}$. Then those are used, together with unitarity, to formulate an optimization problem which is solved numerically with the semi-definite problem solver SDPB \cite{Simmons-Duffin:2015qma}. This involves truncating the derivative expansion \reef{ansatzIntro} to some finite order $k_\text{max}$ (corresponding to $2 k_\text{max}+4$ derivative order); the numerical bounds get stronger with  increasing $k_\text{max}$.

\begin{figure}[t]
\begin{center}
\begin{tikzpicture}
	\node (image) at (0,0) {\includegraphics[width=\textwidth]{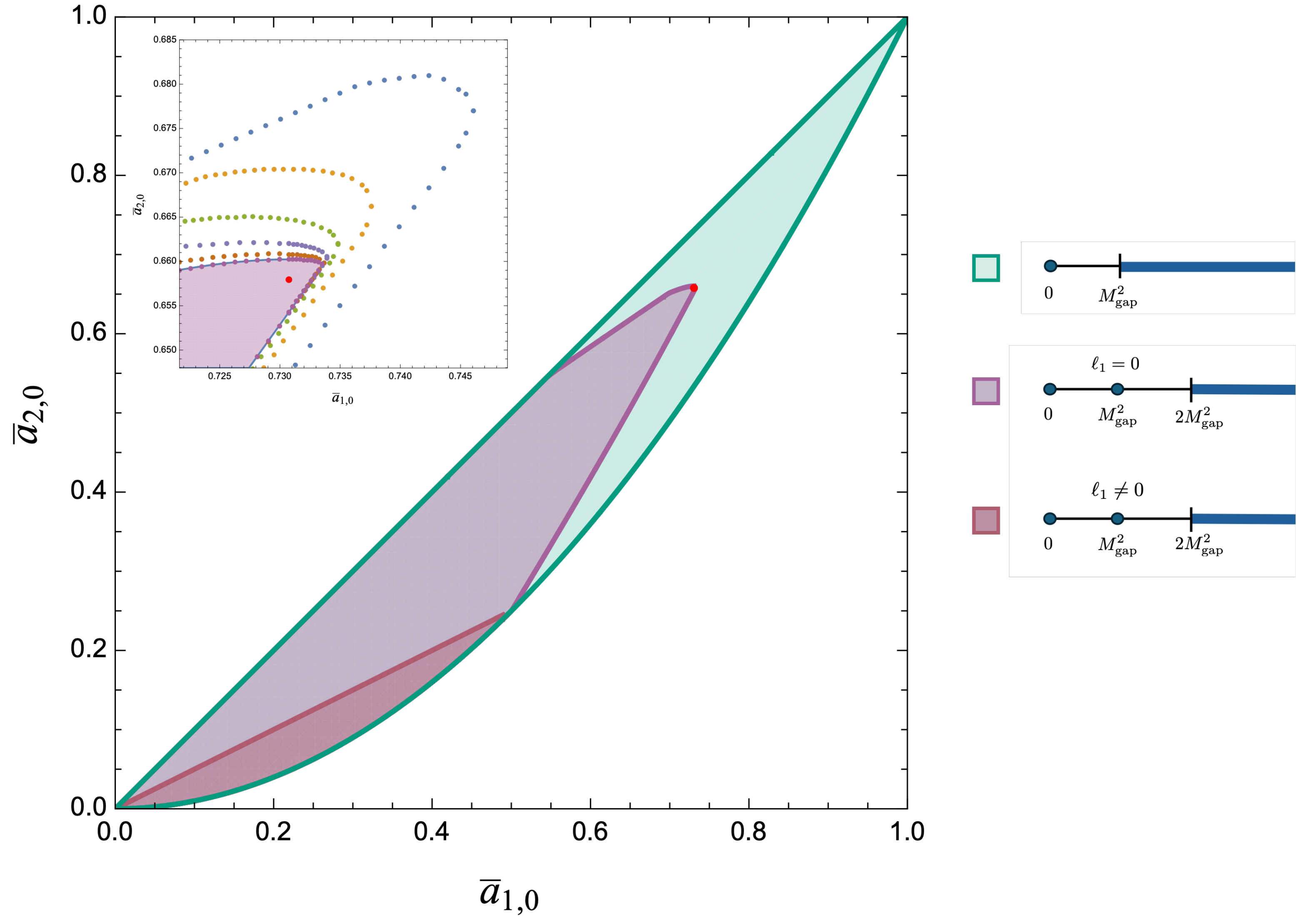}};
\end{tikzpicture}
\end{center}
\caption{This plot shows the allowed region for Wilson coefficients  $\bar{a}_{1,0}$ and $\bar{a}_{2,0}$. {\em Teal} shows the region with unrestricted spectrum above the mass gap. {\em Purple} shows the allowed region for a scalar at the mass gap and no assumptions about the spectrum at and above $2M_\text{gap}^2$. In {\em magenta} is shown the allowed region when the cutoff is at $2M_\text{gap}^2$ and there is a non-zero spin state or no state at $M_\text{gap}^2$. 
The red dot corresponds to the Veneziano amplitude values for these coefficients and it sits very close to the corner in the purple bounds.  
The inset zooms in near the tip of the purple allowed region and shows the dependence of the bounds on the truncation parameter $k_\text{max}$ for $k_\text{max}=4,6,8,10,12,$ and $14$.}
\label{fig:a10a20scalarinp}
\end{figure}

As context for the new bounds, the universal bounds in the $(\bar{a}_{1,0},\bar{a}_{2,0})$ plane (the coefficients of $\tr D^2F^4$ and of one of the $\tr D^4F^4$ operators, respectively) are simply 
\cite{Arkani-Hamed:2020blm,Berman:2023jys}:  
\be
 \label{unibnds}
 \bar{a}_{1,0}^2 \le \bar{a}_{2,0}\le \bar{a}_{1,0} \,.
\ee
This region (whose bounds happen to be independent of $k_\text{max}$) is shown in Figure \ref{fig:a10a20scalarinp} in teal. 
The region has two corners: the corner at $(0,0)$ corresponds to a model in which the $\tr F^4$ coupling dominates all other EFT couplings ($a_{0,0} \gg a_{k,q}$ for all $k>0$) and the $(1,1)$ corner is an (unphysical) model with an infinite spin tower at the mass gap.

The simplest assumption we can make about the massive spectrum is that the lowest mass states contribute to the 4-point amplitude $A(s,u)$ via simple pole exchanges, e.g.~
\be
\label{polediagram}
\raisebox{-1cm}{\includegraphics[width=4cm]{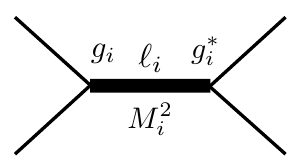}} 
\ee
for a particle with mass $M_i^2$, spin $\ell_i$, and  coupling $g_i$ to the massless external states.\footnote{Specifying the spin of the internal state, which must be part of a supermultiplet, only makes sense when the external states of $A(s,u)$ are chosen. This will be done in Section \ref{s:review}.}  To begin with, we only put in a single state, then subsequently, inspired by \cite{Albert:2023bml}, two states. 

\vspace{2mm}
{\bf Single-State Input.}
Suppose that the amplitude $A(s,u)$ has a pole at the mass gap $M_\text{gap}^2$ from the exchange of a massive spin-$\ell_1$ particle. We assume that, apart from this state, there are no other massive states until the  ``cutoff scale'' $\mu_c M_\text{gap}^2$ for some $\mu_c > 1$:
\be
\label{basicSpec}
\raisebox{-8.5mm}{\includegraphics[width=6cm]{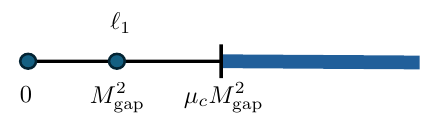}}
\ee
In \reef{basicSpec}, the circle at $s=M_\text{gap}^2$ indicates the simple pole in the $s$-channel of $A(s,u)$ while the thick blue line starting at $\mu_c M_\text{gap}^2$ indicates that we are completely agnostic about what occurs at or above the ``cutoff scale'' $\mu_c M_\text{gap}^2$: it can be poles, branch cuts, or both, so long as it is unitary. 

We first consider the choice of spin $\ell_1$. When $\ell_1 = 1,2,3,4,5$, we find that for $\mu_c > 1$ there is no effect from adding in the spin state at the mass gap: the bounds will be the same as if no states were at the mass gap at all. Further, when we look at the maximal allowed relative couplings $|g_{\ell_1}|^2/a_{0,0}$ for an isolated state with spin $\ell_1 > 0$, we find that they decrease exponentially with $k_{\max}$, corresponding to including constraints from higher derivative operators. That selects the scalar as the only well-motivated choice for the lowest-mass state.

Setting $\ell_1 = 0$, the bounds depend on the choice of cutoff scale $\mu_c$. 
When $\mu_c \to 1$, we recover the full allowed parameter space, the teal region in Figure \ref{fig:a10a20scalarinp}. However, as we increase $\mu_c$, the allowed region shrinks in size. 
Thinking of $M_\text{gap}^2$ as $1/\alpha'$, we make the string-inspired choice $\mu_c = 2$, so that the input is 
\be
\label{stringspect1}
\raisebox{-8.5mm}{\includegraphics[width=6cm]{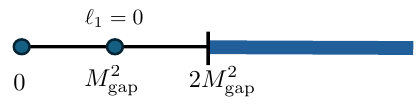}}
\ee
The resulting allowed region, shown in purple in Figure \ref{fig:a10a20scalarinp}, has a new corner at
\be
  \bar{a}_{1,0} \approx 0.7336\,, 
 ~~~~
  \bar{a}_{2,0} \approx 0.6598 \,.
\ee 
(These are the values for $k_\text{max} =14$; the change from $k_\text{max}=12$ to 14 is of order $10^{-4}$.)
The corner values are close to the values for the Veneziano amplitude \reef{scalarVenezianoExp}, 
\be
 \label{Venezianoa10a20}
 \bar{a}^\text{str}_{1,0} = \frac{\zeta_3}{\zeta_2} \approx
 0.7308\,,
 ~~~~
 \bar{a}^\text{str}_{2,0} = \frac{\zeta_4}{\zeta_2} \approx
 0.6580\, ,
\ee
shown as a red dot in Figure \ref{fig:a10a20scalarinp}.
Thus, we have found a new corner on the EFT bounds very close to the Veneziano amplitude! The only string-motivated input was the choice of gap $\mu_c=2$ in \reef{stringspect1}.

\vspace{2mm}
{\bf Single-State Input With Coupling.} 
The lowest-mass state of the  Veneziano amplitude is a scalar that couples to the massless states with (normalized) coupling 
\be
   \label{strcoup}
   \bar{g}_0^2
   \equiv \frac{|g_0^2|}{a_{0,0}}
   = \frac{1}{\zeta_2} \,.
\ee
Let us pretend that this is a number that we have `measured'. If we enter this into the EFT bootstrap with spectrum assumptions, how much freedom is left in the EFT? Naively, one might think that fixing the spin and coupling of the lowest state does not amount to much information, but the EFT bootstrap says otherwise.

\begin{figure}[t]
\centering
\includegraphics[width=0.9\textwidth]{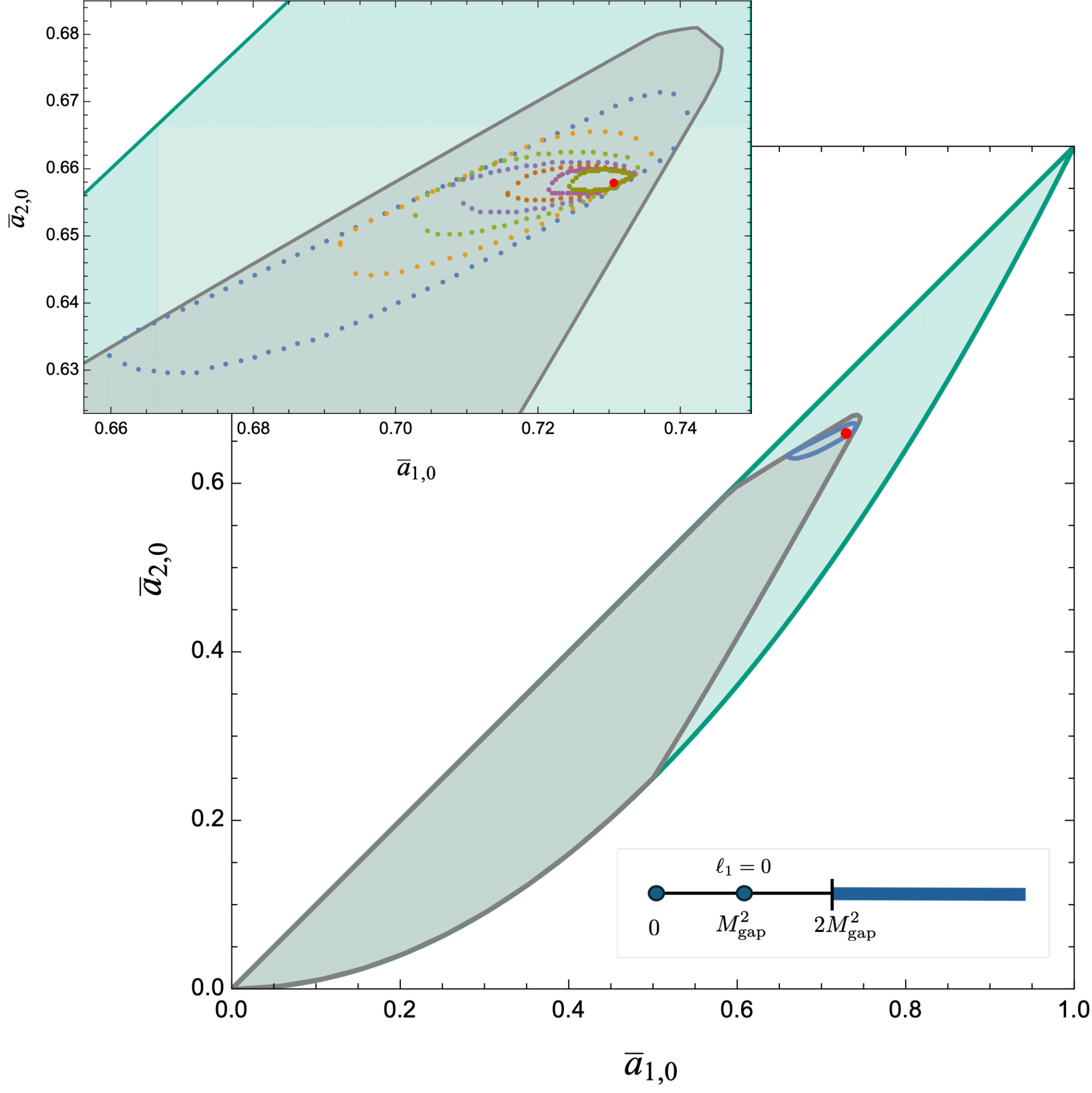}
\caption{Island bounds bootstrapping the Veneziano amplitude (red dot). The gray region shows the $k_\text{max}=4$ bounds without the scalar coupling fixed. The blue island is $k_\text{max}=4$ bounds with the scalar coupling fixed to the string value \reef{strcoup}. The zoomed inset shows that the island bounds continue to shrink for $k_{\max} = 4, 6, 8, 10, 12, 14, 16$, narrowing in on the Veneziano amplitude.
}
\label{fig:SSisland}
\end{figure}

First, we find that the cutoff scale $\mu_c$ above the lowest mass state (cf.~the spectrum \reef{basicSpec}) now has a maximum which we find converges to  $\mu_c = 2$ as $k_\text{max}$ increases (see Figure \ref{fig:maxmu2}). 
Next, choosing this extremal value of the cutoff, $\mu_c=2$, 
we find that the allowed region
becomes a small island around the Veneziano amplitude! This
is illustrated for the $(\bar{a}_{1,0},\bar{a}_{2,0})$ plane in Figure \ref{fig:SSisland}. Moreover, as we increase the value of $k_\text{max}$, the size of the island shrinks, see the zoomed inset in Figure \ref{fig:SSisland} for $k_\text{max} = 4,6,8,10,12,14$, and $16$. (In the main text, we show the allowed islands for other Wilson coefficients.) 
Pushing to $k_{\max}=18$, we find 
\be 
  \label{lubounds}
   0.7261 
   < \bar{a}_{1,0} <    
   0.7333\,,
   ~~~~~~
~~~~~~~
  0.6569 
   < \bar{a}_{2,0} < 
   0.6598\,.
   ~~~~~~
\ee
These values are within about 1\% and 0.5\% of the Veneziano string values \reef{Venezianoa10a20}, respectively. 
Thus, even with the minimal physical input of the lowest mass state, there is very little room for anything else in the maximally supersymmetric EFT.

\begin{figure}[t]
\begin{center}
\begin{tikzpicture}
	\node (image) at (0,0) {\includegraphics[width=0.75\textwidth]{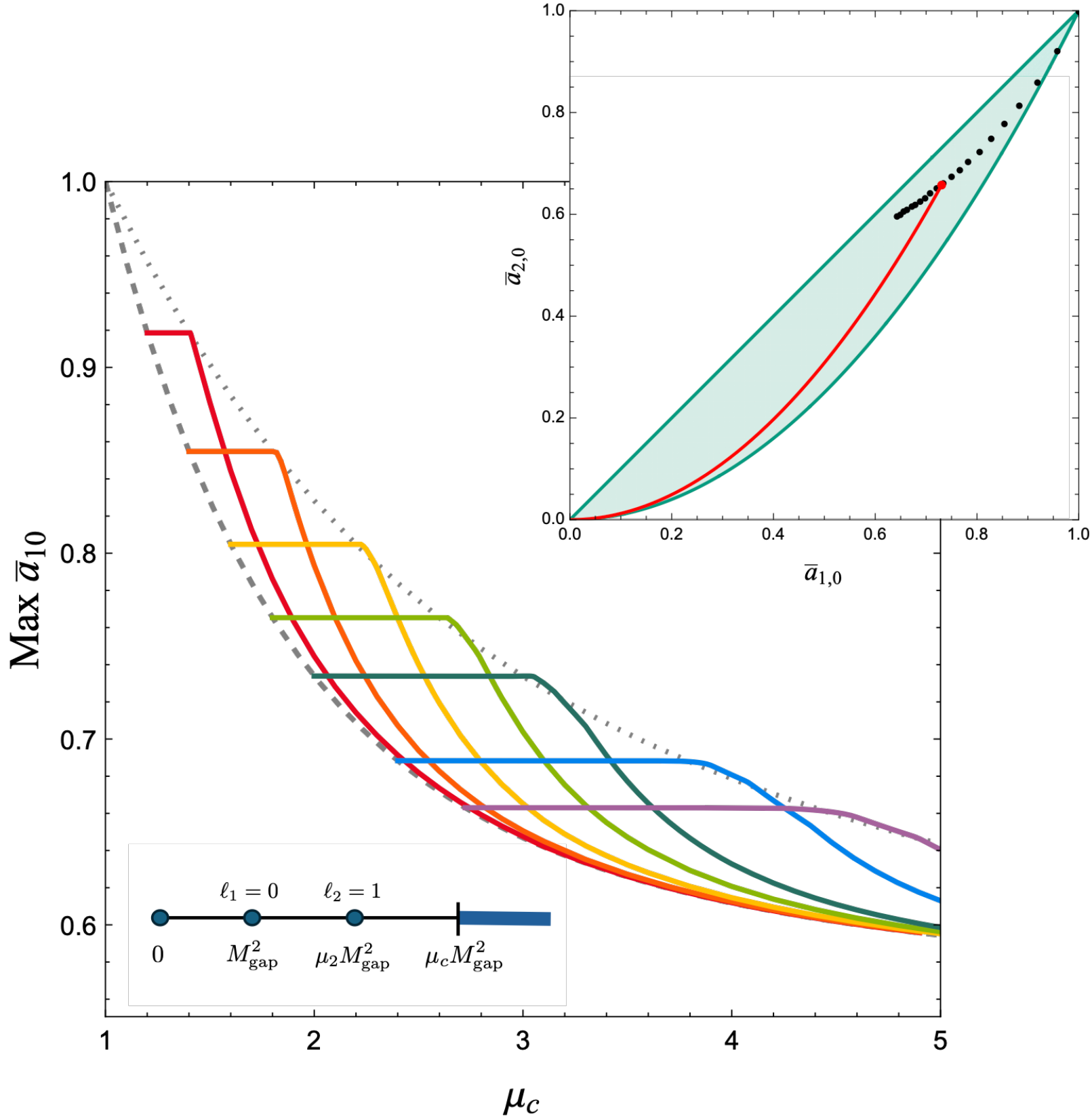}};
        \node at (-3,2.5) {\footnotesize{$\mu_2 = 1.2$}};
        \node at (-2,1.2) {\footnotesize{$\mu_2 = 1.4$}};
        \node at (-1.1,0.2) {\footnotesize{$\mu_2 = 1.6$}};
        \node at (-0.2,-0.5) {\footnotesize{$\mu_2 = 1.8$}};
        \node at (0.55,-1.15) {\footnotesize{$\mu_2 = 2$}};
        \node at (2.3,-1.9) {\footnotesize{$\mu_2 = 2.4$}};
        \node at (3.22,-2.25) {\footnotesize{$\mu_2 = (1.65)^2$}};
        \end{tikzpicture}
\end{center}
\caption{Max($\bar{a}_{1,0}$) vs.~the cutoff scale $\mu_c$ for $\mu_2=1$, $1.2$, $1.4$, $1.6$, $1.8$, $2$, $2.4$ and $(1.65)^2$ computed with $k_\text{max} =10$. Each curve has a corner near $\mu_c = 2\mu_2-1$, suggesting a one-parameter family of models with a state of mass-squared $(2\mu_2-1)M_\text{gap}^2$. 
The dashed line corresponds to the upper bound on $\bar{a}_{1,0}$ with $ \mu_c=\mu_2$, 
while the dotted line shows the max($\bar{a}_{1,0}$) for $\mu_c = 2\mu_2-1$. The inset shows the location of the ``corner theories'' in the $(\bar{a}_{1,0},\bar{a}_{2,0})$-plane for a selection of $\mu_2$ values and with $\mu_c = 2 \mu_2 -1$. The red curve in the inset  corresponds to the Veneziano amplitude with varying $0\leq \a'M_{\gap}^2 \leq 1$ and the red dot is for $\a'M_{\gap}^2=1$. Note that varying $\alpha'M_{\gap}^2$ is distinct from changing the Regge slope by varying $\mu_2$.}
\label{fig:varymu2}
\end{figure}
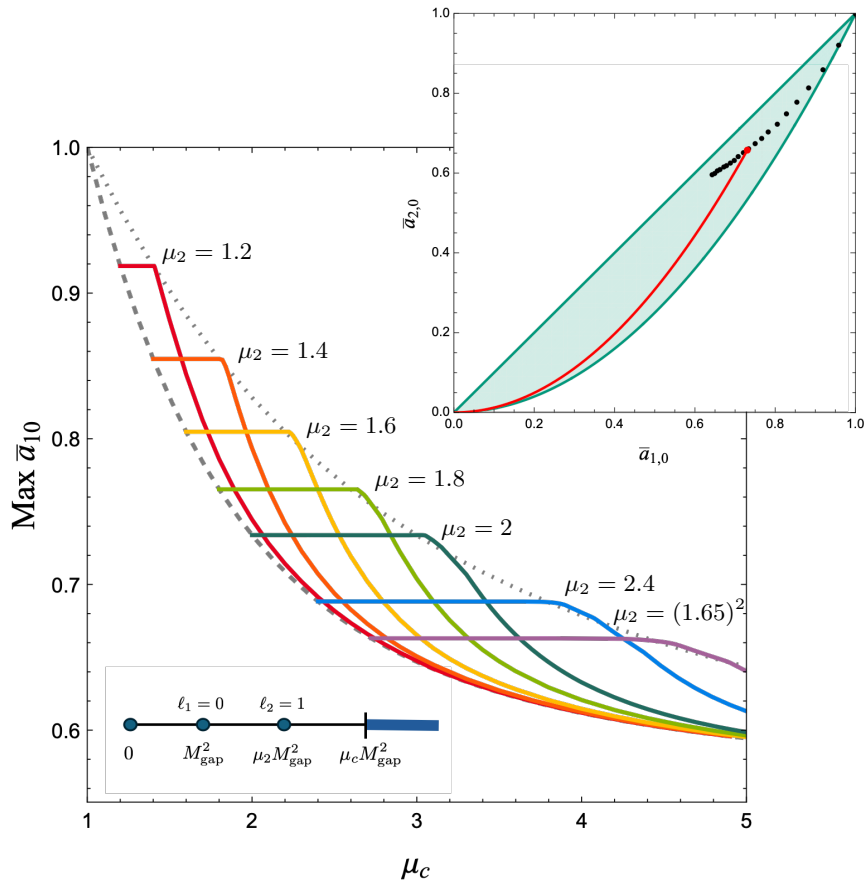

\vspace{2mm}
{\bf Two-States Input.} 
Let us approach the bootstrap of the open string from a different angle that does not require fixing any couplings. 
Consider the input of two massive states instead of one:
\be
\label{our2statespec}
\raisebox{-8.5mm}{\includegraphics[width=6cm]{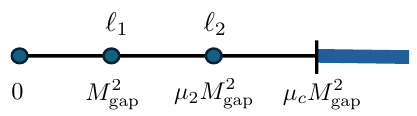}}
\ee
We find that the only non-trivial options for the spins are
\be
 \label{spin12choice}
 \ell_1 = 0 
 ~~~~\text{and}~~~~
 \ell_2 =1
 \,.
\ee
With this choice, we compute, for given choice of  $\mu_2$, 
the maximum allowed value of the Wilson coefficient $\bar{a}_{1,0}$ and find that its value stays constant as $\mu_c$ increases from $\mu_2$ up to $\mu_c \approx 2\mu_2 -1$ where it begins to drop off and reduce the allowed coupling space. This is illustrated in the plot of Figure \ref{fig:varymu2} for a selection of $\mu_2$ choices, including the string-case of $\mu_2 = 2$ shown in dark green. 
The sudden drop in max($\bar{a}_{1,0}$) suggests that when $\mu_c$ is taken larger than $2\mu_2 -1$ some amplitude with a massive state at $(2\mu_2 -1)M_\text{gap}^2$ is ruled out. 
 This would be compatible with a model whose spectrum has a linear Regge trajectory 
\be
 M^2_{\ell} = (\mu_2 -1) \ell + 1\,.
\ee
Thus the bootstrap 
`discovers'  a one-parameter family of potentially interesting models with linear Regge trajectories (at least for  $1<\mu_2 \le 2$ where the corner is quite sharp). The case with $\mu_2=2$ is the 
open string, but in this analysis it is not entirely clear what singles it out among the other ones. Apart from the case of Veneziano with $\mu_2=2$, we do not know a closed form of the generic `corner theory' amplitudes. 

The curves for 
 $\text{max}(\bar{a}_{1,0})$ vs.~$\mu_c$ appear to have another set of corners for 
 $\mu_c > 2\mu_2 -1$ (see Figure \ref{fig:a10max10D4D}.). If any theories `live' there, they would have non-linear Regge trajectories, similar to those of mesons in real-world QCD and in the pion EFT studied in \cite{Albert:2023bml}. If we select $\mu_2$ to take the same value as the mass gap between the $\rho$ and $f_2$ mesons, namely $(1.65)^2$, we find a corner in the bounds (when done in 4D) quite similar to that found in the pion model \cite{Albert:2023bml}.
There are some practical and qualitative similarities between our maximally supersymmetric model and the pion EFT of \cite{Albert:2022oes,Albert:2023jtd,Albert:2023bml} that are likely responsible for this coincidence of numbers. 

\vspace{2mm}
In this work, we have sought to assume only a minimum of well-motivated physical information about the spectrum. The guideline for the input has been to think about what data would be experimentally available had this been a ``real-world'' model. From that perspective, it seems reasonable to assume that the lowest EFT coefficient $a_{0,0}$ can been measured and that a scattering experiment can determine the mass, spin, and coupling $g_0^2$ of the lowest massive state. (This is certainly the case for  pion scattering where much more information about the meson spectrum is also available.) For the maximally supersymmetric YM EFT,
this basic data was sufficient to reduce the allowed space of Wilson coefficients to a small island around the Veneziano amplitude. Such strong constraints may not be found in bootstraps of generic EFTs with less symmetry. Also, experimentally, there would be error bars on the measurements of masses and couplings, so one would have to smear over any islands resulting from the bootstrap to take these uncertainties into account. Nonetheless, one can think of our setup as a very simple toy model for the application of the S-matrix bootstrap to more realistic cases.

\vspace{2mm}
{\bf Outline.} In Section \ref{s:disprep}, we discuss the constraints of maximal supersymmetry, the dispersive representation, and some simple analytic bounds. The input of spectral information is described and put in the context of known UV completions. We also outline the numerical implementation.

Section \ref{s:singlestate}
begins with a bootstrap of the possible spins for the state with the lowest mass in the spectrum. We then discuss the string at the corner of the bounds (e.g.~Figure \ref{fig:a10a20scalarinp}) and  the string islands (Figure \ref{fig:SSisland}.)

Next, in Section \ref{s:MSinput}, we argue that input of specific states can give corners in the bounds of Wilson coefficients and we show how that motivates particular spectrum input for the bootstrap. We examine this first for the  bootstrap of the Veneziano amplitude and next more generally to find the one-parameter of corner theories from Figure \ref{fig:varymu2}. We notice that, for larger mass gaps, the bounds have secondary corners which would correspond to models with non-linear Regge trajectories. We compare one such corner with the pion-bootstrap. 

In Section \ref{s:Regge}, we show that the bounds obtained with one- and two-state input  are very similar to those computed by imposing a fixed leading Regge trajectory on the full spectrum. 

We discuss the results and  outlook in Section \ref{s:disc}. 
The appendices contain  additional details about the numerical implementation (Appendix \ref{app:numerics}) as well as some results of bootstrapping the Veneziano amplitude using  information about the two lowest  massive states and their couplings (Appendix \ref{app:2stateVen}).

\textbf{Note Added:} As this paper was being completed, we learned about similar ideas being pursued in the forthcoming paper \cite{AKRboot} by Albert, Knop, and Rastelli with 
complementary results for the open string bootstrap.


\section{SUSY Constraints and the Dispersive Representation}\label{s:disprep}
In this section, we show how to implement the constraints of maximal supersymmetry and 
generalize the derivation of the dispersive represention in 
\cite{Berman:2023jys} to $D$-dimensions. 
We then describe the general set up for adding the lowest-massive states into the spectrum. Finally we discuss the spectrum and couplings of the massive states exchanged in the open string tree amplitude.

\subsection{Supersymmetry Constraints}
\label{s:review}

In $D=4$ dimensions, maximal super Yang-Mills theory has $\mathcal{N}=4$ supersymmetry and the self-dual supermultiplet of massless states consists of the gluons, the four gluinos, and three pairs of complex scalars. 
The $\mathcal{N}=4$ SUSY Ward identities imply that all 4-point single-trace amplitudes are proportional to each other. Thus, without loss of generality, we can focus on the color-ordered scalar amplitude 
\be
  \label{scalarAmp}
  A(s,u) = A[zz\bar{z}\bar{z}]\,,
\ee
where $z$ and $\bar{z}$ is any  pair of conjugate complex scalars of the  $\mathcal{N}=4$ massless supermultiplet. 
It was shown in \cite{Berman:2023jys}  that the $\mathcal{N}=4$ SUSY Ward identities requires the amplitude \reef{scalarAmp} to take the form 
$A(s,u) = s^2 f(s,u)$, where $f(u,s) = f(s,u)$. This is a {\em SUSY version of crossing symmetry}. 
The most general ansatz for the EFT expansion of this amplitude is then\footnote{For simplicity, we scale all amplitudes to be dimensionless and have no explicit coupling for the leading pole term $-s/u$. Since we only bound ratios of couplings, this has no impact on the results. Also, our  4-point Mandelstam variables are
$s= -(p_1+p_2)^2$, $t = -(p_1+p_3)^2$, and $u = -(p_1+p_4)^2$, treating all momenta as outgoing.} 
\begin{equation}
\label{ansatz}
A^\text{EFT}(s,u)  =
-\frac{s}{u}+s^{2}\sum_{0 \le q\le  k} a_{k,q} \, s^{k-q}\, u^{q} \, 
~~~~
\text{with}
~~~~
a_{k,k-q} = a_{k,q}\,.
\end{equation}
The constraints $a_{k,k-q} = a_{k,q}$ follows from the SUSY crossing condition $f(u,s) = f(s,u)$.
The first term is the gluon-exchange in the $u$-channel of the leading order 2-derivative SYM theory. 
The polynomial terms in the Mandelstam variables $s$ and $u$ are in 1-to-1 correspondence with (linear combinations of) the on-shell $\mathcal{N}=4$ compatible local operators of the schematic form $\tr(D^{2k+4}z^2\bar{z}^2) \xleftrightarrow{\text{SUSY}}
\tr(D^{2k}F^4)$. 
The $a_{k,q}$ are Wilson coefficients for these operators, e.g.~$a_{0,0}$ is the coefficient of $\tr(F^4)$, 
$a_{1,0}=a_{1,1}$ is the coefficient of $\tr(D^2F^4)$ and so on, as outlined in \reef{opcoeff}. The subscript $k$ on $a_{k,q}$ is associated with the derivative order and $q$ labels distinct operators of the same order.

Consider now  $2 \to 2$ scattering amplitudes in maximally supersymmetric Yang-Mills theory in $D>4$ spacetime dimensions. If we restrict the external states to a 4D subspace, the states can be decomposed into the massless $\mathcal{N}=4$ supermultiplet and the amplitudes obey the 4D $\mathcal{N}=4$ SUSY Ward identities. We can then reuse all the above 4D results.

Thus, in $D$ dimensions, we consider a four-point amplitude $A(s,u)$ whose restriction to a 4D subspace gives the scalar amplitude \reef{scalarAmp}. 
The dependence on the spacetime dimension $D$  enters via the 
partial wave decomposition 
\begin{align}\label{partwaveexp}
    A(s,u) = \sum_{\ell = 0}^{\infty}\,n_{\ell}^{(D)}a_{\ell}(s)\,G_{\ell}^{(D)}\Big(1+\frac{2u}{s}\Big) \, ,
\end{align}
in which $a_{\ell}(s)$ is the spectral density, $n_{\ell}^{(D)}$ is a dimension-dependent normalization \cite{Correia:2020xtr,Caron-Huot:2020cmc},
\begin{align}
    n_\ell^{(D)} = \frac{(4\pi)^{D/2} (D+2\ell-3)\Gamma(D+\ell-3)}{\pi\Gamma(\frac{D-2}{2})\Gamma(\ell+1) }
    \, ,
\end{align}
and 
 $G_{\ell}^{(D)}$ are the $D$-dimensional Gegenbauer polynomials, 
which can be written in terms of the hypergeometric function ${}_{2}{F}_1$ as
\be
\label{GegenB}
G^{(D)}_\ell(x)\equiv {}_{2}{F}_1\left(-\ell, \ell + D - 3, \frac{D-2}{2},\frac{1-x}{2}\right)\,.
\ee 
Thus, 
the internally exchanged particles with spin `know' that they `live' in $D$ dimensions.

\subsection{Dispersive Representation}

The EFT S-matrix bootstrap assumptions are  unitarity (specifically positivity since we are working at tree-level), analyticity, and the Froissart-like bounds that $A(s,u)/s^2 \to 0$ for $|s| \to \infty$ at fixed $u<0$ (and similarly at fixed $t<0$).
In addition, we assume a mass gap, large rank of the gauge group (e.g.~only single-trace operators considered), and weak couplings so that any running of the couplings is suppressed and we can  work at tree-level in the EFT. More detailed statements of the assumptions are given in Section 3 of \cite{Berman:2023jys}.

Using a standard contour deformation argument,  it was shown in \cite{Berman:2023jys} that all the Wilson coefficients $a_{k,q}$ of the amplitude \reef{ansatz} have a dispersive representation. Generalizing to $D$-dimensions and redefining all $a_{k,q}$ to be dimensionless, we have
\begin{equation}\label{projectivepro}
a_{k,q}=\sum_{\ell=0}^{\infty}\int_{M^{2}_{\textrm{gap}}}^\infty \frac{dM^{2}}{M^2}\, n_\ell^{(D)}\rho_{\ell}(M^{2})\left  ( \frac{M_{\gap}^2}{M^{2}}\right )^{k+D/2}v^{(D)}_{\ell,q}  \,,
~~~~~
\rho_{\ell}(M^{2}) \ge 0\, .
\end{equation}
Unitarity ensures positivity of the spectral density $\rho_{\ell}(s) = s^{(D-4)/2}\text{Im}\big(a_{\ell}(s)\big)$. 
In $D=4$, the
$v^{(4)}_{\ell,q}$ are the coefficients of the Legendre polynomials $P_{\ell}\big(1+2\delta\big)=\sum_{q=0}^{\ell}v_{\ell,q}
\delta^{q}$. In general $D$ dimensions, 
the numbers
$v_{\ell,q}^{(D)}$ are similarly derived from the Gegenbauers \reef{GegenB} as
\be
  v_{\ell,q}^{(D)} = 
  (-1)^q \frac{1}{q!}
  \frac{\partial^q}{\partial x^q}\,
  {_2}F_1\left(
  -\ell, \ell + D - 3, \frac{D - 2}{2}, x
  \right) \bigg|_{x=0} \,.
\ee
Importantly, all $v^{(D)}_{\ell,q}$ are non-negative. Our analysis is in $D=10$, so henceforth we drop the superscript ${}^{(D)}$ with the understanding that $v_{\ell,q}$ are the 10-dimensional coefficients unless otherwise stated.

Changing integration variable to 
$y = M^2/M_{\gap}^2$ in \reef{projectivepro}, we find
\begin{equation}
\label{finalequ}
a_{k,q}\,=\,\sum_{\ell=0}^{\infty}\int_{1}^\infty dy \,f_{\ell}(y) \,y^{-k} v_{\ell,q}  
~=~ 
\Big\<y^{-k} v_{\ell,q}  \Big\>_1 \,
\ ,
\end{equation}
where $f_{\ell}(y) = y^{-(D/2+1)} n_\ell^{(D)}\rho_{\ell}(M_{\gap}^2y) \ge 0$ and we have introduced the compact notation
\be 
 \label{defbracket}
\Big\< Q\Big\>_\mu
= \sum_{\ell=0}^{\infty}\int_{\mu}^\infty dy \,f_{\ell}(y) \,Q(\ell,y) \,.
\ee
If $\mu > 1$, the  dispersion integral starts above the mass gap, i.e.~the lowest mass-scale to enter the dispersive integral is $\mu M_\text{gap}^2$.

\vspace{2mm}
{\bf Null Constraints.} 
The SUSY crossing constraints $a_{k,k-q} = a_{k,q}$ from \reef{ansatz} place constraints on the spectral density. It is further constrained by relations derived from the dispersion relations at constant $t$ (as opposed to constant $u<0$ used to derive \reef{projectivepro}). 
These constraints are collectively referred to as the ``null constraints'' and they play a key role for the numerical computation of the bounds. We provide the explicit expressions for the null constraints in Appendix \ref{app:null}.

\subsection{Simple Analytic Bounds}\label{s:analyticbds}

It follows directly from the integral \reef{finalequ} and the non-negativity of the $v_{\ell,q}$ that  $a_{k,q} \ge 0$ for all $k$ and $q$, and that
\be
   a_{k,q}  \geq a_{k',q}
   ~~~~
   \text{for}
   ~~~~
   k < k' \,.
\ee
Together with the SUSY crossing constraint, $a_{k,k-q} = a_{k,q}$, one can show \cite{Berman:2023jys}
that $a_{k,q} \leq a_{0,0}$ for all $k$ and $q$. Thus, the lowest-dimension operator, $\tr F^4$, has an effective coupling $a_{0,0}$ that, in units of the mass gap $M_\text{gap}^2$, is greater than all other Wilson coefficients. It is therefore natural to bound the higher-derivative couplings relative to $a_{0,0}$: 
\be
  \label{basicInEq}
  \bar{a}_{k,q} \equiv \frac{a_{k,q}}{a_{0,0}}
   ~~~\text{for which}~~~
  0 \le 
  \bar{a}_{k,q}
 \le 1 \,.
\ee
One can derive the analytic bounds  
\cite{Berman:2023jys}  
\begin{align}\label{ak0akp0bd}
    \bar{a}_{k,0}^{k'/k} 
\leq \bar{a}_{k',0} \leq \bar{a}_{k,0}
~~~~
\text{for~~ $k \leq k'$}\,.
\end{align}
The case of $k=1$ and $k'=2$ was given in 
\reef{unibnds} and shown as the teal region in Figure \ref{fig:a10a20scalarinp}.

\vspace{2mm}
\noindent {\bf Scaling of Bounds.} 
The bounds in \reef{basicInEq} and \reef{ak0akp0bd} were derived from the dispersive representation \reef{finalequ}. If we change the lower bound on the dispersion integral from $1$ to $\mu > 1$, i.e.~if we define 
\begin{align}
 \label{basicscaling}
    a_{k,q}^{(\mu)} &= \Big\<y^{-k} v_{\ell,q}\Big\>_{\mu}\,,
\end{align}
then it follows from a scaling of the dispersion integral that 
\begin{align}\label{mu1rescale}
    \max \big(\bar{a}_{k,q}^{(\mu)}\big) = \frac{1}{\mu^{k}}\max \big(\bar{a}_{k,q}^{(1)}\big)
    ~~~~\text{with}~~~~
    \bar{a}_{k,q}^{(\mu)}
    = \frac{a_{k,q}^{(\mu)}}{a_{0,0}^{(\mu)}}
    \,.
\end{align}
The bounds \reef{basicInEq} and \reef{ak0akp0bd} then become
\begin{align}\label{ak0akp0bdmu1}
0 \le \bar{a}_{k,q}^{(\mu)}
\le \mu^{-k}
~~~~\text{and}~~~~
(\bar{a}_{k,0}^{(\mu)})^{k'/k} 
\leq \bar{a}_{k',0}^{(\mu)} \leq \mu^{k-k'}\bar{a}_{k,0}^{(\mu)}
~~~
\text{for $k \leq k'$}\,.
\end{align}
For example, with $\mu=2$ we get
\be
  \label{onehalfa10a20}
  0 \le \bar{a}_{1,0}^{(2)} \le \frac{1}{2}
  ~~~~\text{and}~~~~
  (\bar{a}_{1,0}^{(2)})^2 \le \bar{a}_{2,0}^{(2)} \le \frac{1}{2} \,\bar{a}_{1,0}^{(2)}\,.
\ee
This is relevant for the projection of the allowed region
to the $(a_{1,0},a_{2,0})$ plane 
and is shown as the magenta-colored region in Figure \ref{fig:a10a20scalarinp}.

\subsection{Spectrum Input}\label{s:specinp}

In the Introduction, we discussed the specification of explicit low-mass states in the EFT bootstrap. 
We now detail how this is implemented, reviewing the ideas proposed in \cite{Albert:2022oes,Albert:2023bml}. 
The contribution from a state with spin $\ell_i$ and mass $M_i$
to an $s$-channel pole is captured by the residue of the 4-point amplitude as
\be
  \label{residueG}
  \raisebox{-7mm}{\includegraphics[width=2.8cm]{Figures/poleterm.pdf}}
  ~~~\longrightarrow~~~
 \text{Res}_{s=M_i^2} A_4
 = - 
 |g_{i}|^2 \,G_{\ell_i}\Big(1+ \frac{2u}{M_i^2}\Big)\,,
\ee
where $g_i$ is the coupling between the massive state and the two massless external states $z$.\footnote{Technically, our bootstrap cannot distinguish whether there is a single spin $\ell_i$ particle with coupling $|g_i|^2$ or multiple spin $\ell_i$ particles with couplings that add up to $|g_i|^2$.}

\begin{figure}[t]
	\centering
  \begin{tikzpicture}
  [
    decoration={%
      markings,
      mark=at position 0.5 with {\arrow[line width=1pt]{>}},
    }
  ]
  \draw [help lines,->] (-2,0) -- (6,0) coordinate (xaxis);
  \draw [help lines,->] (0,-1) -- (0,1) coordinate (yaxis);
  \node [right] at (xaxis) {Re};
  \node [left] at (yaxis) {Im};

  \node[red] at (0,0) {$\times$};
  \node[red] at (1,0) {$\times$};
  \node[red] at (2,0) {$\times$};
  \node[red] at (3,0) {$\times$};
  \node[red] at (4,0) {$|$};

  \draw[line width=0.8pt, red] (4,0) [decorate, decoration=zigzag] --(6,0);

  \node[below=0.25cm,red] at (1,0) {$\mu_{1}$};
  \node[below=0.25cm,red] at (2,0) {$\mu_{2}$};
  \node[below=0.25cm,red] at (3,0) {$\mu_{N}$};
  \node[below=0.25cm,red] at (4,0) {$\mu_{c}$};
  \node[below=0.25cm,red] at (2.5,0) {\ldots};

 \node[draw] at (6,1) {$s/M_{\gap}^2$};

\end{tikzpicture}
\caption{The analytic structure of the amplitude $A(s,u)$ with $N$ poles located on the positive real $s$-axis at  $s=\mu_n M_\gap^2$ for $n=1,2,\dots,N$. 
For $s \ge \mu_c M_\gap^2$ we are agnostic about the form of the spectral density.
}
\label{fig:specinp}
\end{figure}
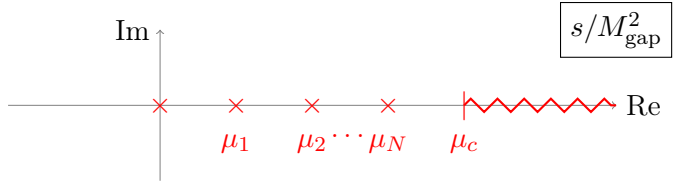

The residue appears in the spectrum as a delta-function, i.e.~
\begin{align}\label{spectocoup}
    (M^2)^{(4-D)/2}\rho_{\ell}(M^2) 
    \,\supset\,
|g_i|^2 \d\Big(M^2-M_i^2\Big) \,.
\end{align}
Thus, including an explicit spectrum up to a given cutoff scale $\mu_c M_\text{gap}^2$ means we can analytically integrate \reef{finalequ} for each Wilson coefficient $a_{k,q}$ up to the scale set by $\mu_c$. For the spectrum of simple poles illustrated in Figure \ref{fig:specinp}, we get 
\begin{align}
\label{inputdisp}
    a_{k,q} = \sum_{n=1}^{N}\sum_{\ell=0}^{\infty}\frac{|g_{\ell,\mu_n}|^2}{\mu_n^{k+3}}v_{\ell,q}+\Big\<y^{-k}v_{\ell,q}\Big\>_{\mu_{c}} \, .
\end{align}
The couplings $|g_{\ell,\mu_n}|^2$ enter the bootstrap as additional parameters in the optimization problem. They can be specified explicitly, extremized, or left as free parameters of the optimization. The numerical implementation is briefly discussed in Section \ref{s:num} with further details relegated to Appendix \ref{app:numerics}.

As an example,
for a spin $\ell_1$ state with coupling $g_0$ at the mass gap, i.e.~$\mu_1 = 1$, and a spin $\ell_2$ state with coupling $g_1$ at $\mu_2$ gives
\begin{align}\label{inputdispN2}
    a_{k,q} = |g_{0}|^2v_{\ell_1,q}+\frac{|g_{1}|^2}{\mu_2^{k+3}}v_{\ell_2,q}+\Big\<y^{-k}v_{\ell,q}\Big\>_{\mu_{c}} \, .
\end{align}
The Gegenbauer coefficients $v_{\ell,q}$ vanish unless $\ell \ge q$. Naively, that means that a spin $\ell_i$ state only contributes to Wilson coefficients with $q \le \ell_i$; however, the SUSY crossing constraints $a_{k,k-q} = a_{k,q}$ makes this more subtle. 
For example, a spin 0 state contributes explicitly to $a_{1,0} = |g_0|^2 + \Big\<y^{-k}v_{\ell,0}\Big\>_{\mu_{c}}$, but also indirectly  to $a_{1,1} = \Big\<y^{-k}v_{\ell,1}\Big\>_{\mu_{c}}$ via the SUSY crossing constraints $a_{1,0} = a_{1,1}$.

The factors of $\mu_n^{-k-3}$  in \reef{inputdisp} and \reef{inputdispN2} arise from the redefined spectral function, $f_{\ell}(y) \propto  y^{-(D/2+1)} \rho_{\ell}\big(M_{\gap}^2y\big)$, the factor of $(M^2)^{(4-D)/2}$ that appears in the relationship between $\rho_{\ell}$ and the couplings \reef{spectocoup}, and the integrand factor of $y^{-k}$ with $y = M_\text{gap}^2/M^2$. Their appearance implies that the contributions from higher-mass states are significantly suppressed, especially for higher values of $k$. This offers some intuition of why the input of the lowest-mass state(s)  has a substantial effect on the bootstrap bounds. We discuss this further in Section \ref{s:singlestate}.

It should be noted that any explicitly input spin-$\ell$ state for the $s$-channel of the amplitude $A[zz\bar{z}\bar{z}]$ is part of a supermultiplet. The other states of that supermultiplet are exchanged in other component amplitudes proportional to $A[zz\bar{z}\bar{z}]$ by supersymmetry.

\subsection{UV Models: Veneziano, IST, and SSE} \label{s:venspec}

\subsubsection*{Veneziano Spectrum and Couplings}

\begin{figure}[t]
\begin{displaymath}
\begin{array}{c|ccccccc}
  n \backslash \ell 
  & 0 & 1 & 2 & 3 & 4 & 5 \\
  \hline
  1 & 1 & 0 & \cdots \\
  2 & 0 & 2 & 0 & \cdots \\
  3 & 0 & 0 & 3 & 0  &\cdots \\
  4 & 0 & \frac{4}{33} & 0 & \frac{128}{33} & 0 & \cdots \\
  5 & \frac{5}{2376} & 0 & \frac{625}{1404} & 0 & \frac{15625}{3432} & 0 & \cdots \\
  \vdots
\end{array}
\hspace{1.3cm}
\raisebox{-3cm}{
\includegraphics[width=6cm]{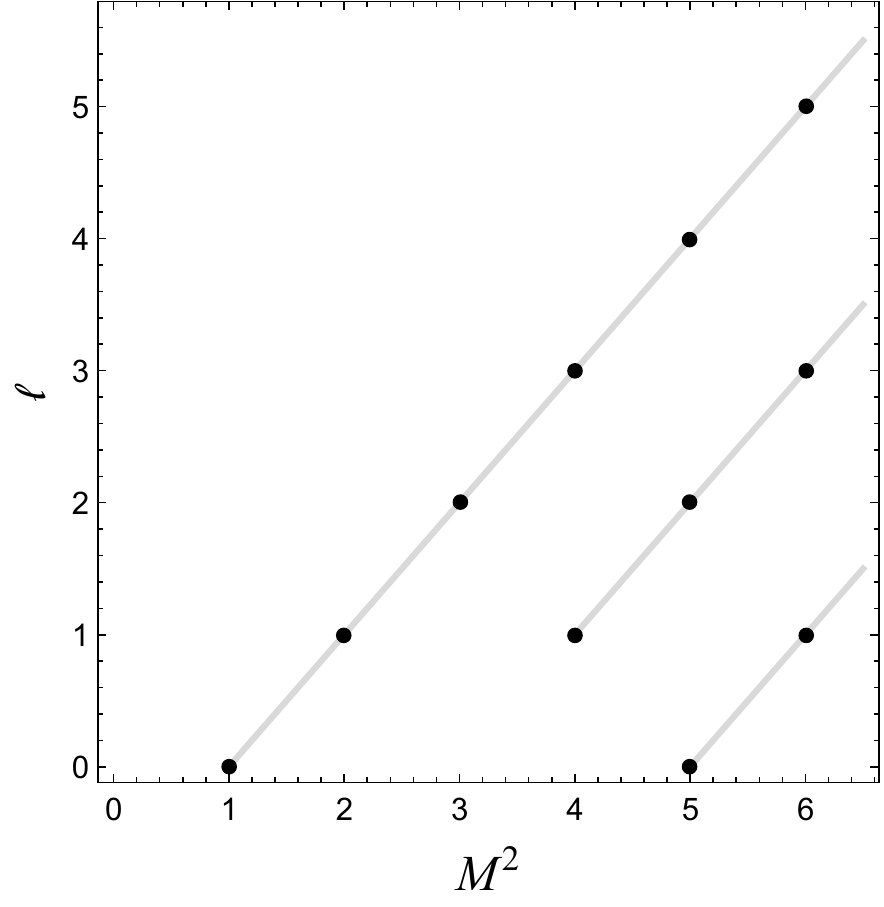}}
\end{displaymath}
\caption{The table shows the 10D values of $|g_{\ell,\mu_n}|^2$, as defined in \reef{residueG}, in units of $1/\alpha'$ for the scalar Veneziano amplitude \reef{scalarVeneziano}. $\ell$ labels the spin and the $n$th level has mass $n/\alpha'$. The plot of $M^2$ vs.~spin shows the Regge trajectories for the Veneziano amplitude. Note that the scalar coupling at mass level $n = 3$ is $\frac{3(10-D)}{8(D-1)\alpha'}$, so it vanishes in $D=10$.}
\label{fig:venSpec}
\end{figure}

Consider the Veneziano amplitude in $D \ge 4$ dimensions. With the external states restricted to a 4D subspace, the Veneziano amplitude can be written 
\be
  \label{venezianoAgain}
   A^\text{str}(s,u) = A^\text{str}[zz \bar{z}\bar{z}]
   = -(\alpha's)^2 \frac{\G(-\a's)\G(-\a'u)}{\G(1-\a'(s+u))}\,,
\ee
which was the form presented in \reef{scalarVeneziano}.
The low-energy expansion of the Veneziano amplitude was given in \reef{scalarVenezianoExp}.

The gamma-functions in the numerator of \reef{venezianoAgain} give simple poles whenever their arguments are 0 or a negative integer. In the $s$-channel, the poles are at $s = m_n^2 =  n/\alpha'$ for $n=0,1,2,3,\dots$, but the $s=0$ pole is eliminated by the overall SUSY factor $s^2$; in $\mathcal{N}=4$ SYM, there is no massless pole in the $s$-channel of the amplitude $A^\text{str}[zz \bar{z}\bar{z}]$ (but there is a gluon exchange in the $u$-channel).

The residues of the first few massive poles show that the exchanged states are as follows: a scalar at $1/\alpha'$, a vector at $2/\alpha'$, a spin 2 particle at $3/\alpha'$ etc. The tower of spin $n$ states at $n/\alpha'$ lie on the first Regge trajectory, illustrated in Figure \ref{fig:venSpec}. In addition, there are towers of daughter trajectories starting at $4/\alpha'$. Also listed in Figure \ref{fig:venSpec} are the couplings $g_{\ell,n}$ of each state as computed via \reef{residueG} for $D=10$.\footnote{The precise value of these couplings depends on the choice of normalization for the Gegenbauer polynomials.}
When we compare bootstrap results to the Veneziano amplitude, we take the mass gap to be given by the lowest-mass state of the Venaziano amplitude, i.e.~$M_\text{gap}^2 = 1/\alpha'$.

\subsubsection*{Infinite Spin Tower (IST)} 
The amplitude 
\be
A_{m^2}^{\text{IST}}(s,u)
= -\frac{s}{u}+\frac{s^2}{\left(m^2- s\right)\left(m^2- u\right)}\, ,
\label{ISTampl}
\ee
 has a infinite tower of spins, all with the same mass $m^2$. This is not expected to be physical, but it is also not ruled out by the universal EFT bounds. With $m^2 = \mu M_\text{gap}^2$, the low-energy expansion gives effective couplings,
\be
  \bar{a}_{k,q}^\text{IST} = \frac{1}{\mu^{k-2}}\, ,
\ee  
for all $k$ and $q$. Thus, in the $(\bar{a}_{1,0},\bar{a}_{2,0})$-plane, the IST amplitudes saturate the lower bound in \reef{unibnds} as they have $\bar{a}_{2,0}=\bar{a}_{1,0}^2$. Hence, the lower bound on the teal region in Figure
\ref{fig:a10a20scalarinp} is matched by IST amplitudes with mass $m^2 = \mu M_\text{gap}^2$. The corner at the $(1,1)$ in that plot is the IST amplitude with $m^2 = M_\text{gap}^2$, whereas the corner at (1/2,1/4) of the purple region in Figure
\ref{fig:a10a20scalarinp} is the IST with $m^2 = 2M_\text{gap}^2$.

\subsubsection*{SUSY Scalar Exchange (SSE)} 
The amplitude 
\be
A_{m^2}^{\text{SSE}}
(s,u)
= -\frac{s}{u}+\frac{s^2}{2M_\text{gap}^2}\left(\frac{1}{m^2-s}+\frac{1}{m^2-u}
\right)\,
\label{SSEampl}
\ee
grows as $s^2$ for $|s| \to \infty$ and therefore just barely fails to satisfy our Froissart bound. A modification of the amplitude (see \cite{Albert:2022oes}) can remedy this and, as such, we can consider $A^{\text{SSE}}$ as a borderline case. 
The low-energy expansion identifies the values of the Wilson coefficients as $a_{0,0}=1$, $a_{k,0}=a_{k,k}=1/(2m^{2(k+1)}M_{\gap}^2)$ for $k>0$, and $a_{k,q}=0$ for $0<q<k$. 
In Figure \ref{fig:a10a20scalarinp}, the SSE amplitude with $m^2 = M_{\gap}^2$ is located at $(\bar{a}_{1,0},\bar{a}_{2,0})=(1/2,1/2)$ on the diagonal upper bound on the allowed region. The rest of the diagonal is the SSE amplitude with mass greater than $M_\text{gap}^2$ (for $\bar{a}_{1,0} < 1/2$) or a linear combination of the SSE and the IST (for $1/2 <\bar{a}_{1,0} < 1$). 

\subsubsection*{Coulomb Branch} 
On the Coulomb branch, the massive states couple quadratically to the massless states, so they cannot enter as tree-level exchange of the 4-point amplitudes, but they contribute via loops. Thus, the simplest Coulomb branch amplitude has a branch cut starting at $4m^2$, where $m$ is the mass of the massive $W$-supermultiplet. Our spectrum input assumes poles 
from the lowest-mass states and explicitly excludes branch cut below the cutoff $\mu_c M_\text{gap}^2$. For that reason, the Coulomb branch does not play any significant role in the bootstrap analysis of this paper.


\subsection{Numerical Implementation}
\label{s:num}
The method for turning the dispersive representation of the Wilson coefficients with positivity bound $\rho_{\ell}(M^2)\geq 0$ into a linear optimization problem has been discussed in detail in the literature; see for example \cite{Caron-Huot:2020cmc,Albert:2022oes,Berman:2023jys}. Here, we briefly describe the essential components so we can fix notation, following Section 4 in \cite{Berman:2023jys}. To start, we allow only a finite number of spins contained in a spin vector $\vec{\ell}$ to appear in our dispersive sum and write a vector equation
\begin{align}
\vec{V}&=\sum_{\ell\in \vec{\ell}}\int_{0}^{1} dy \,f_{\ell}(y) \,\vec{E}_{\ell,y} \label{vectorequation} 
\end{align}
where $\vec{V}$ contains, as its first element, $a_{0,0}$, and $a_{k,q}$ as its second element when $\bar{a}_{k,q}$ is the coefficient we wish to extremize. The subsequent  entries contain any expression we wish to set to zero, that is, $a_{k',q'} - Ra_{0,0}$ to fix the value fo $\bar{a}_{k',q'}$ to $R$, or $|g_{\ell,\mu_n}|^2 - G a_{0,0}$ fixing $|\bar{g}_{\ell,\mu_n}|^2 = |g_{\ell,\mu_n}|^2/a_{0,0}$  to the value $G$. 
The vector $\vec{V}$ also includes all linearly independent null constraints from SUSY crossing (given explicitly in Appendix \ref{app:null}) up to some $k_{\max}$ which truncates the derivative expansion to $2k_\text{max}+4$ derivative order. Emperically, we find that the bounds only change substantially at even $k_{\max}$, and so do not give bounds at odd values.
The entries of $\vec{E}_{\ell,y}$ are chosen so that \reef{vectorequation} matches the dispersive representation of the coefficients and null constraints. The vertex representation \reef{vectorequation} can brought to the standard form of a linear or semi-definite optimization problem \cite{Caron-Huot:2020cmc,Albert:2022oes,Berman:2023jys}.

As noted, the spin sum has to be truncated to a finite list of spins. Rather than including all spins up to some $\ell_{\max}$, it turns out to be computationally advantageous to 
instead include all spins up to some cutoff between 50 and 200, then use a sparse set of spins that includes both even and odd spins up to some much higher maximum. The bounds depend on the spin list $\vec{\ell}$, which is  chosen empirically for each $k_{\max}$ by ensuring that the bounds do not change more than an acceptable amount (say, less than $\mathcal{O}(10^{-6})$) when 
the maximum spin is increased or the spin list includes a denser set of spins. 
Using this kind of spin vector was initially suggested in \cite{Albert:2022oes} and greatly reduces the computational time needed to  bound coefficients at large $k_{\max}$.


\section{Single State Input}
\label{s:singlestate}
In this section, we consider bounds on amplitudes with a single massive state at the  mass gap and then no other contributions up to a second cutoff scale. We assume the state contributes to the amplitude $A(s,u)$ via a tree-level exchange, as in Section \ref{s:specinp}. 
First, we show that the single state at the mass gap has to be scalar. We then 
 describe how leveraging this information, along with one additional piece of string-inspired input, 
 reduces the allowed space in a way that singles out the Veneziano amplitude as special.

\subsection{Bootstrapping the Spectrum at the Mass Gap} \label{s:bdspex}

Consider a spin $\ell_1$ particle at the mass gap. Using \reef{inputdisp}, 
we have
\begin{align} \label{singleinpdisp}
\raisebox{-7.5mm}{\includegraphics[width=5.5cm]{Figures/InputMassGen.pdf}}
\hspace{1cm}
a_{k,q} = |g_{\ell_1,1}|^2 v_{\ell,q}+\Big\<y^{-k}v_{\ell,q}\Big\>_{\mu_c} \, .
\end{align}
The coupling 
$|g_{\ell_1,1}|^2$ is a variable that can be optimized on the same footing as the Wilson coefficients. 

To examine the different options for the choice of spin $\ell_1$, we compute the maximum allowed value for the coupling of the massive state to the massless states relative to the $\tr F^4$ coupling $a_{0,0}$. Specifically, we maximize 
\be
|\bar{g}_{\ell_1,1}|^2 
= \frac{|g_{\ell_1,1}|^2}{a_{0,0}} \,.
\ee
To enforce that the spin $\ell_1$ state is the only state at the mass gap, we  set the cutoff $\mu_c M_\text{gap}^2$ to be above the mass gap, e.g.~$\mu_c > 1$. 
For non-zero spin, $\ell_1 >  0$,  we find that the maximum allowed value of  $|\bar{g}_{\ell_1,1}|^2$ decreases quickly as the  cutoff $\mu_c$ increases. 
It is expected that all non-zero spin couplings go to zero as $\mu_c \to \infty$ because there is no single-spin $\ell > 0$ exchange amplitude (analogous to \reef{SSEampl}) compatible with the Froissart bound.

\begin{figure}[t]
\begin{center}
\begin{tikzpicture}
	\node (image) at (0,0) {\includegraphics[width=\textwidth]{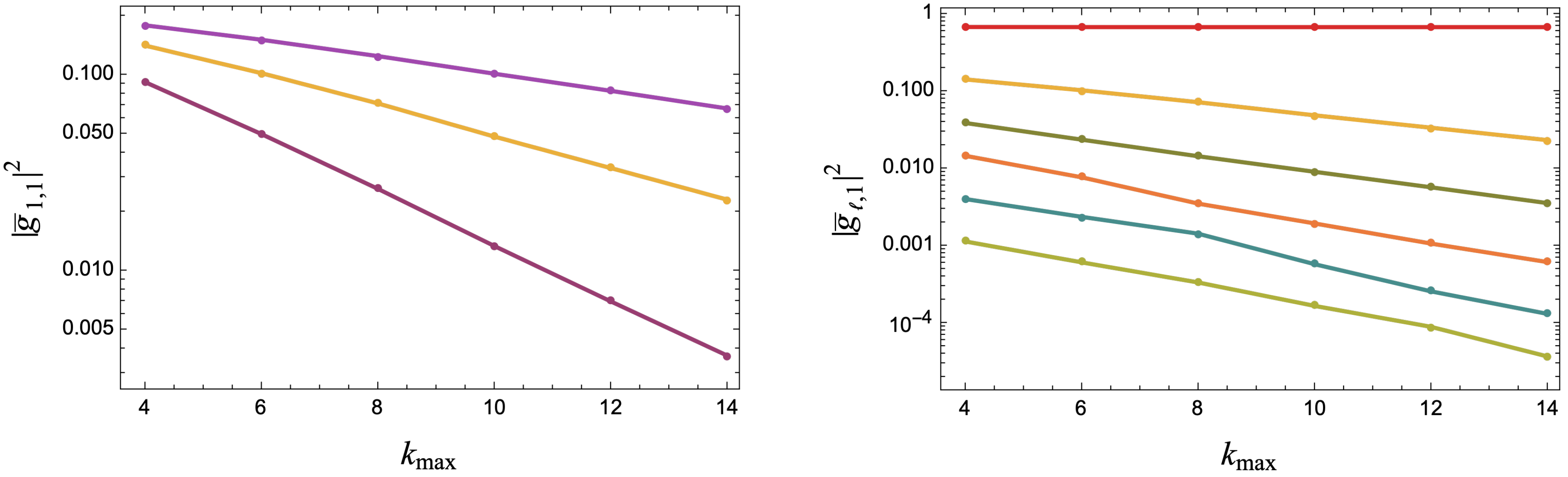}};
        \node [rotate = -8] at (-1.5,1.7) {$\mu_c = 1.05$};
        \node [rotate = -18] at (-1.9,0.4) {$\mu_c = 1.1$};
        \node [rotate = -28] at (-2.2,-0.8) {$\mu_c = 1.2$};
        \node at (6.9,1.8) {$\ell = 0$};
        \node [rotate = -5] at (6.85,1.3) {$\ell = 1$};
        \node [rotate = -8] at (6.7,0.7) {$\ell = 2$};
        \node [rotate = -9] at (6.6,0.15) {$\ell = 3$};
        \node [rotate = -11] at (6.5,-0.35) {$\ell = 4$};
        \node [rotate = -12] at (6.3,-1.1) {$\ell = 5$};
        \end{tikzpicture}
\end{center}
\caption{\textbf{Left:} The maximum of the coupling $|\bar{g}_{1,1}|^2$ of a spin-1 particle at the mass gap with three different values of $\mu_c$ as a function of $k_\text{max}$. \textbf{Right:} The maximum of $|\bar{g}_{\ell,1}|^2$ for the given spins and $\mu_c = 1.1$. Both plots illustrate that a state with spin $\ell_1 > 0$ at the mass gap $M_{\gap}$ has a coupling that is suppressed exponentially with increasing $k_{\max}$. }
\label{fig:spincoup}
\end{figure}
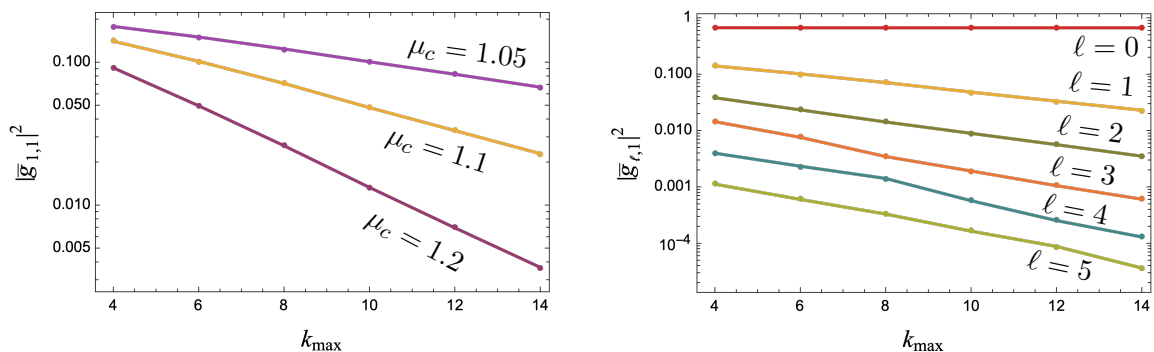

However, even just above the mass gap, e.g.~for $\mu_c =1.1$, we find that $\text{max}\big(|g_{\ell_1,1}|^2\big)$ decreases towards 0 exponentially fast with increasing $k_\text{max}$. This is illustrated in Figure \ref{fig:spincoup} for $\ell_1 = 1,2,3,4,5$, indicating that in the limit of including constraints from arbitrarily high orders in the derivative expansion, $k_\text{max} \to \infty$, having a single state with non-zero spin is not allowed: its coupling to the massless states is exponentially suppressed. This implies that bounds on the Wilson coefficients $\bar{a}_{k,q}$ will not be sensitive to a non-zero spin state at the mass gap.

In contrast, for a spin 0 state, the maximum allowed coupling $|\bar{g}_{0,1}|^2$ is constant as a function of $k_\text{max}$ and has no suppression, as also shown in Figure \ref{fig:spincoup} for $\mu_c =1.1$. Thus, we conclude that if the spectrum only has one state at the mass gap $M_\text{gap}^2$, that state has to be a scalar!

We also checked the  generalized scenario in which we allowed two states at the mass gap, a scalar and either a spin one or two exchange. There, with $\mu_c = 1.1$, the maxima of the allowed couplings for the  spin 1 and 2 states are larger than those given on the right side of Figure \ref{fig:spincoup}, but they are still are decreasing exponentially with $k_{\max}$. 

Studying the scalar coupling as a function of $\mu_c$ shows that the maximal value of $|\bar{g}_{0,1}|^2$ decreases monotonically. At   $\mu_c=1$, 
it takes the value 
$(777-1120\ln(2)) \approx 0.6752$, which matches the scalar coupling of the IST model \reef{ISTampl} and as $\mu_c  \to \infty$, $|\bar{g}_{0,1}|^2$ asymptotes toward $1/2$, which is the normalized scalar coupling of the SSE model \reef{SSEampl}. There are no particular features in that plot that points to any special values of the cutoff $\mu_c$. 

\subsection{Cornering Veneziano}
\label{s:Vcorner}
A single scalar exchanged at the mass gap exactly matches the first massive state at $M_\text{gap}^2 = 1/\alpha'$ in the open string spectrum. We know from Section \ref{s:venspec} there are no other states that contribute to the Veneziano amplitude until $2M_\text{gap}^2 = 2/\alpha'$, so we now impose this string-inspired cutoff $\mu_c = 2$:
\begin{align} \label{scalarinpdisp}
\raisebox{-7mm}{\includegraphics[width=5.5cm]{Figures/SingleMassInput.pdf}}
\hspace{1cm}
        a_{k,q} = |g_{0}|^2 v_{0,q}+\Big\<y^{-k}v_{\ell,q}\Big\>_{2} \, .
\end{align}
Leaving $|g_{0}|^2$ as a free parameter for the maximization/minimization of the Wilson coefficient $\bar{a}_{2,0}$ as a function of $\bar{a}_{1,0}$, we obtain  Figure \ref{fig:a10a20scalarinp}. 
The purple region corresponds to the bounds on the $\bar{a}_{2,0}$ vs.~$\bar{a}_{1,0}$ region with this spectrum. 
The region outside the purple but within the teal bounds is where there must be more than  a scalar in the spectrum below $M^2 = 2M_{\gap}^2$. Moreover, the region within the purple space but outside of the magenta region must have a nontrivial contribution from a scalar at the mass gap $M_\text{gap}^2$.

As discussed in the Introduction, the string values of $a_{k,q}$ are close to the boundary of the allowed region.\footnote{These bounds are computed in 10D. For lower dimensions,  $4 \le D < 10$, the bounds are not as sharp and a bit further from the string. The Veneziano amplitude is unitary only in $\leq 10$ dimensions (for a recent discussion, see \cite{Arkani-Hamed:2022gsa}) and our numerical bootstrap can also exclude $D \ge 11$. Since $D=10$ is also the critical dimension of the superstring, it is natural do carry out the analysis in 10D.} Quantitatively, the maximal values of $\bar{a}_{1,0}$ and $\bar{a}_{2,0}$ differ from the string coefficients by just $2.8\times10^{-3}$ and $2.3\times10^{-3}$ for $k_{\max} \geq 14$.

In Figure \ref{fig:a10a20scalarinp}, the inset shows that, upon zooming in close to the string value, 
the bounds sharpen as we increase $k_{\max}$. Note that the lower bound on $\bar{a}_{2,0}$ appears to be moving extremely slowly with $k_{\max}$. Though the space does become slightly more constrained, there is no clear evidence that it is shrinking fast enough that the string would lie directly on the boundary even as $k_\text{max}$ is taken very large. It is possible that further physics input would be needed to truly corner the string and we discuss examples of this in the following sections. 

We picked the gap  $\mu_c = 2$ solely based on input from the string spectrum. It is natural to ask what happens for other values of $\mu_c$. 
We discuss this further in Section \ref{s:newlinregge}.

\subsection{Veneziano Island} \label{s:strisl}
Next we experiment with fixing the coupling $g_0$ of the massive scalar to the massless states from \reef{scalarinpdisp}. We let the cutoff $\mu_c M_\text{gap}^2$ be general, but pick the string-value (see table in Figure \ref{fig:venSpec}) for the coupling: 
\be
 \label{g0strvalue}
|\bar{g}_0^{\str}|^2 = 
 \frac{|{g}_0^{\str}|^2}{a_{0,0}^\text{str}}
 =\frac{1}{\zeta_2} \,,
\ee 
We then have
\begin{align}\label{dispsscoup}
\raisebox{-8mm}{\includegraphics[width=6cm]{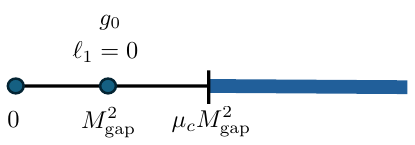}}
\hspace{1cm}
        \bar{a}_{k,q} = \frac{v_{0,q}}{\zeta_2}+\frac{1}{a_{0,0}}\Big\<y^{-k}v_{\ell,q}\Big\>_{\mu_c} \, .
\end{align}

Since $v_{0,q} = 1$ for $q = 0$ and is zero for $q > 0$, only the $a_{k,0}$ coefficients `see' the contribution from the scalar directly.

\begin{figure}[t]
\centering
\includegraphics[width=0.8\textwidth]{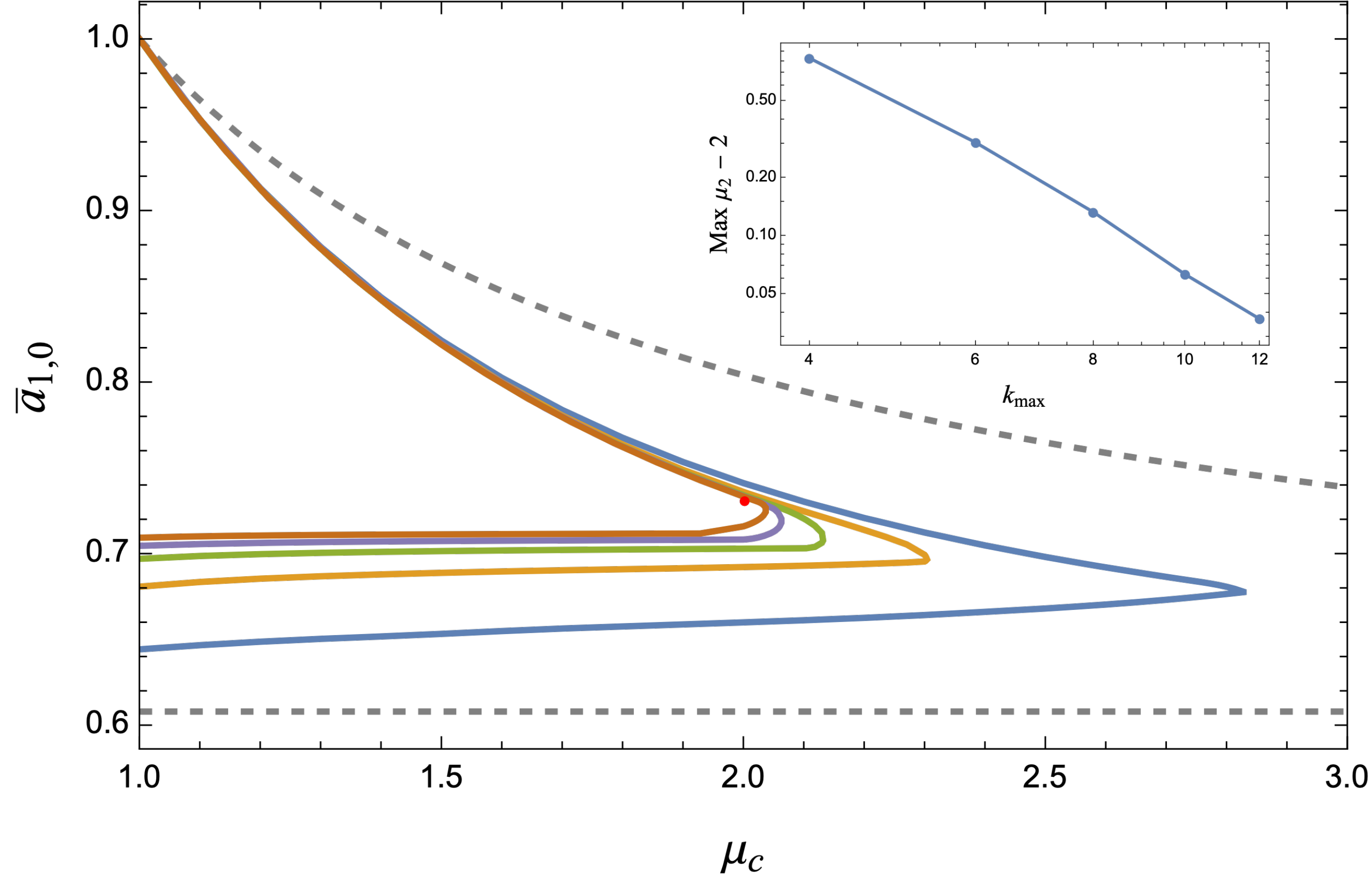}
\caption{$\bar{a}_{1,0}$ vs.~$\mu_2$ with $|\bar{g}_0|^2 = 1/\zeta_2$ for $k_{\max} = 4, 6, 8, 10, 12$. The gray dashed lines are the naive bounds on $\bar{a}_{1,0}$ from \reef{naivecoupbds}. The inset shows on a log-log scale the maximum allowed value of the cutoff $\mu_2$ converging to 2 as $k_{\max}$ is increased. }
\label{fig:maxmu2}
\end{figure}

{\bf Analytic Bounds.} 
Due to this explicit contribution to the $a_{k,0}$ coefficients, fixing the coupling to the string value as in \reef{g0strvalue} gives a non-zero lower bound on 
$\bar{a}_{k,0}$ since the high energy integral in \reef{dispsscoup} must be non-negative:
\begin{align}\label{naivelower}
    \bar{a}_{k,0} \geq \frac{1}{\zeta_2} \approx 0.608 \, .
\end{align}
We can also extract an analytic expression for a new upper bound on $\bar{a}_{k,0}$. 
The high energy integral obeys scaling arguments like those discussed in Section  \ref{s:analyticbds}, i.e.~
\begin{align}\label{ssscale}
{\max}\Big(\Big\<y^{-k}v_{\ell,0}\Big\>_{\mu_c}\Big) = {\max}\Big(\frac{1}{\mu_c^{k}}\Big\<y^{-k}v_{\ell,0}\Big\>_{1} \Big) \, .
\end{align}
Normally, $\frac{1}{a_{0,0}}\Big\<y^{-k}v_{\ell,0}\Big\>_{1}$ would be bounded from above by 1, but because we have already subtracted out a contribution of size $1/\zeta_2$, we must find that in the $\mu_c \to 1$ limit of \reef{dispsscoup},
\begin{align} \label{ssupper}
    \frac{1}{a_{0,0}}\Big\<y^{-k}v_{\ell,0}\Big\>_1 \leq 1-\frac{1}{\zeta_2}\, .
\end{align}
Combining \reef{naivelower} and  \reef{ssscale}, we arrive at the two-sided bounds 
\begin{align}\label{naivecoupbds}
    \frac{1}{\zeta_2} \leq \bar{a}_{k,0} \leq \frac{1}{\zeta_2}+\frac{1-1/\zeta_2}{\mu_c^{k}}.
\end{align}

Thus, for any $\mu_c>1$, the value of $\bar{a}_{k,0}$ is increasingly squeezed from above and below for larger values of $k$ and  
$\lim_{k\to\infty}a_{k,0} = 1/\zeta_2$. Notably, for the Veneziano amplitude, we have
\be 
  \bar{a}_{k,0} = \frac{\zeta_{k+2}}{\zeta_2}
  \to \frac{1}{\zeta_2} 
  ~~~
  \text{for}
  ~~~k \to \infty\,.
\ee
It is rather surprising that simple low-energy input, as  the mass, spin, and coupling of the lowest-mass state in the spectrum has such significant impact on the coupling of high-dimension operators: at large $k$, we find that only the string values of the $\bar{a}_{k,0}$ coefficients are allowed!

{\bf Numerical Bounds.} \
For lower values of $k$, analytic bounds \reef{naivecoupbds} are not as constraining as for higher $k$. However, it turns out that that the SDPB bounds are substantially stronger. 
Figure \ref{fig:maxmu2} shows the upper and lower bounds on $\bar{a}_{1,0}$ as a function of $\mu_c$. 
The analytic bounds \reef{naivecoupbds} are shown for comparison as the dashed curves. 
Importantly, the upper and lower numerical bounds show that there are {\em no solutions} to the bootstrap above a certain $k_\text{max}$-dependent value of the cutoff $\mu_c$. The existence of this maximal cutoff means that there \textit{must} be a state either at or below that cutoff in order for there to be a unitary theory with the chosen $|\bar{g}_0|^2$. The inset gives evidence that the maximum allowed value of $\mu_c$ is 2 in the limit of large $k_\text{max}$. 
We take this maximum to be the ``bootstrap'' choice for $\mu_c$. Thus, by fixing the coupling to the scalar, we have bootstrapped the choice of cutoff $\mu_c=2$ from  Section \ref{s:bdspex}.

\begin{figure}[t]
\centering
\includegraphics[width=\textwidth]{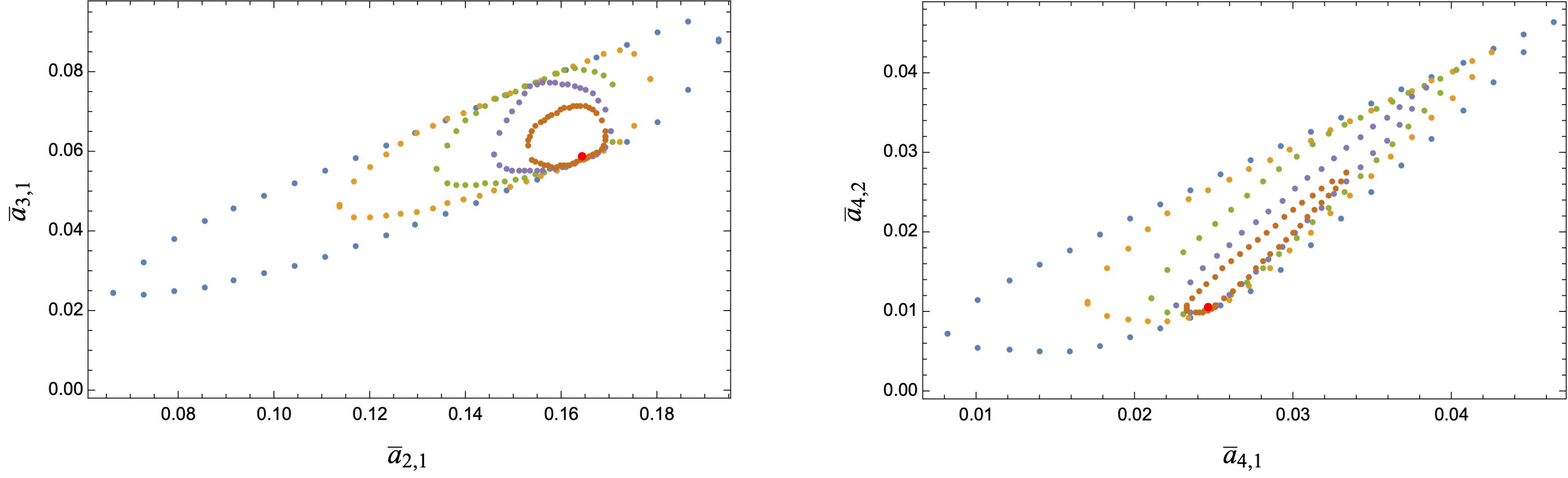}
\caption{Bounds on the allowed regions in the $(\bar{a}_{2,1},\bar{a}_{3,1})$ and $(\bar{a}_{4,1},\bar{a}_{4,2})$ planes for $k_{\max} = 4, 6, 8, 10, 12$. We assume a scalar at $M_\gap^2$ with coupling $|\bar{g}_0|^2 = 1/\zeta_2$ 
and no other state until $2M_\gap^2$.}
\label{fig:othSSisland}
\end{figure}

\begin{figure}[t]
\centering
\includegraphics[width=0.97\textwidth]{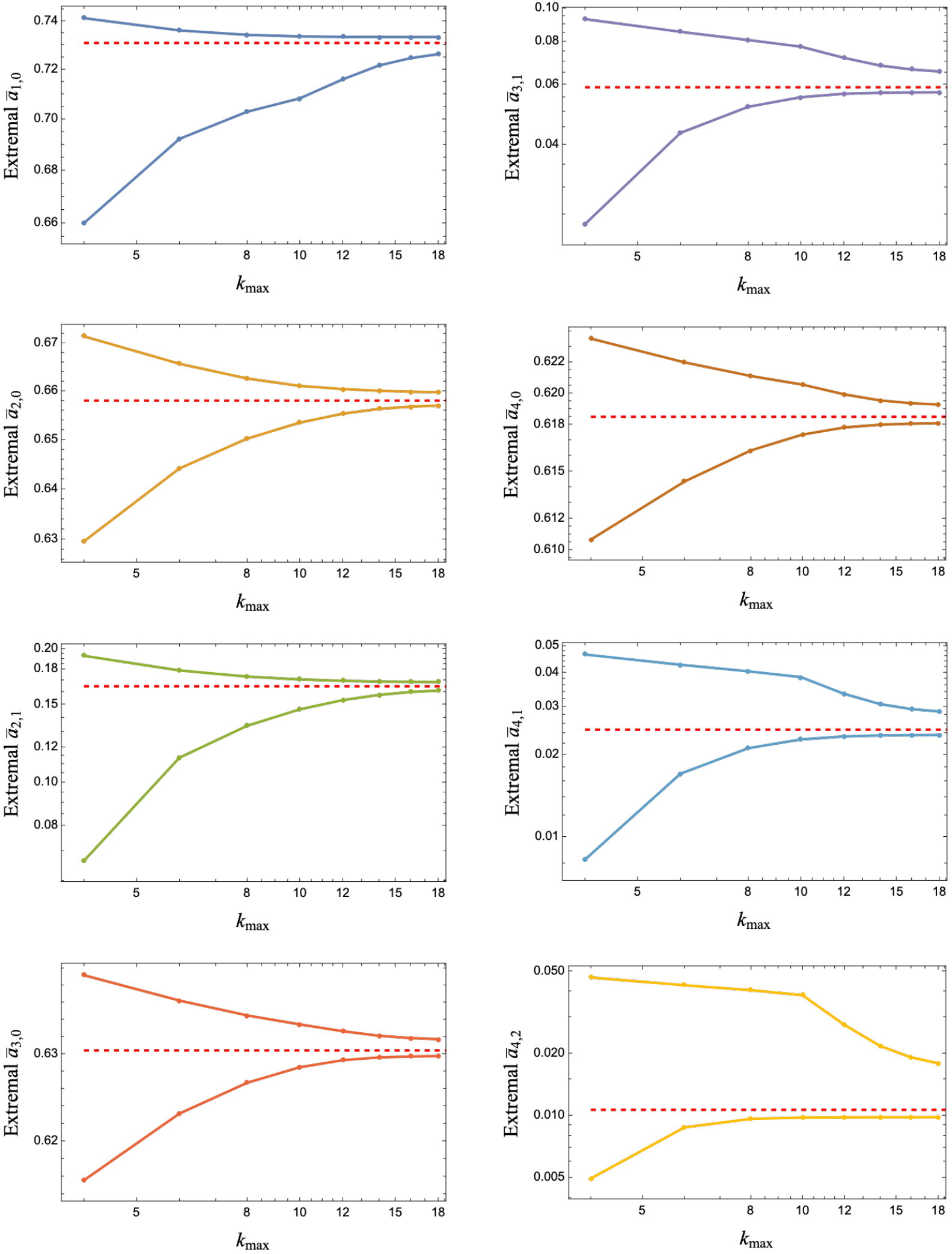}
\caption{Two-sided bounds on the eight lowest-dimension  Wilson coefficients $\bar{a}_{k,q}$ computed with a
single scalar input at $M_\text{gap}^2$, with the coupling $|\bar{g}_0|^2$ fixed to the string value \reef{g0strvalue}, and no other states until $2M_{\gap}^2$. The bounds indicate convergence towards the string values (red dashed lines) as $k_{\max}$ increases, shown here for $k_{\max}$ between $4$ and $18$.
}
\label{fig:extakq}
\end{figure}

Continuing with $\mu_c=2$, we find that the allowed region in the $(\bar{a}_{1,0}, \bar{a}_{2,0})$ coupling space is reduced to the shrinking islands displayed in Figure \ref{fig:SSisland} of the Introduction. As noted, the scalar input affects the only the $\bar{a}_{k,0}$ coefficients directly, but we also find islands in other projections, for example for the $(\bar{a}_{2,1},\bar{a}_{3,1})$-plane and the 
$(\bar{a}_{4,1},\bar{a}_{4,2})$-plane shown in Figure  \ref{fig:othSSisland}.
In all these cases, the allowed coupling region shrinks to an island around the string, much smaller than the region that was allowed by our universal bounds or even the bounds with the naive constraint in \reef{naivecoupbds}. More generally, we find that the upper and lower bounds for all coefficients $\bar{a}_{k,q}$ with $k\le 4$ narrow in on their string values with increasing $k_{\max}$, as illustrated in Figure \ref{fig:extakq}.

The fact that $\mu_c = 2$ appears to be the maximal value for $\mu_c$ for the choice of $|\bar{g}_0|^2 = 1/\zeta_2$ implies that there must be some contribution to the high energy spectrum that lives at $M^2 = 2M_{\gap}^2$. We can determine what that contribution is by performing similar tests to those described by Figure \ref{fig:spincoup}. We assume the most simple input, that there is a single particle exchange of spin $\ell$ at $\mu_2 = 2$, set the cutoff mass to various values slightly above $\mu_2 = 2$, then evaluate the maximal $|\bar{g}_{\ell,2}|^2$. 
We find that at $k_{\max} = 10$, there are no unitary solutions to the optimization problem unless the single particle input is a vector. Once we know that, we can insert the vector, then test whether other particles can live at $\mu_2 = 2$ by again evaluating the maximal $|\bar{g}_{\ell,2}|^2$ (still with the scalar at the mass gap input with its coupling fixed by \reef{g0strvalue}, but with the  vector at $\mu_2 = 2$ having unfixed  coupling). We find strong evidence that the couplings to states with $\ell > 2$ would vanish at large $k_{\max}$. For the scalar, though, the maximal $|\bar{g}_{0,2}|^2$ appears to decrease more slowly at higher $k_{\max}$, at least to the point we tested, so there is no obvious bootstrap-inspired reason to rule out its existence. The Veneziano amplitude has no scalar contribution at $\mu_2 = 2$, so it is not clear the bootstrap can directly determine the Veneziano spectrum without either more input or higher $k_{\max}$ information. In the next section, we study how the low-mass spectrum affects the bounds.

\newpage

\section{Multiple State Input}\label{s:MSinput}

We now take a step back from the Veneziano-centric analysis in Sections \ref{s:Vcorner} and \ref{s:strisl} in order to understand better how state input affects the coupling bounds and selects different ``bootstrap trajectories''.

\subsection{Bifurcation from State Input}
\label{s:bifurcations}
Consider the following 
input to the bootstrap:
\begin{align}\label{2spindisprep}
\raisebox{-7.1mm}{\includegraphics[width=5.5cm]{Figures/IntroSpectrum1.pdf}}
\hspace{0.8cm}
    a_{k,q} = |g_{0}|^2v_{\ell_1,q}+\frac{|g_{1}|^2}{\mu_2^{k+3}}v_{\ell_2,q}+\Big\<y^{-k}v_{\ell,q}\Big\>_{\mu_{c}} \, .
\end{align}
We do not fix the couplings $g_0$ or $g_1$; they are variables in the optimization problem.

The Wilson coefficient $a_{1,0}$ is sensitive to all spins, so we study the maximum allowed value of $\bar{a}_{1,0}$ as a function of $\mu_2$ and the cutoff $\mu_c$. We allow $\mu_c$ to extend all the way down to the mass gap, $\mu_c = 1$, in order to track the effect of the state insertion at $\mu_2$. The results we describe here are computed with $\mu_2 = 2$ but are qualitatively the same for other values of $\mu_2$, as will be discussed further in Section \ref{s:newlinregge}.

\begin{figure}[t]
\centering
\begin{center}
\begin{tikzpicture}
	\node (image) at (0,0) {\includegraphics[width=\textwidth]{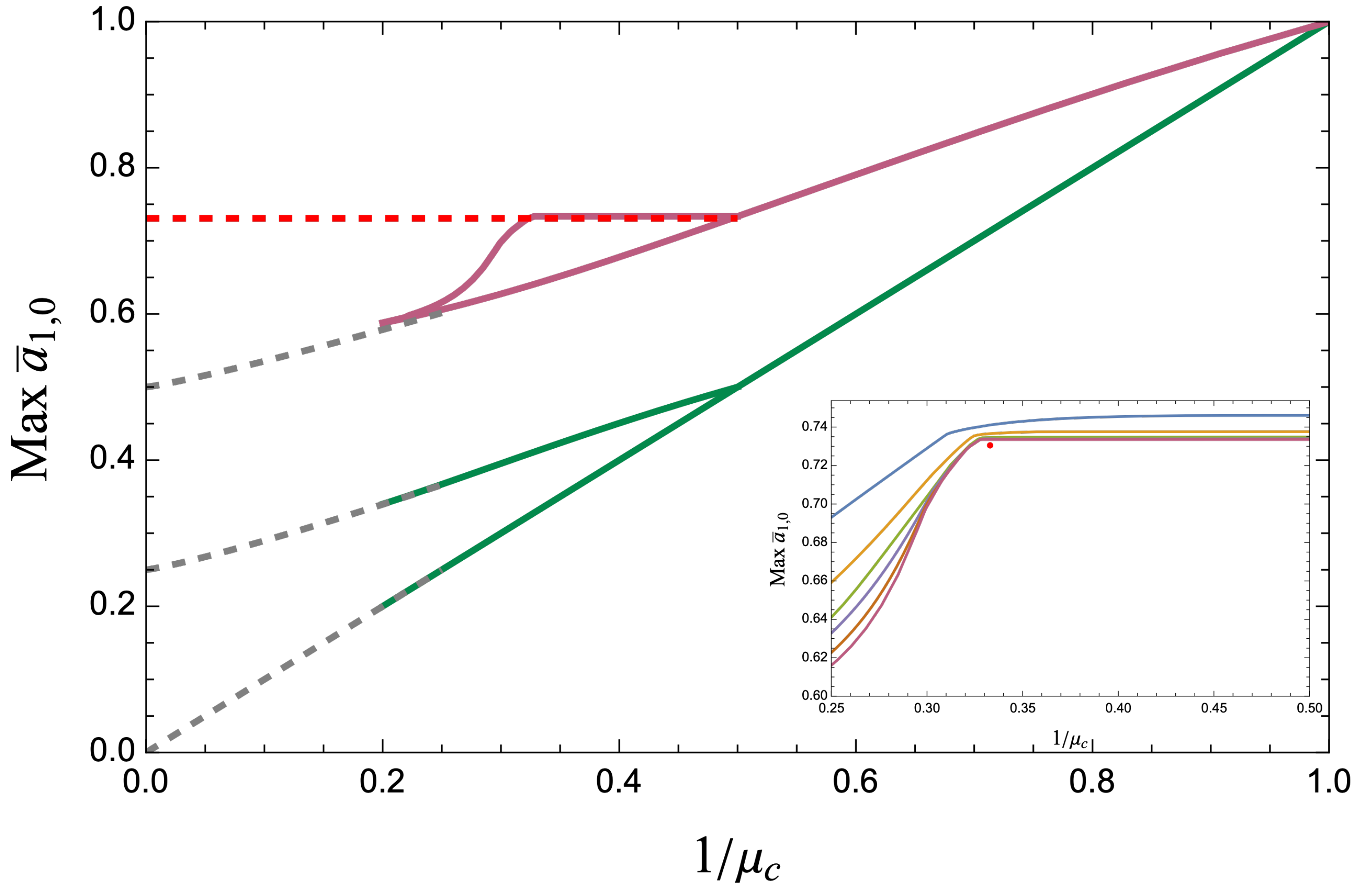}};
        \node [rotate = 15] at (2.2,3.5) {$\ell_1 = 0$};
        \node [rotate = 35] at (2.2,2) {any $\ell_1 \neq 0$};
        \node at (-0.43,2.85) {$\ell_2 = 1$};
        \node at (0.7,3.6) {$\mu_c = 2$};
        \node at (-1.56,3.6) {$\mu_c = 3$};
        \node [rotate = 19] at (-0.7,1.8) {any $\ell_2 \neq 1$};
        \node [rotate = 15] at (-1.2,0.4) {$\ell_2 = 0$};
        \node [rotate = 35] at (-1.2,-0.9) {any $\ell_2 \neq 0$};
        \node at (-4.3,-0.2) {{\small SSE at $2M_\text{gap}^2$}};
        \node at (-4.5,1.65) {{\small SSE at $M_\text{gap}^2$}};
        \draw[->,thick] (-5.2,1.4) -- (-5.75,0.8);
        \draw[->,thick] (0.7,3.4) -- (0.7,2.65);
        \draw[->,thick] (-1.56,3.4) -- (-1.56,2.65);
        \draw[->,thick] (-5.2,-0.5) -- (-5.75,-1.2);
        \node at (-6.5,2.55) {{\small $\zeta_3/\zeta_2$}};
\end{tikzpicture}
\end{center}
\caption{Max($\bar{a}_{1,0}$) vs.~the cutoff scale $\mu_c$ at $k_{\max} = 14$ for listed state insertions. Starting at the upper right corner, the  bound on $\bar{a}_{1,0}$ follows two different trajectories depending on whether a scalar is assumed at $M_{\text{gap}}^2$ or not. A similar bifurcation happens based on what states are allowed at $2M_{\text{gap}}^2$. Near 
$3M_{\text{gap}}^2$ there is a corner in the bounds, indicating that another bifurcation is possible when a specific state is inserted at that point. We tested other spins up to $\ell = 5$ and found results similar to Figure \ref{fig:spincoup}.
The dashed gray lines extrapolate the bounds to $\mu_c\to \infty$ (in the large $k_{\max}$ limit). As indicated, two of these curves are expected to go to the Single Scalar Exchange (SSE) amplitude from Section \ref{s:venspec} at $M_\text{gap}^2$ and $2M_\text{gap}^2$, respectively.}
\label{fig:invertedSH}
\end{figure}
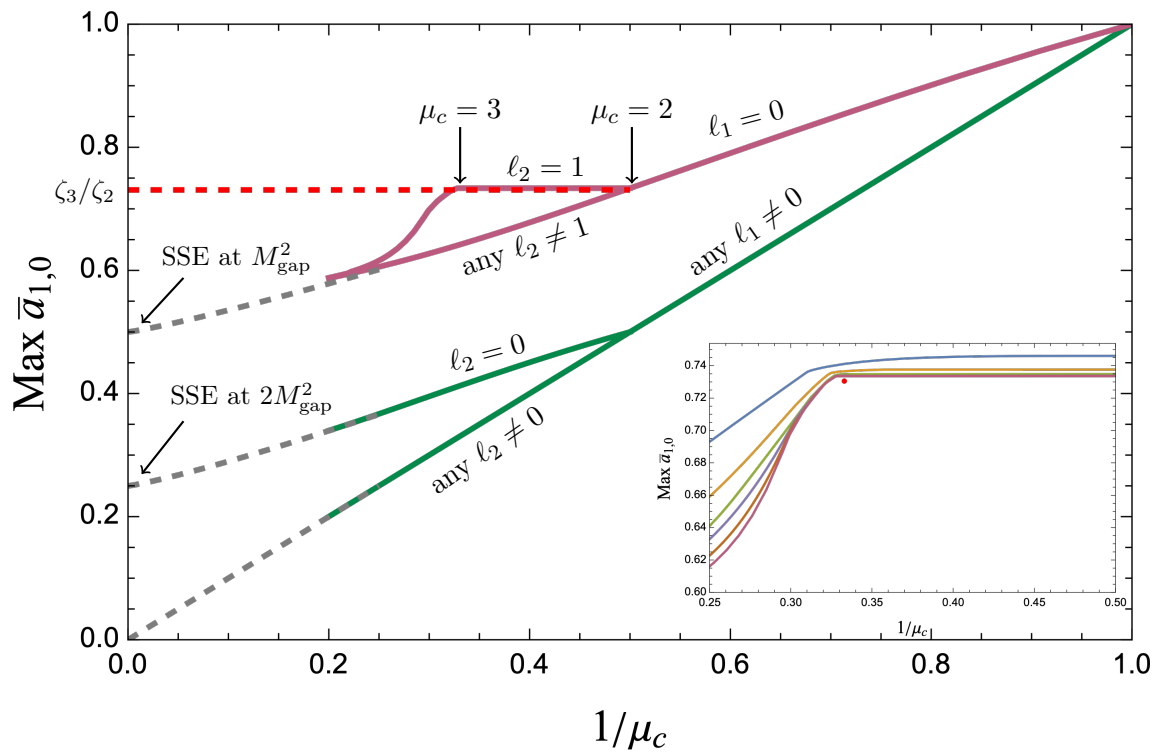

Figure \ref{fig:invertedSH} shows the maximum value of $\bar{a}_{1,0}$ vs.~$1/\mu_c$. We start in the upper right corner where 
with $\text{max}(\bar{a}_{1,0}) =1$ for $\mu_c=1$; this is the maximum from the basic universal bounds \reef{basicInEq}. The diagonal line from (1,1) to (0,0) corresponds to the basic scaling from eq.~\reef{mu1rescale}, which gives $\text{max}(\bar{a}_{1,0}^{(\mu_c)}) = \frac{1}{\mu_c} \text{max}(\bar{a}_{1,0}^{(1)})) = 1/\mu_c$ when there are no states at all below $\mu_c$. This upper bound on $\bar{a}_{1,0}$ is saturated by the Infinite Spin Tower (IST) amplitude from Section \ref{s:venspec}. 

As we increase $\mu_c$, we find two separate paths: one for which the state at the mass gap is a scalar and one when it is not. 
The latter is simply the diagonal in Figure \ref{fig:invertedSH}; hence including a spin $\ell \ne 0$ state at the mass gap is the same as not allowing it at all! This is equivalent to the finding in Section \ref{s:bdspex} that non-scalar input at the mass gap have highly suppressed couplings. In contrast, a scalar at the mass gap gives the higher trajectory in Figure \ref{fig:invertedSH}.

When $\mu_c$ reaches $\mu_2=2$, there is a bifurcation due to the explicit state input at that point. Following first the diagonal path, a new trajectory  splits off for $\ell_2 = 0$ only. This is simply a repeat of the split of paths at (1,1). For the upper path with a scalar at the mass gap ($\ell_1=0$), the bifurcation is more interesting. The maximum of $\bar{a}_{1,0}$ is insensitive to any other state than a vector $\ell_2=1$. When we input a vector state at $\mu_2=2$, the maximum of $\bar{a}_{1,0}$ stays nearly constant\footnote{Numerically, the difference between the maximal $\bar{a}_{1,0}$ at $\mu_c = 2$ and $\mu_c = 3$ is less than $7\times 10^{-6}$ at $k_{\max}=14$ and that difference is even smaller for higher $k_{\max}$.} 
until close to $\mu_c = 3$.  Around $\mu_c = 3$, the maximum suddenly decreases and asymptotes back to the trajectory of having only the scalar input at the mass gap.  
The inset on the lower right of Figure \ref{fig:invertedSH} zooms in on the curve near the corner and illustrates its dependence on increasing $k_{\max}$. It shows that the constant value corner  is nearly saturated by $k_{\max} = 10$. 

The fact that $\text{max}(\bar{a}_{1,0})$ stays  nearly constant and then suddenly decreases is a sign of another potential bifurcation point. As seen at $\mu_2=2$, bifurcations and the resulting corners in the bounds are associated with state inputs. 
However, unlike the other splittings, this new corner  does not occur at a place that we have explicitly inserted a state. Instead, it appears naturally as we increase the cutoff, so it can be interpreted as the bootstrap discovering that a state is  ``missing''  at $3M_{\gap}^2$! 

From the Veneziano amplitude, we know that the missing state at $3M_{\gap}^2 = 3/\alpha'$ is a spin 2 state. For further comparison with the string, 
the dashed red line in Figure \ref{fig:invertedSH} shows the value, $\bar{a}_{1,0}^\text{str}=\zeta_{3}/\zeta_{2}$ for Veneziano. This is close to the nearly flat bound with the scalar and vector input. 
If we include a spin 2 state at $3M_{\gap}^2$, 
then  $\text{max}(\bar{a}_{1,0})$ has to stay above the string value, so this generates a new path which is nearly horizontal until the cutoff reaches the next ``missing'' state. 

One could continue adding states this way to understand which states and where to insert them preserves the nearly horizontal behavior of the  $\text{max}(\bar{a}_{1,0})$. Rather than adding more states in by hand, we now discuss the bootstrap with two states input.

\subsection{Veneziano Bootstrap With 2-State Input}
\label{s:2stVen}

A main take-away from the previous section is that with a scalar at $M_\text{gap}^2$ and a vector at $2M_\text{gap}^2$, the bootstrap tells us we should maximally push the cutoff to  
$\sim 3M_\text{gap}^2$ if no other states are inserted. Setting $\mu_c=3$, we can then proceed to compute the resulting allowed regions for the Wilson coefficients. This gives a somewhat sharper corner in the $(\bar{a}_{1,0},\bar{a}_{2,0})$-plane near the Veneziano amplitude than for the single scalar input in Figure \ref{fig:a10a20scalarinp}. Such plots are shown in Appendix \ref{app:2stateVen} where we also experiment with fixing both the scalar and vector couplings to get even smaller islands than in Figure \ref{fig:SSisland}.

\begin{figure}[t]
\centering
\includegraphics[width=0.9\textwidth]{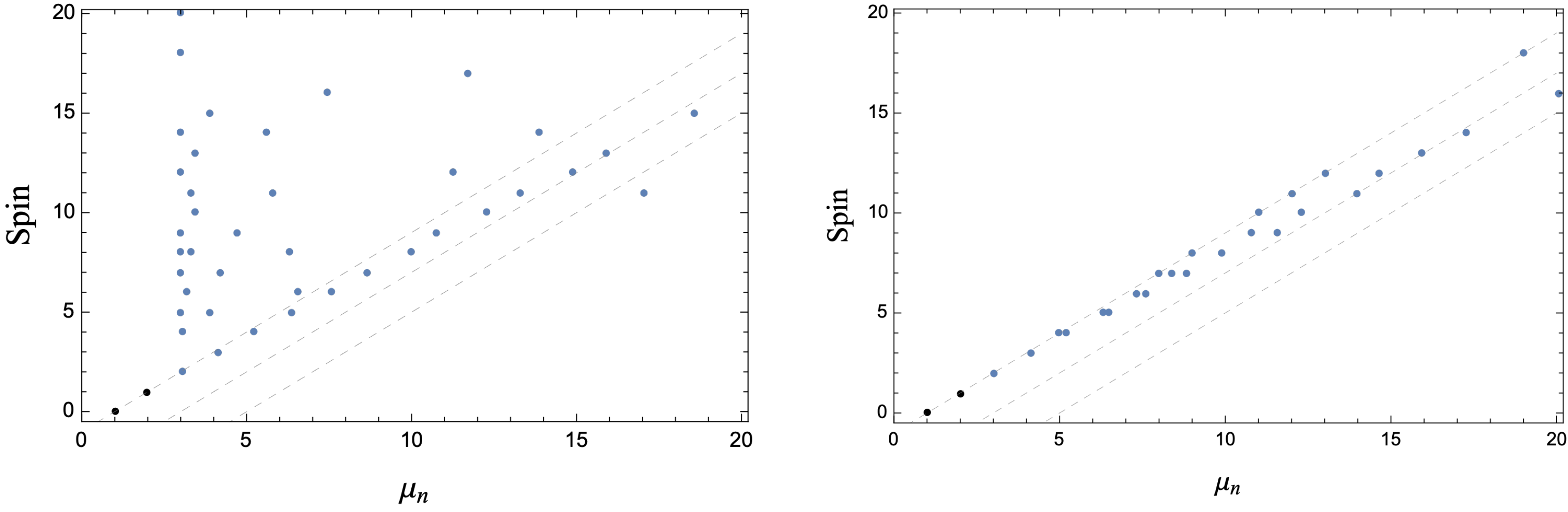}
\caption{The $k_{\max} = 14$ extremal spectrum for the maximal $\bar{a}_{1,0}$ with the string spectrum inserted up to $\mu_c = 3$ with (right) and without (left) additional maximal spin assumptions. The first two states in the string spectrum are enforced (black) and the rest of the spectrum is generated by SDPB (blue). The gray dashed lines correspond to the lowest Regge trajectories of the string spectrum.}
\label{fig:stringspec}
\end{figure}

Rather than using the approach from Figure \ref{fig:invertedSH} to determine higher-mass state insertions, 
we can use the built-in SDPB tool\footnote{This was used in \cite{Albert:2023bml} to find that extremal spectra near corners in an EFT of massless pions at large-$N$ and they found that it looked strikingly similar to the experimental spectrum of QCD for the lowest states, but differed more at higher masses.} that can extract the spectrum for a theory, assuming it has only tree-level exchanges. 
Given that we find the maximal $\bar{a}_{1,0}$ to be nearly flat, we can expect it to track the actual value the bootstrapped model, and hence SDPB should be able to extract the spectrum from maximizing $\bar{a}_{1,0}$. 
As shown on the left in Figure \ref{fig:stringspec}, 
for any $\mu_c$ between two and three, we find that the SDPB spectra all share the same important feature: a spin two state near $3 M_\gap^2$, a spin three state around $4M_{\gap}^2$, and a spin four state close to $5M_{\gap}^2$, following closely the leading string Regge trajectory.
At higher masses, we find that the SDPB states are not quite the string trajectory, but instead there are states that are slightly larger in mass than the string spectrum. 

SDPB does not appear to find the ``daughter'' Regge trajectories. 
Instead, the SDPB spectra tends to have many low-mass states with large spin, i.e.~states that lie below the first ``physical'' linear Regge trajectory and mimic infinite spin tower theories. (This was also observed in \cite{Albert:2023bml}.) These are similar to the states which have maximal couplings that decrease exponentially as we increase $k_{\max}$, as we discussed in Section \ref{s:bdspex}. 
We can \textit{ad hoc} eliminate such ``too high spin'' states by requiring that there be a minimal mass allowed for states with a particular spin, a condition recently studied in \cite{Haring:2023zwu}. In particular, we assume that the spectrum obeys a Regge-like constraint
\begin{align}\label{spincond}
    \ell \leq \g \left(\frac{M^2}{M_{\gap}^2}-1\right) \, ,
\end{align}
where $\g$ is the first Regge slope. Upon imposing this condition, the spectrum, shown on the right hand side of Figure \ref{fig:stringspec}, is much cleaner and has more states along the first Regge trajectory and a few states along the second Regge trajectory, yet it also has many additional states that do not lie at stringy locations. This may be a finite-$k_\text{max}$ effect.


\subsection{New Linear Regge Trajectories}\label{s:newlinregge}

We have seen that we obtain nontrivial constraints from the bootstrap when we insert a scalar at the mass gap $M_\text{gap}^2$ and a spin 1 state at twice that mass gap, i.e.~at $M^2/M_\text{gap}^2=\mu_2 = 2$. However, this choice of mass for the second state is not unique. 

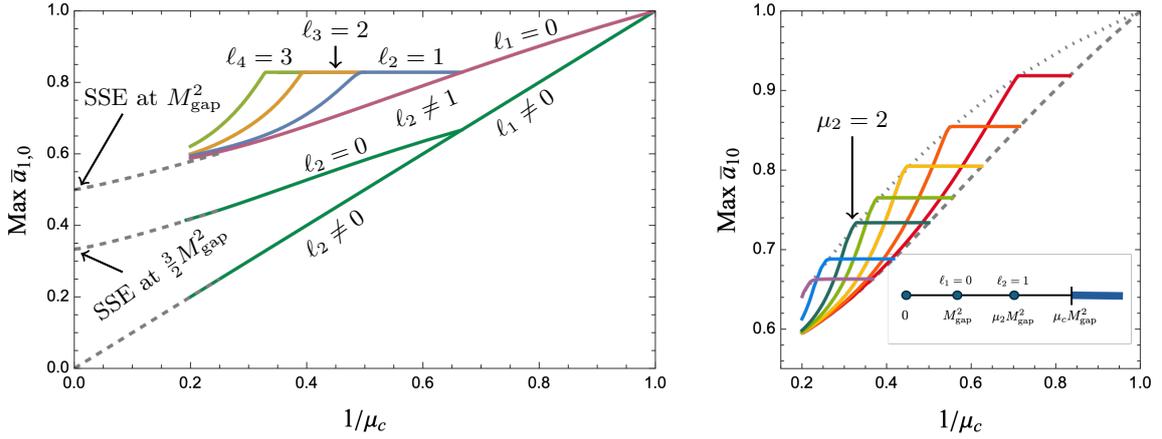
\begin{figure}[t]
\centering
\begin{center}
\begin{tikzpicture}
	\node (image) at (0,0) {\includegraphics[width=\textwidth]{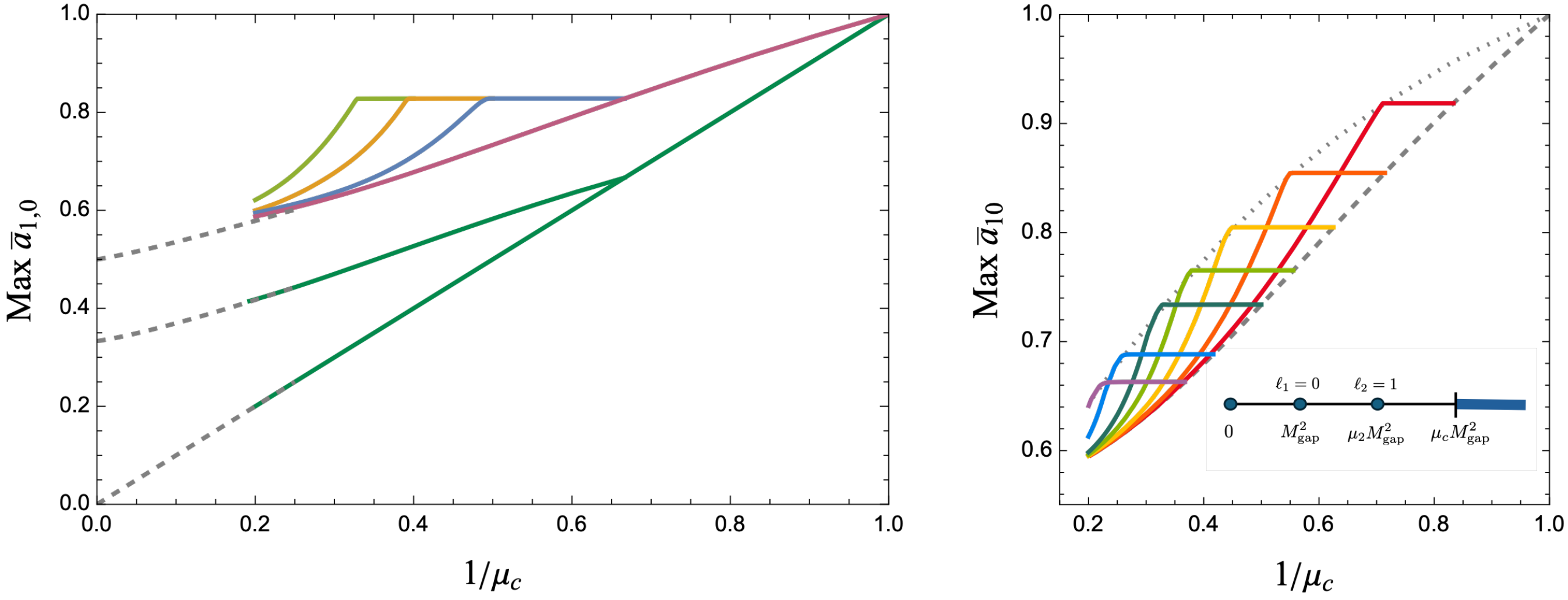}};
        \node [rotate = 15] at (-0.7,2.45) {\footnotesize{$\ell_1 = 0$}};
        \node [rotate = 35] at (-0.7,1.4) {\footnotesize{$\ell_1 \neq 0$}};
        \node at (-2.2,2.15) {\footnotesize{$\ell_2 = 1$}};
        \node at (-3.2,2.5) {\footnotesize{$\ell_3 = 2$}};
        \node at (-4.2,2.15) {\footnotesize{$\ell_4 = 3$}};
        \node [rotate = 23] at (-2,1.45) {\footnotesize{$\ell_2 \neq 1$}};
        \node [rotate = 19] at (-3.2,0.85) {\footnotesize{$\ell_2 = 0$}};
        \node [rotate = 35] at (-3.2,-0.2) {\footnotesize{$\ell_2 \neq 0$}};
        \node [rotate = 35] at (-5.5,-0.7) {{\footnotesize{SSE at $\frac{3}{2}M_\text{gap}^2$}}};
        \node at (-5.6,1.6) {{\footnotesize{SSE at $M_\text{gap}^2$}}};
        \draw[->,thick] (-6,1.4) -- (-6.55,0.45);
        \draw[->,thick] (-6,-0.7) -- (-6.55,-0.45);
        \draw[->,thick] (-3.2,2.3) -- (-3.2,2);
        \node at (3.6,1.3) {\footnotesize{$\mu_2 = 2$}};
        \draw[->,thick] (3.6,1.1) -- (3.6,0);
\end{tikzpicture}
\end{center}
\caption{\textbf{Left:} Max($\bar{a}_{1,0}$) vs. the cutoff mass $\mu_c$ at $k_{\max} = 10$ for  $\mu_2 = 3/2$. As in Figure \ref{fig:invertedSH}, dashed gray lines indicate expected behavior of these curves as $\mu_c\to \infty$ in the large $k_{\max}$ limit. \textbf{Right:} Max($\bar{a}_{1,0}$) vs.~$1/\mu_c$ at $k_{\max} = 10$ for $\mu_2=1$, $1.2$, $1.4$, $1.6$, $1.8$, $2$, $2.4$ and $(1.65)^2$. This figure is the same as Figure \ref{fig:varymu2}, but the horizontal axis is inverted.}
\label{fig:invertedSH3o2}
\end{figure}

As an example, we choose the vector to be at $1.5 M_\gap^2$ instead of  $2 M_\gap^2$, i.e.~pick $\mu_2 = 1.5$. The result of repeating the analysis from Section \ref{s:bifurcations} is shown in  Figure \ref{fig:invertedSH3o2}, which is qualitatively similar to Figure \ref{fig:invertedSH}. The horizontal path now has a corner at $\mu_c = 2$. We find that inserting a spin two state at $2M_\text{gap}^2$ is the choice that allows the horizontal trajectory to continue to be flat for higher $\mu_c$, but now there is a new dropoff at $\mu_c$ near 2.5. Figure \ref{fig:invertedSH3o2} includes the path with a spin 3 state allowed at $2.5M_\text{gap}^2$. This set of states corresponds to a linear Regge trajectory with slope $1/2$ instead of the string choice of 1.

More generally, we can examine the behavior of $\text{max}(\bar{a}_{1,0})$ for the spectrum 
\begin{align}
\raisebox{-7.1mm}{\includegraphics[width=5.5cm]{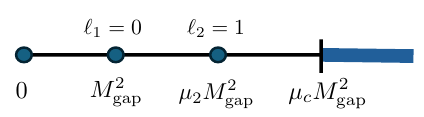}}
\end{align}
as a function of $\mu_c$. In that case, we find corners near $2 \mu_2 -1$. This is illustrated at on the right of Figure \ref{fig:invertedSH3o2}, which is the same as Figure \ref{fig:varymu2}, though with $1/\mu_c$ on the horizontal axis. There is a sharp corner for the lower values of $\mu_2$, but it becomes more rounded as $\mu_2$ increases. 

\begin{figure}[t]
\centering
\includegraphics[width=0.9\textwidth]{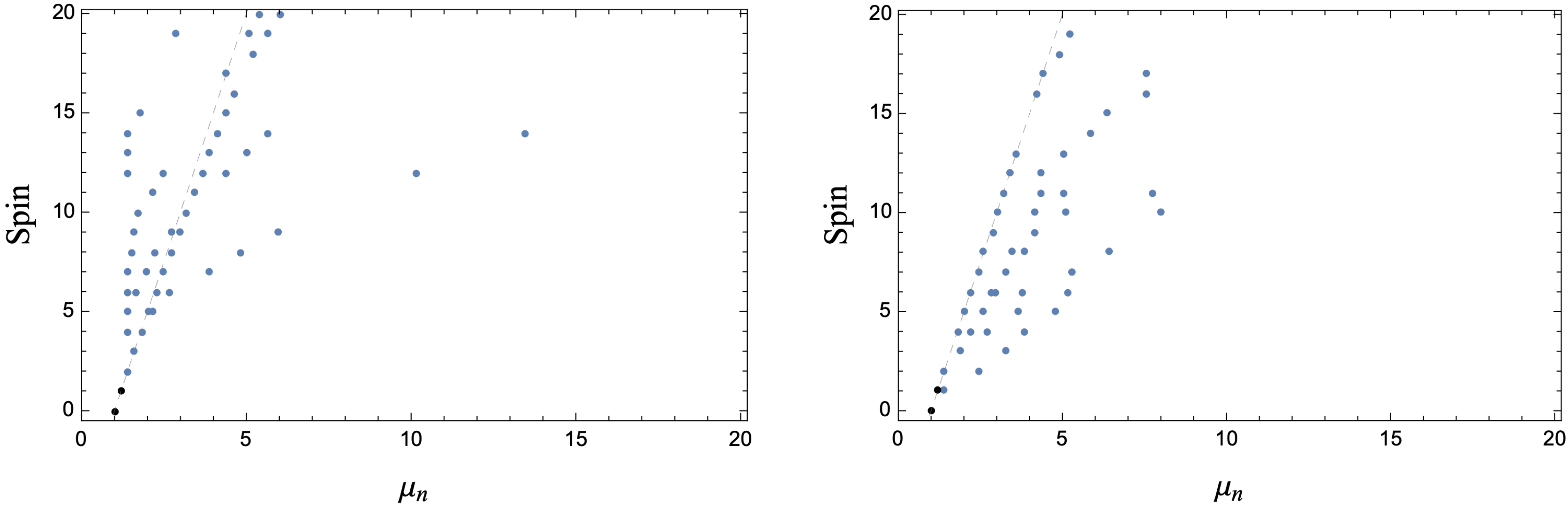}
\caption{The $k_{\max} = 14$ extremal spectrum for the maximal $\bar{a}_{1,0}$ with a scalar at $\mu_1$, a vector at $\mu_2 = 1.2$ and no other state up to $\mu_c = 1.4$ when the maximal spin assumptions are (right) and are not (left) enforced. The first two states are input (black) and the rest of the spectrum is generated by SDPB (blue). The gray dashed line corresponds to a linear trajectory with slope 5.}
\label{fig:nsspec}
\end{figure}

The corner at $2 \mu_2 -1$ suggests, as we saw for the $\mu_2 = 1.5$ case, that some model has a spin 2 state at $(2 \mu_2 -1)M_\text{gap}^2$. This indicates a linear Regge trajectory of the form:
\be
  \label{linReggeFam}
  \frac{M^2_{\ell}}{M_\text{gap}^2} = (\mu_2 -1) \ell + 1\, .
\ee
This family of linear Regge trajectories has a spin $0$ state at the mass gap, a spin one state at $\mu_2$, and a spin two state at $2\mu_2 -1$, and predicts a spin 3 state at $3\mu_2 -2$. 
Extracting the spectrum from SDPB for $\mu_2=1.2$ corroborates this linear Regge trajectory, as shown in the left of Figure \ref{fig:nsspec}. 
As in Figure \ref{fig:stringspec}, the SDPB spectrum contains several other states with high spin. We can input the Regge constraint \reef{spincond}  with $\g = 5$ to match the $\mu_2 = 1.2$ trajectory.
The result is shown on the right in Figure \ref{fig:nsspec}, but it is a bit unclear how exactly to interpret these spectrum plots. 

We briefly turn to the question of which amplitudes may correspond to the 1-parameter family of corner theories. 
Parameterizing them by their Regge slope with $\g = 1/(\mu_2-1)$, we know three explicit cases:
$\gamma=1$ is the Veneziano amplitude. For $\g = 0$, we must have  $\ell\leq 0$ for all finite mass, so only the scalar exchange is allowed in the $s$-channel. This corresponds to the scalar exchange amplitude \reef{SSEampl}.
Finally, for the  $\g \to \infty$ limits, particles of all spin are allowed for any $M^2 \geq M_{\gap}^2$, so the maximal $\bar{a}_{1,0}$ will match the generic bounds, that is, will be $\bar{a}_{1,0} = 1$. The same will be true for all other Wilson coefficients, so the corner theory corresponds to the Infnite Spin Tower amplitude  \reef{ISTampl}. 
The corners in Figure \ref{fig:varymu2} indicates that there could be a 1-parameter family of unitary 4-point amplitudes that connect these three cases. However, for no other choices of $\g$ do we have a closed form expression for the amplitude that corresponds to the corner. 

There has been recent progress in studying general variations on the Veneziano amplitude \cite{Cheung:2022mkw,Geiser:2022exp,Cheung:2023adk,Cheung:2023uwn}, but these amplitudes do not have linear Regge trajectories with slope $\ne 1$ and they are not supersymmetrizable. One possible way forward is something similar to the proposal in \cite{Cheung:2023uwn} for an amplitude with a fully customizable spectrum, but the discussion there is based on the non-supersymmetric open string amplitude and so does not appear to be compatible with our bounds.

Other possible variations involve modifying the Veneziano amplitude in the style of the Lovelace-Schapiro amplitude \cite{Lovelace:1968kjy,Shapiro:1969km,Bianchi:2020cfc}:
\begin{align}
   A(s,u) \sim
    -(\a's)^2\frac{\G(\a_0-\a's)\G(\a_1-\a'u)}{\G(b+\a'(s+u))} \, ,
\end{align}
so the slope is  controlled by $\a'/\a_0$. However, any such simple modification results in an infinite tower of negative norm states or tachyons. 
One could, in principle, try to subtract off the negative norm states from the amplitude, but this would still not necessarily lead to an expression for the amplitude any more exact than trying to read off Wilson coefficients from these corner plots. Further, it is difficult to see how an amplitude with such a form could approach the scalar exchange or infinite spin tower amplitudes in their appropriate limits.


\subsection{Non-Linear Regge Trajectories?}
\label{s:pions}

One interesting feature of the maximally supersymmetric model is that, while it has nothing to do with 4D real-world QCD or the large-$N$ pion EFT, the optimization problems we solve are almost identical to those solved for the pions in \cite{Albert:2022oes,Fernandez:2022kzi,Albert:2023bml}. The large-$N$ pion model has no massless poles and an Adler zero, so, instead of \reef{ansatzIntro}, the generic ansatz for the low-energy color-ordered amplitude is
\begin{align}\label{ansatzPion}
    A^\text{pion}(s,u) = b_{1,0}(s+u) + b_{2,0}(s^2+u^2)+b_{2,1}su+\ldots
\end{align}
Other than the absence of the $a_{0,0}$ coefficient, the amplitude has the same degrees of freedom as \reef{ansatzIntro}. Where the pion amplitude has crossing symmetry $A^\text{pion}(u,s) = A^\text{pion}(s,u)$, the SYM amplitude has the SUSY-induced crossing relation $A^\text{SUSY}(s,u) = s^2 f(s,u)$ with $f(u,s) = f(s,u)$. The null constraints from these crossing relations are equivalent, and, since these numerical bootstrap procedures relies on imposing null constraints, the 
optimization problems 
are mathematically very similar. 
The only technical difference  
is that the additional factor of $s^2$ in $A^\text{SUSY}$ means that we can write convergent dispersion relations for \textit{all} of our Wilson coefficients, while in the large $N$ pion bootstrap, the $b_{k,k}$ coefficient is inaccessible at every level $k$. Therefore, crossing relations such as $b_{k,k} - b_{k,0} = 0$ cannot be imposed. These simple equalities are related to the 
$\mathcal{X}_{k,q}^{\ell,y}$ and $\mathcal{Y}_{k,q}^{\ell,y}$ null relations described in Appendix \ref{app:null}, so while we can impose the relations for all $k$ and $q$, only some can be enforced in the pion model. At $k_{\max} = 2$, for example, crossing symmetry implies the following high energy integrals vanish,
\begin{align}
    \left\langle\mathcal{X}^{\ell,y}_{1,0}\right\rangle \, ,
    \left\langle\mathcal{X}^{\ell,y}_{2,0}\right\rangle \, ,
    \left\langle\mathcal{Y}^{\ell,y}_{0,0}\right\rangle \, , \left\langle\mathcal{Y}^{\ell,y}_{1,0}\right\rangle\, ,
    \left\langle\mathcal{Y}^{\ell,y}_{2,0}\right\rangle\, ,
    \left\langle\mathcal{Y}^{\ell,y}_{2,1}\right\rangle = 0 \,.
\end{align}
In the large-$N$ pion bootstrap, though, only a single linear combination of these null relations has a convergent dispersion relation:
\begin{align}
\left\langle2\mathcal{Y}^{\ell,y}_{2,0}-\mathcal{Y}^{\ell,y}_{2,1}\right\rangle = 0 \, .
\end{align}
In general, we have at least one additional null constraint at every level compared with the pion bootstrap. Importantly for the discussion here, these additional null constraints prevent the ``scalar-subtracted'' versions of amplitudes that are allowed in \cite{Caron-Huot:2020cmc,Fernandez:2022kzi,Albert:2023bml} from always living in our bounds because the $0 = a_{k,k} - a_{k,0}$ null constraints access information about the scalar. Nevertheless, the allowed regions for Wilson coefficients still share many qualitative features.

\begin{figure}[t]
\centering
\includegraphics[width=0.96\textwidth]{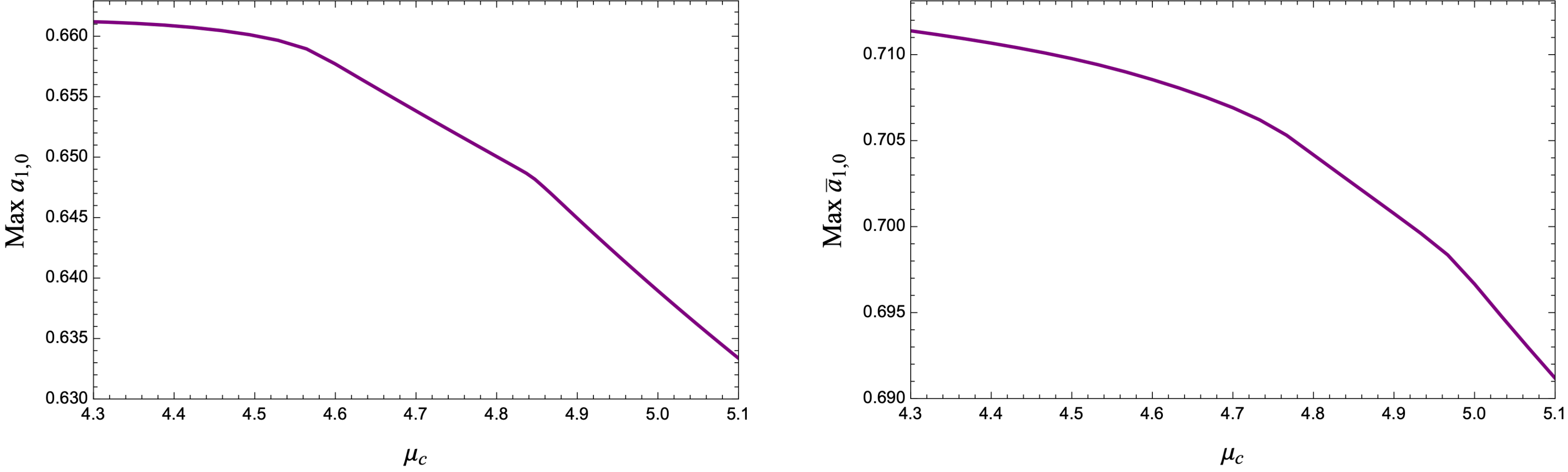}
\caption{Maximum $\bar{a}_{1,0}$ vs.~$\mu_c$ for
$\mu_2 = (1.65)^2$ computed at $k_{\max} = 12$ in 10D (left) and in 4D (right).}
\label{fig:a10max10D4D}
\end{figure}

Albert, Henriksson, Rastelli, and Vichi studied spectrum assumptions in the pion model in  \cite{Albert:2023bml}. Motivated by the experimentally observed meson spectrum, they input as the two-lowest mass states a spin $1$ particle (the $\rho$-meson $m_\rho =770$\,MeV/$c^2$) at $\mu_1=1$ and the $f_2$-meson ($1270$\,MeV/$c^2$) with spin 2 at $\mu_2=(1270/770)^2\approx(1.65)^2$. They find a corner in the maximum of the $f_2$-meson coupling $\tilde{g}_{f_2}^2$ near  
$\mu_c=4.748$. This translates to a mass of $\sqrt{4.748} \,m_\rho \approx 1678$\,MeV/$c^2$, remarkably close to that of the spin 3 $\rho_3$ meson whose mass is $1690$\,MeV/$c^2$. Similarly, an SDPB spectrum calculation gives a few more of the next states near the leading QCD Regge trajectory, see Figure 1 of \cite{Albert:2023bml}. Unlike our corner theories in Section \ref{s:newlinregge}, the QCD meson spectrum is not linear. 

The results of \cite{Albert:2023bml} inspired us to extend our plots beyond $\mu_2 = 2$ and to look carefully at the $\mu_2= (1.65)^2$ case as well. 
We find that the $\text{max}(\bar{a}_{1,0})$ vs.~$\mu_c$ curves
have hints of  two  corners at masses greater than $(2\mu_c -1) M_\text{gap}^2$. 
For $\mu_2=(1.65)^2$, these ``corners'' are located at $\mu_c \approx 4.56$ and $\mu_c \approx 4.85$, neither of which match the pion model's particularly well. Of course, this analysis is in 10D and with spins that are shifted by one compared with those of QCD. Re-analyzing the bounds in $D = 4$, shown on the right hand side of Figure \ref{fig:a10max10D4D}, the first corner
is roughly at $\mu_c = 4.77$, nearly matching the QCD spectrum! The other, more prominent corner moves to higher values, approximately $\mu_c = 4.97$.

\begin{figure}[t]
\centering
\includegraphics[width=0.5\textwidth]{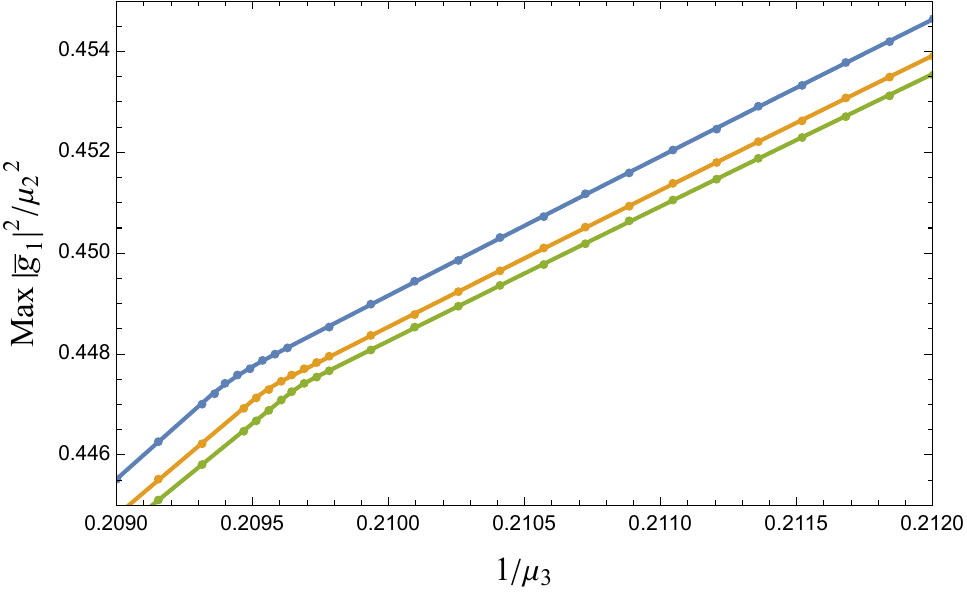}
\caption{Maximum coupling, $|\bar{g}_{1}|^2$, for the vector at $\mu_2 = (1.65)^2$ vs.~$1/\mu_c$ computed at $k_{\max} = 12$, $14$, and $16$ in 4D. The horizontal axis matches Figure 6 of \cite{Albert:2023bml} for direct comparison, and we find a corner at $\mu_c$ within $\sim 0.02$ of theirs. We normalize $|\bar{g}_1|^2$ by $\mu_2^2$ to remove differences in definition from $\tilde{g}_{f_2}$ in \cite{Albert:2023bml}.}
\label{fig:g1max4D}
\end{figure}

To compare more directly, we also plot the maximum of $|\bar{g}_{1}|^2$, the coupling to the vector state at $\mu_2 = (1.65)^2$, in Figure 
\ref{fig:g1max4D}.
There is a clear corner for $\mu_c \approx 4.770$ (for $k_{\max} = 16$). (We do not see any features in the $\text{max}(|\bar{g}_{1}|^2)$ curve associated with the $\mu_c \approx 4.97$ corner of the Figure \ref{fig:a10max10D4D}.) 
The quantity $|\bar{g}_{1}|^2$ is analogous to the 
coupling $\tilde{g}_{f_2}^2$ 
 of the spin-two $f_2$ state of the pion model, shown in Figures 5 and 6 of \cite{Albert:2023bml}. We use the same range of $\mu_c$ as the zoomed-in Figure 6 from \cite{Albert:2023bml}.
At $n_{\max} = 17$ (their equivalent of $k_{\max}$), the corner in the pion bootstrap appears to be at $\mu_c \approx 4.747$, while for our bootstrap the $k_{\max} = 16$ corner is closer to $4.770$. While not exact, the agreement is 
surprising 
for two ostensibly unrelated problems. The values of the maximal couplings near the corner, though, are not clearly directly related. This discrepancy is not unexpected since we are studying different models.

\newpage

\section{Regge Bounds}
\label{s:Regge}

\begin{figure}[t]
\centering
\includegraphics[width=0.9\textwidth]{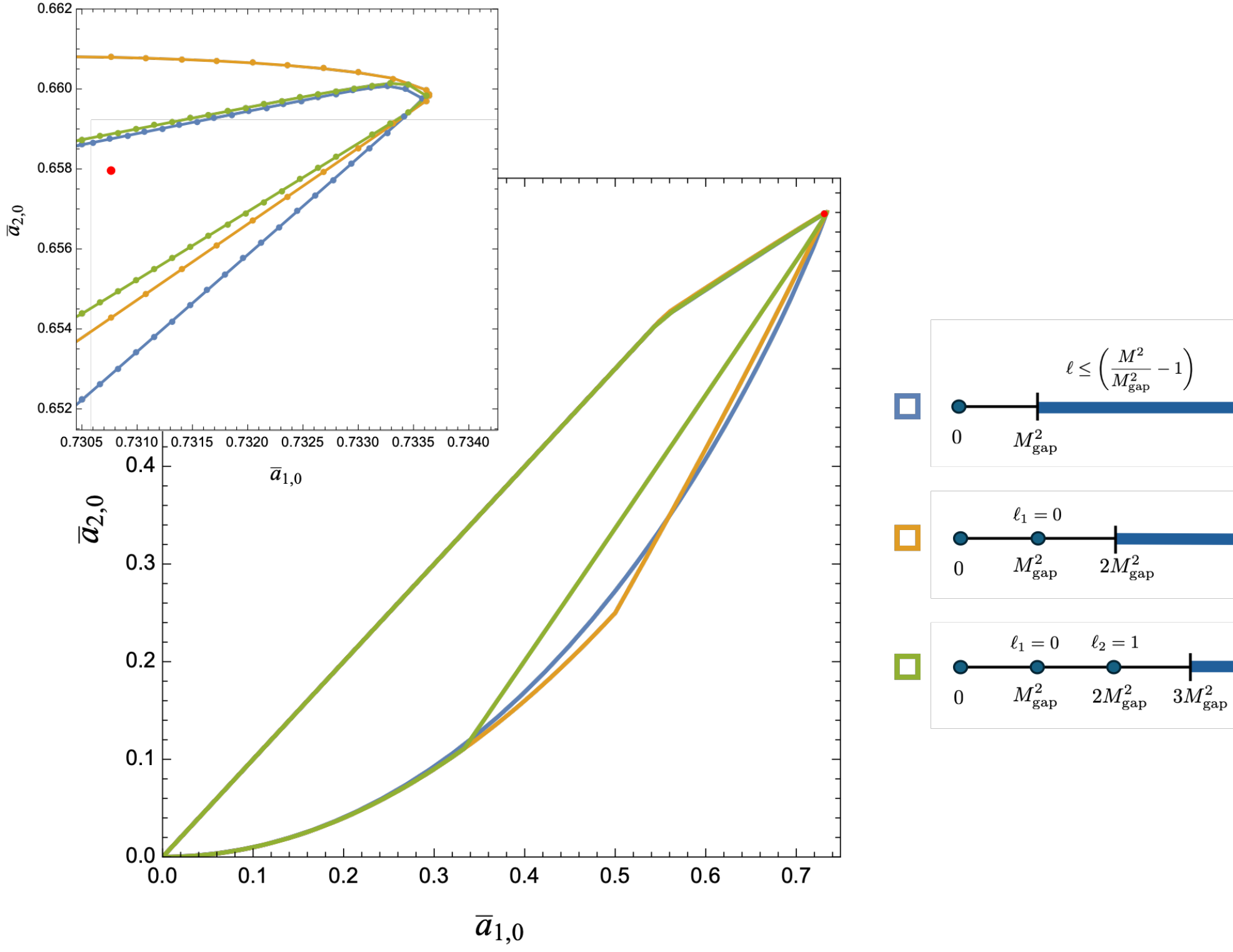}
\caption{Outer bounds on the $k_{\max} = 12~~(\bar{a}_{1,0},\bar{a}_{2,0})$ region with the Regge maximal spin requirement (blue), the scalar input at the mass gap with $\mu_c = 2$ (orange), and the scalar at the mass gap and vector at $\mu_2 = 2$ input with $\mu_c = 3$.}
\label{fig:maxspin}
\end{figure}

Imposing Regge-bounds, such as \reef{spincond}, together with the S-matrix bootstrap constraints has been pursued
for both the closed and open string in 
\cite{Haring:2023zwu}. 
Let us compare the spectrum restrictions resulting from  imposing  Regge slope 1 versus the single-state input at the mass gap $\mu_c =2$:
\be
\begin{array}{lcl}
\text{Regge:}~~
    \ell \leq \left(\frac{M^2}{M_{\gap}^2}-1\right) 
&~\text{vs.}~~~~~~&
    \raisebox{-5mm}{\includegraphics[width=5cm]{Figures/SingleMassInput.pdf}}
\\[2.5mm]
    \text{1.~A state at the mass gap}
    &&
    \text{1.~At the mass gap, couplings }
    \\
    \text{must be a scalar:}
    &&
    \text{of states with }\ell_1 \le 0~\text{are suppressed}
    \\
    M^2 = M_\text{gap}^2 \implies \ell = 0 
    &&
    M^2 = M_\text{gap}^2 \implies \ell_1 = 0 
\\[2.5mm]
    \text{2.~Scalars allowed }
    &&
    \text{2.~No states allowed }
    \\
    \text{for $M_\text{gap}^2 \le M^2 < 2 M_\text{gap}^2$}
     &&
     \text{for $M_\text{gap}^2 <M^2 < 2 M_\text{gap}^2$}
     \\[2.5mm]
     \text{3.~Spin of states with $M^2 \ge 2 M_\text{gap}^2$} 
     && \text{3.~Spin of states with $M^2 \ge 2 M_\text{gap}^2$} 
     \\
     \text{restricted by Regge}
     &&
     \text{unrestricted}
\end{array}    
\ee
A priori, the bounds on Wilson coefficients resulting from these two different sets of constraints do not obviously have anything to do with each other. Surprisingly, we find that they are largely identical. This is illustrated in Figure \ref{fig:maxspin}, which for comparison also includes the bounds from inputting both the scalar state at the mass gap $M_\text{gap}^2$ and the vector at $2M_\text{gap}^2$.

It is surprising that what we consider a rather mild, low-spectrum constraint --- a scalar at $M_\text{gap}^2$  and the string-inspired gap to the next state --- yields essentially the same constraints on the $\bar{a}_{k,q}$ as the imposing the linear Regge behavior at all orders (up to the maximum spin considered for the numerical implementation) of the spectrum.  

\begin{figure}[t]
\centering
\includegraphics[width=0.9\textwidth]{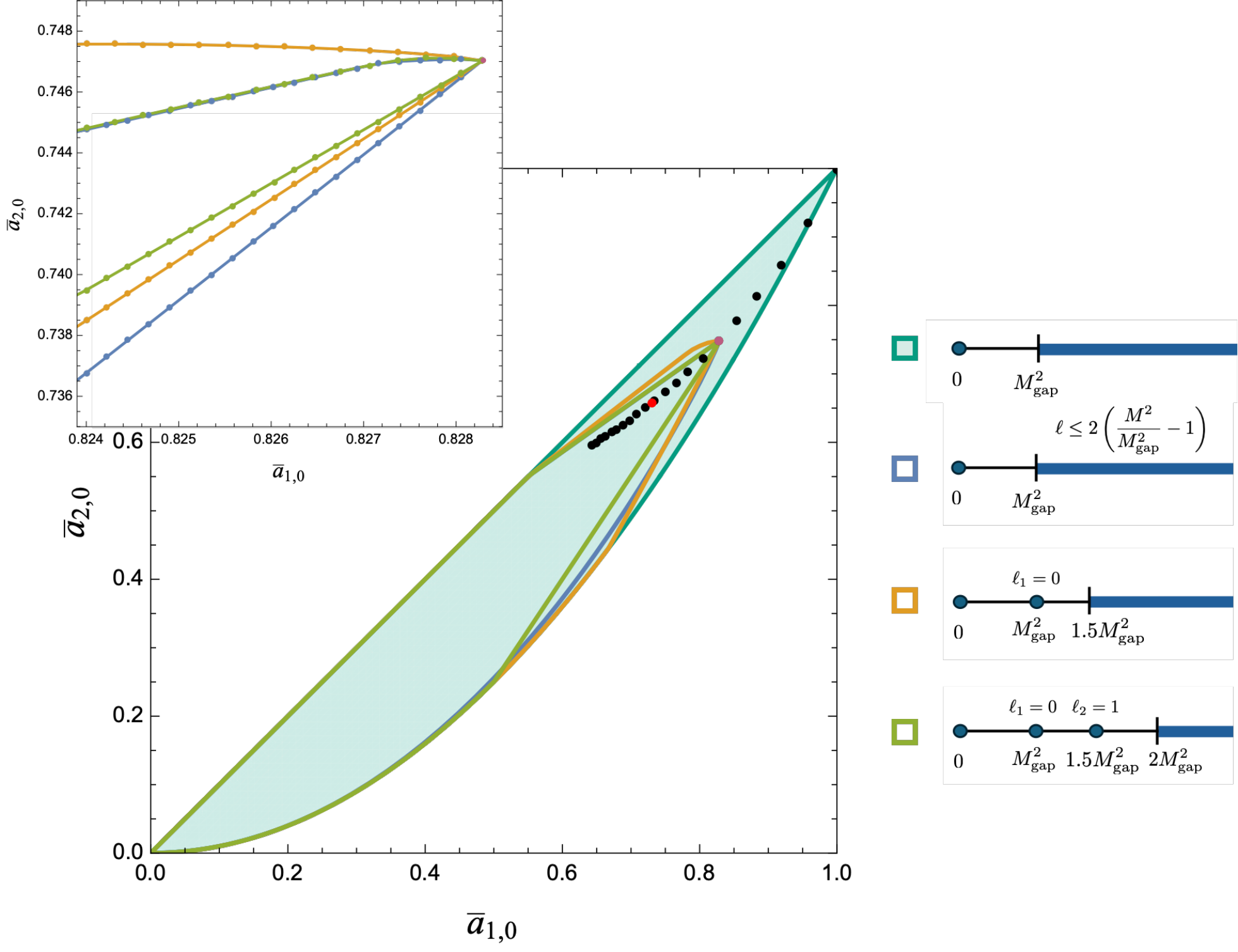}
\caption{Outer bounds on the $k_{\max} = 12~~(\bar{a}_{1,0},\bar{a}_{2,0})$ region with the Regge slope $\g = 2$ maximal spin requirement (blue), the scalar input at the mass gap with $\mu_c = 3/2$ (orange), and the scalar at the mass gap and vector at $\mu_2 = 3/2$ input with $\mu_c = 2$. The black points indicate the  locations of the tip of the allowed region for the ``corner theories'' parameterized by the first gap parameter $\mu_2$  (see Section \ref{s:newlinregge}). The brown dot is the case with $\mu_2=3/2$, and as shown it is right at the tip of the allowed region.}
\label{fig:maxspinhalf}
\end{figure}

By considering different values of $\mu_2$, we found corners in the bounds corresponding to a 1-parameter family of models with linear trajectories $M^2_{\ell} = (\mu_2 -1) \ell + 1$. We can extend the analysis to compare the bounds from 
\be
\text{Regge:}~~\ell \leq \frac{1}{\mu_2-1}\left(\frac{M^2}{M_{\gap}^2}-1\right)
\ee
to bounds obtained with the spectrum input
\be
\raisebox{-7.1mm}{\includegraphics[width=5.5cm]{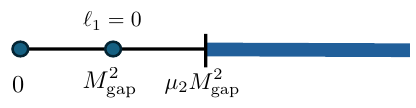}}
~~~~~\text{and}~~~~~
\raisebox{-7.1mm}{\includegraphics[width=5.5cm]{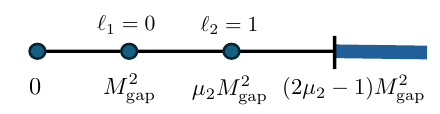}}
\ee
The result for $\mu_2 = 1.5$ is shown in Figure \ref{fig:maxspinhalf}. These bounds again give very similar constraints and they all have a sharp corner. This corner is where we expect to find the models with linear Regge trajectories. 

In Section \ref{s:MSinput}, we discussed how imposing the Regge bound can help give a cleaner SDPB spectrum. We have also tested how the combined constraints of Regge plus lowest mass state coupling input affect the bounds; details are given in Appendix \ref{app:2stateVen}. In short, the outcome is that for the single state input with the scalar coupling fixed at the string-value, we get   significantly smaller islands around the Veneziano amplitude when the Regge slope conditions are imposed. This is shown in Figure \ref{fig:addislands}. However, for the two-state input, there is hardly any difference between the islands found from fixing the couplings of both the scalar and the vector to their string values versus those with the Regge slope condition added.


\section{Discussion} \label{s:disc}

We have shown that basic, low-energy input to maximally supersymmetric YM EFT generates  novel, physically interesting features in the space of Wilson coefficients consistent with a local, unitary UV theory. We found that when there is a single state at bottom of the massive spectrum, it has to be a scalar. 
When we enforce the existence of a scalar at the mass gap $M_\text{gap}^2$, a vector at $\mu_2 M_\text{gap}^2$, and no other states until a cutoff scale $\mu_c M_\text{gap}^2$, the maximal values of the $\bar{a}_{1,0}$ Wilson coefficient remains almost exactly the same until $\mu_c \approx 2\mu_2 - 1$, at which point it begins falling off rapidly. The dramatic change in be\-havior suggests the existence of an amplitude with a contribution from a spin two state at $\mu_3 M_\text{gap}^2= (2\mu_2-1)M_\text{gap}^2$. This would correspond to a theory with a linear Regge trajectory.

If, instead of explicitly requiring the vector at $\mu_2 M_\text{gap}^2$, we require that the scalar at $M_\text{gap}^2$ has a coupling to the massless states equal to that known from the Veneziano amplitude, then the maximal size of the second mass scale was found to be $\mu_2 = 2$. Assuming there are no states until $2M_\text{gap}^2$, the allowed region of Wilson coefficients shrinks to a small island. As more constraints are included from the derivative expansion,  the islands were found to shrink in size, indicating that perhaps the  only allowed point corresponds to the Veneziano amplitude in the $k_{\max} \to \infty$ limit. If the island did indeed shrink all the way, that would mean that to bootstrap Veneziano, one only needs two pieces of low-energy information:
\begin{enumerate}
\item that there is only one state at the lowest mass and it contributes via a pole exchange to the 4-point amplitude (the bootstrap then requires it must be a scalar), 
\item the ratio, $|\bar{g}_0|^2=|g_0|^2/a_{0,0}$, between the massive scalar's coupling to the massless external states and the $\tr F^4$ Wilson coefficient  $a_{0,0}$.
\end{enumerate}
In a practical scenario, these two inputs might necessarily become three because it would likely be difficult to determine $|\bar{g}_0|^2$  without measuring both $g_0$ and $a_{0,0}$ individually. 
It is still surprising, though, how little information is needed to bootstrap the Veneziano amplitude. It would be interesting to understand to what extent this is a consequence of supersymmetry and/or crossing. 

Consider now what happens when we fix $|\bar{g}_0|^2$ to a something different from the string value. There are two qualitatively different cases to discuss, depending on whether $|\bar{g}_0|^2$ is greater or smaller than 1/2. Starting with the former, we find that the maximum allowed value of $|\bar{g}_0|^2$ is $(777-1120 \log(2))= 0.675158$, which occurs for the Infinite Spin Tower and requires the cutoff to be taken all the way down to the mass gap, $\mu_c =1$.
The Single Scalar Exchange model (SSE, Section \ref{s:venspec}) has 
 $|\bar{g}_0|^2 = 1/2$ and cutoff $\mu_c \to \infty$. Any value of  $|\bar{g}_0|^2$ between these extremes, i.e.~$1/2<|\bar{g}_0|^2\le 0.675158$, will bootstrap the cutoff $\mu_c$ to a maximum allowed value between 1 and $\infty$. The resulting bounds are expected to be islands around the black points in Figure \ref{fig:maxspinhalf}. These are the models with linear Regge slopes $1/(\text{max}(\mu_c)-1)$ found as the corner theories in Section \ref{s:newlinregge}. It would be interesting to understand if these models have realizations as generalized versions of the Veneziano amplitude, as explored in for example \cite{Cheung:2022mkw,Cheung:2023adk,Cheung:2023uwn}.
If they exist, such new amplitudes would presumably have to interpolate between the SSE amplitude, the Veneziano amplitude, and the Infinite Spin Tower.

To actually compute this family of islands, one needs a very precise determination of the  cutoff scale $\mu_c$ for a given value of $|\bar{g}_0|^2$; since the islands are going to be small, slightly different values of  $\mu_c$ 
can give mutually excluding islands. For any finite value of  $k_{\max}$, one can determine the maximum allowed value of $\mu_c$, but one would then need to extrapolate that to 
$k_{\max}\to \infty$. Alternatively, one can specify the cutoff $\mu_c$ and seek to bootstrap the value of $|\bar{g}_0|^2$ by a maximization principle.

From a theoretical perspective there is no reason to necessarily favor coupling input over input of the cutoff mass. Imagine doing a bootstrap like this in an context where the input comes from actual experimental data. 
One may then view the coupling choice as more well-motivated  than the gap input because in an experimental situation that directly produces a high energy particle, we would be able to determine both the spin and coupling of that state. On the other hand, it would be impossible to experimentally determine the exact gap to the next state without having enough energy to produce it, at which point, the spin and coupling of that state could be used as bootstrap input. 
However, an experiment would not give us an exact value for any measurement, so the ratio $|\bar{g}_0|^2$ 
would be determined with some error. The bootstrapped maximal $\mu_c$ value would then give the approximate scale at which new physics must appear. To get a reasonable ``island'' in such a scenario, one would have to compute islands for a selection of values of  $|\bar{g}_0|^2$ within the  experimental bounds and their corresponding largest cutoff scale. The full allowed region of Wilson coefficients would then be obtained from smearing of these islands. This smeared region would likely not shrink to a single point with increasing $k_\text{max}$, but would still be far more constraining than any naive bound.

Consider now the range of couplings $0 \le |\bar{g}_0|^2 \le 1/2$. In this case, there is no maximal cutoff mass $\mu_c$. Similarly to the case of SSE with $|\bar{g}_0|^2 = 1/2$, we only expect islands in the  $\mu_c \to \infty$ limit.  Models with $0 \le |\bar{g}_0|^2 \le 1/2$ can be obtained as linear combinations of SSE amplitudes with different choices of $m^2$, for example (with $M_\text{gap}=1$ for simplicity)
\begin{align}
\lambda A_1^{\text{SSE}} + (1-\lambda)A_{m^2}^{\text{SSE}} \, .
\end{align}
For this amplitude, $|g_0|^2 = \lambda/2$ because when $m^2>1$ its only contribution is from the $A_1^{\text{SSE}}$ part of the amplitude. The coefficient $a_{0,0}$, on the other hand, is given by $a_{0,0} = (\lambda +(1-\lambda)/m^2)$, and so the ratio becomes
\begin{align}
|\bar{g}_0|^2 = \frac{\lambda}{2(\lambda+(1-\lambda)/m^2)} \,.
\end{align}
By then setting $m^2 = r/\lambda$, we find that in the  limit of $\lambda \to 0$ and $r$ fixed,
\begin{align}
\lim_{\lambda\to 0}|\bar{g}_0|^2 = \frac{r}{2r+2} \,.
\end{align}
Therefore, $0 < |\bar{g}_0|^2 < 1/2$ is allowed and gives examples of islands in a limiting sense.\footnote{Limiting, because the SSE amplitudes are borderline cases for the Froissart behavior.}  

Returning to the bootstrap of the Veneziano amplitude, one potential path to stronger bounds is through the fact that, in the Regge limit with fixed $u < 0$, the Veneziano amplitude \reef{scalarVeneziano} actually scales with large $|s|$ as
\begin{align}
\lim_{|s|\to\infty}\frac{A^{\str}(s,u)}{s} \sim s^{u} \to 0\, .
\end{align}
This is one power of $s$ stronger than we assume with the Froissart bound that $A(s,u)/s^2 \to 0$. Therefore, one could restrict to amplitudes that have this improved Regge behavior and derive additional null constraints that correspond to the fact that $a_{k,k+1} = 0$ for all $k \geq -1$. This rules out, for example, the SSE amplitude, so it could improve our ability to corner and isolate the string. The implementation of these null constraints is discussed in \cite{AKRboot} using techniques developed in \cite{Caron-Huot:2021rmr}.

The type of bounds computed in this paper are so-called ``dual'' bootstrap bounds. Any point outside of our bounds does not have a unitary UV completion (with the specified assumptions), but points within our bounds may or may not have them. An important aspect of the bootstrap program not considered in this work is the ``primal'' formulation of the S-matrix bootstrap, which has been studied for permutation symmetric scalars, pions, photons, and gravitons \cite{Guerrieri:2020bto,Guerrieri:2021ivu,Chen:2022nym,Haring:2022sdp,Haring:2023zwu,Li:2023qzs}. In the primal version of the bootstrap, unitarity is imposed numerically on an ansatz that manifestly satisfies both locality and crossing symmetry, and points within their bounds necessarily have a unitary UV complete amplitude,\footnote{But this does not guarantee a UV complete {\em theory}.} but points outside could as well. The primal bootstrap has the advantage of being clearly applicable beyond the strict perturbative limit. Further, as shown in \cite{Haring:2023zwu,Li:2023qzs}, there can be reasonably good agreement between the dual and primal bounds. It would be interesting to know whether the primal type bounds could rule-in parameter space such that we can test if the scalar-input islands somehow do not continue to shrink to include the string alone.

Finally, the fact that low-energy input leads to important new features of the space of allowed Wilson coefficients in both the pion and $\mathcal{N} = 4$ SYM models suggests that such features might also appear in less mathematically similar bootstrap problems. 
If the appearance of novel features is generic, one might hope to eventually apply these principles to more phenomenologically relevant models as well.


\section*{Acknowledgements}
We would like to thank Jan Albert, Cliff Cheung, Prudhvi Bhattiprolu, Alan Shih-Kuan Chen, Nick Geiser, Aidan Herderschee, Aaron Hillman, Loki Lin, Brian McPeak, Leonardo Rastelli, Grant Remmen, and Yuan Xin
for useful comments and discussions. We  also  thank Jan Albert, Waltraut Knop, and Leonardo Rastelli for sharing their upcoming draft \cite{AKRboot}. This research was supported in part through computational resources and services provided by Advanced Research Computing at the University of Michigan, Ann Arbor. HE and JB are supported in part by Department of Energy grant DE-SC0007859. In addition, JB was supported by the 
Cottrell SEED Award number 
CS-SEED-2023-004 from the Research Corporation for Science Advancement.

\appendix

\section{Implementation as an SDP}
\label{app:numerics}
\subsection{Null Constraints}
\label{app:null}
There are two sets of null constraints: the $\mathcal{X}$-constraints are due to the basic SUSY crossing condition $a_{k,q} - a_{k,k-q} = 0$, and the $\mathcal{Y}$-constraints come from reconciliation of  dispersive representations of the $a_{k,q}$ derived for  fixed $u$ and fixed $t$. These two sets of conditions impose constraints on the spectral density and they were derived in detail in 
 \cite{Berman:2023jys}. They are
\begin{equation}
\label{crossingsymmold}
\begin{split}
&\forall ~ k,q\!:
~~~
\sum_{\ell=0}\int_{0}^{1}dx\, p_{\ell}(x)\,\mathcal{X}^{\ell,x}_{k,q} =0
~~~\text{with}~~~
\mathcal{X}^{\ell,x}_{k,q}=
x^{k}
\big[v_{\ell,q}-v_{\ell,k-q}\big]  \,\\
&\forall ~ k,q\!:
~~~ \sum_{\ell}\int_{0}^{1} dx\, p_{\ell}(x)\,\mathcal{Y}^{\ell,x}_{k,q} =0
\\
&
\text{with}~~~
\mathcal{Y}^{\ell,x}_{k,q}= x^{k}\left [v_{\ell,q}-(-1)^{\ell}\sum_{q'=0}^{k}(-1)^{q'}v_{\ell,q'}\left (\binom{q'}{k-q}+\binom{q'}{q}\right )\right ] \, ,
\end{split}
\end{equation}
where 
\begin{align}
    x \equiv \frac{M_{\gap}^2}{M^2} = y^{-1} 
    \quad
    \text{and}
    \quad
    p_\ell(x) \equiv x\rho_\ell\big(M_{\textrm{gap}}^{2}/x\big) = y^2f_{\ell}(y) \, .
\end{align}
We can rewrite these in terms of $y$ and $f_{\ell}(y)$ to match the notation here by simply making a change of variables from $x \to y$. Doing this variable replacement takes
\begin{align}
x^k p_{\ell}(x) dx = y^{-k} (y^{-3}\rho(M_{\gap}^2 y))dy = y^{-k} f_{\ell}(y) dy\, .
\end{align}
Thus, we find that in terms of $y$ and $f_{\ell}(y)$, the null constraints become
\begin{equation}
\label{crossingsymmnew}
\begin{split}
&\forall ~ k,q\!:
~~~
\sum_{\ell=0}\int_{1}^{\infty}dy\, f_{\ell}(y)\,\mathcal{X}^{\ell,y}_{k,q} =0
~~~\text{with}~~~
\mathcal{X}^{\ell,y}_{k,q}=
y^{-k}
\big[v_{\ell,q}-v_{\ell,k-q}\big]  \,\\
&\forall ~ k,q\!:
~~~ \sum_{\ell}\int_{1}^{\infty} dy\, f_{\ell}(y)\,\mathcal{Y}^{\ell,y}_{k,q} =0
\\
&
\text{with}~~~
\mathcal{Y}^{\ell,y}_{k,q}= y^{-k}\left [v_{\ell,q}-(-1)^{\ell}\sum_{q'=0}^{k}(-1)^{q'}v_{\ell,q'}\left (\binom{q'}{k-q}+\binom{q'}{q}\right )\right ] \, .
\end{split}
\end{equation}

\subsection{Explicit States}
For the practical implementation in SDPB, it is  convenient to set $M_\text{gap} =1$. Recall that the dispersive representation  \reef{finalequ} is
\be
  \label{akq1disp}
   a_{k,q}
   = 
   \sum_{\ell=0}^\infty
    \int_{1}^\infty
    dy \, y^{-k} 
    f_\ell(y)\, 
    v_{\ell,q} \,.
\ee
Consider the case  of  a spin-0 state at the mass gap ($y=1$) and a spin 1 state at $y = \mu_2$. The simple poles correspond to delta functions in the spectral density, so with the definition of $f_\ell(y)$ above equation \reef{defbracket}, we have
\be
  \label{ffull}
  f_\ell(y)
  = \delta(y-1) \delta_{\ell,0} |g_0|^2
  + 
  \delta(y-\mu_2) \delta_{\ell,1} |g_1|^2 \frac{1}{\mu_2^3}
  + f^{(\mu_c)}_\ell(y)\,,
\ee
where $f^{(\mu_c)}_\ell(y)$ only has support for $y \ge \mu_c$. From this, one obtains eq.~\reef{inputdispN2}.

The derivation of the null constraints using the {\em full} $f_\ell(y)$ gives exactly the same results as without spectral input. We denote these crossing constraints jointly as ``null'' in the following. Using the ``bracket-notation'' \reef{defbracket} the  general  vertex representation \reef{vectorequation} is of the form
\be 
 \label{vvv}
 \vec{V} = \big\< \vec{E} \big\>_1\,,
\ee
where $\vec{V}$ and $\vec{E}_{\ell,y}$ are described briefly in Section \ref{s:num}; more explicit expressions can be found in equation (4.2) in \cite{Berman:2023jys}. 
The precise specification of \reef{vvv} 
depends on which quantity we wish to extremize. It is useful to note, for example, that it follows from \reef{ffull} that
\be
\label{Esplit}
\big\< \vec{E} \big\>_1
 = |g_0|^2 \vec{E}_{0,1}
    + |g_1|^2 \frac{1}{\mu_2^3} \vec{E}_{1,\mu_2}
    + \big\< \vec{E} \big\>_{\mu_c}\,.
\ee
This is the key ingredient in the following.

\subsection{Optimization problem: maximizing $a_{k,q}$} \label{app:maxakq}

Define 
\be
  \vec{V} = 
  \begin{pmatrix}
     a_{0,0} \\
     a_{k,q} \\[2mm]
     \< \,\overrightarrow{\text{null}}\, \>_1
  \end{pmatrix}
  ~~~
  \text{and}
  ~~~
  \vec{E}_{\ell,y} = 
  \begin{pmatrix}
     1 \\
     y^{-k} v_{\ell,q } \\[2mm]
     \overrightarrow{\text{null}}_{\ell,y} 
  \end{pmatrix}
  \,,
\ee
where
$\< \,\overrightarrow{\text{null}}\, \>_1$ indicates all the null constraints, which may including also conditions such as $a_{k',q'} = R a_{0,0}$ that sets the question to: what are the extremal values of $a_{k,q}/a_{0,0}$ when 
$a_{k',q'}/a_{0,0}$ is fixed to be $R$.

Now let 
\be 
  \vec{\alpha} = (A, -1 , \vec{\beta})\,.
\ee  
  With the vanishing of the null constraints, it is clear that
$\vec{\alpha} \cdot \vec{V} = A a_{0,0} - a_{k,q}$. Hence, if we assume that 
\be
  \begin{split}
&\vec{\alpha}\cdot\vec{E}_{0,1} \ge 0\,,~~~~
  \vec{\alpha}\cdot\vec{E}_{1,\mu_2} \ge 0\,,~~~~
  \\ 
  &
\vec{\alpha}\cdot\vec{E}_{\ell,y}  \ge 0
  ~~~~\text{for all } \ell=0,1,\ldots,\ell_\text{max}~~~
  \text{and}~~~
  y \ge \mu_c \,,
  \end{split}
\ee
then $\vec{\alpha} \cdot \vec{V}  = \big\< \vec{\alpha} \cdot \vec{E} \big\>_1 \ge 0$ implies that on the vanishing of the null constraints
\be
   A a_{0,0} - a_{k,q} \ge 0
   ~~~~
  \implies
   ~~~~
   A \ge \frac{a_{k,q}}{a_{0,0}}\,,
\ee
so that minimizing $A$ is maximizing $a_{k,q}/a_{0,0}$.
In SDPB, the  conditions 
$\vec{\alpha}\cdot\vec{E}_{0,1} \ge 0$ and $\vec{\alpha}\cdot\vec{E}_{1,\mu_2} \ge 0$ simply mean that we allow for scalar or vector states, respectively, at $y=1$ and $y= \mu_2$. Leaving out one or both of these conditions mean that we disallow all states at the respective masses. 

It is useful to reformulate the problem as follows. Write 
\be
  \label{Vsplit}
  \vec{V} = a_{0,0} \vec{v}_1
  - a_{k,q} \vec{v}_O
  + \vec{V}_\text{null}\,,~~~~
  \vec{V}_\text{null}
  =  \begin{pmatrix}
     0 \\
     0 \\[2mm]
     \< \,\overrightarrow{\text{null}}\, \>_1
  \end{pmatrix}
\ee
for $ \vec{v}_1 = (1,0,0,\dots)$ and $ \vec{v}_O = (0,-1,0,\dots)$.
The vector $\vec{v}_1$ identifies the quantity we optimize in $\vec\alpha$, i.e. $A = \vec{\alpha} \cdot  \vec{v}_1$ whereas the objective vector 
$\vec{v}_O$ ensures the proper normalization of the ``objective'' of our optimization by having 
\be 
  \label{alphanormalization}
  \vec\alpha \cdot \vec{v}_O = 1 \,.
\ee
Next, consider optimization of the couplings $|g_0|^2$ and  $|g_1|^2$.

\subsection{Optimization problem: bounding couplings} 
\label{app:bdg1}
In this Appendix, we describe two methods for optimizing couplings.

\subsection*{Bounding Couplings by Choosing Objective Vectors}
The approach described here was developed in \cite{Albert:2022oes}.  
Define  
\be
  \vec{V} = 
  \begin{pmatrix}
     a_{0,0}  \\[2mm]
     \< \,\overrightarrow{\text{null}}\, \>_1
  \end{pmatrix}
  ~~~
  \text{and}
  ~~~
  \vec{E}_{\ell,y} = 
  \begin{pmatrix}
     1 \\[2mm]
    \overrightarrow{\text{null}}_{\ell,y} 
  \end{pmatrix}
\ee
i.e.~
\be
  \vec{V} = a_{0,0} \vec{v}_1
  + \vec{V}_\text{null}\,,~~~~
  \text{with}~~~
  \vec{V}_\text{null}
  =  \begin{pmatrix}
     0 \\[2mm]
     \< \,\overrightarrow{\text{null}}\, \>_1
  \end{pmatrix}
\ee
This time we pick $\vec{\alpha} = (A, \vec{\beta})$. We now  describe how to use $\vec{V}  = \big\< \vec{E} \big\>_1$ to maximize $|g_0|^2$. Then we show how a similar approach is used to extremize  $|g_1|^2$.
\begin{itemize}
\item Maximize $|g_0|^2$. 

With the help of \reef{Esplit}, we can write $\vec{V}  = \big\< \vec{E} \big\>_1$   as
\be
  \label{coupling1}
  a_{0,0} \vec{v}_1
   - |g_0|^2 \vec{E}_{0,1}+ \vec{V}_\text{null}
  = |g_1|^2 \frac{1}{\mu_2^3} \vec{E}_{1,\mu_2}
    + \big\< \vec{E} \big\>_{\mu_c}\,.
\ee
This can be viewed as an optimization problem like \reef{Vsplit} with objective $|g_0|^2$ and objective vector 
$\vec{v}_O = \vec{E}_{0,1}$. The normalization condition \reef{alphanormalization} then says that we need to have
\be 
  \vec{\alpha} \cdot \vec{E}_{0,1} = 1\,.
\ee
Using that, and imposing the null constraints, we then find that dotting $\vec{\alpha}$ into \reef{coupling1} gives
\be
   a_{0,0} A
   - |g_0|^2 
  = |g_1|^2 \frac{1}{\mu_1^3} \,\vec{\alpha} \cdot \vec{E}_{1,\mu_1}
    + \big\< \vec{\alpha} \cdot \vec{E} \big\>_{\mu_c}\,
\ee
so that if we impose
\be
  \begin{split}
&\vec{\alpha}\cdot\vec{E}_{1,\mu_2} \ge 0\,,~~~~
  \\ 
  &
\vec{\alpha}\cdot\vec{E}_{\ell,y}  \ge 0
  ~~~~\text{for all } \ell=0,1,\ldots,\ell_\text{max}~~~
  \text{and}~~~
  y \ge \mu_c \,,
  \end{split}
\ee
then we get 
\be
  a_{0,0} A
   - |g_0|^2 \ge 0 
   ~~~\implies~~~
   A \ge \frac{|g_0|^2}{a_{0,0}}
\ee
i.e.~minimizing $A$ maximizes $|g_0|^2/a_{0,0}$.


\item Maximize $|g_1|^2$. 

Using \reef{Esplit}, we now write $\vec{V}  = \big\< \vec{E} \big\>_1$   as
\be
  \label{coupling2}
  a_{0,0} \vec{v}_1
   - |g_1|^2 \frac{1}{\mu_2^3} \vec{E}_{1,\mu_2}
   + \vec{V}_\text{null}
  = 
  |g_0|^2 \vec{E}_{0,1}
    + \big\< \vec{E} \big\>_{\mu_c}\,.
\ee
To extremize $|g_1|^2$, we choose the objective vector 
$\vec{v}_O = \frac{1}{\mu_2^3} \vec{E}_{1,\mu_2}$ and the normalization condition \reef{alphanormalization} then becomes 
\be 
  \mu_2^{-3} \,
 \vec{\alpha} \cdot \vec{E}_{1,\mu_2} = 1\,.
\ee
Dotting $\vec{\alpha}$ into \reef{coupling2} and imposing the null constraints then gives
\be
 a_{0,0} A - |g_1|^2 \ge 0
\ee
assuming that 
\be
  \begin{split}
&\vec{\alpha}\cdot\vec{E}_{0,1} \ge 0\,,~~~~
  \\ 
  &
\vec{\alpha}\cdot\vec{E}_{\ell,y}  \ge 0
  ~~~~\text{for all } \ell=0,1,\ldots,\ell_\text{max}~~~
  \text{and}~~~
  y \ge \mu_c \, .
  \end{split}
\ee
As in previous cases, this lets us find the maximum of $|g_1|^2/a_{0,0}$ by minimizing $A$.
\end{itemize}

\subsection*{Bounding Couplings by Solving Null Constraints}
The above approach allows us to bound couplings $|g_{\ell,\mu}|^2$ of the massive exchanged particles to the massless particles relative to $a_{0,0}$. However, it does not allow to fix $|\bar{g}_{\ell,\mu}|^2= |g_{\ell,\mu}|^2/a_{0,0}$ to a specific value. To do so, we use a different method which is to use the null constraints to derive dispersive representations for the couplings $|g_{\ell,\mu}|^2$ that we want to fix. 

As a simple example of this, suppose we want to fix the coupling $|g_0|^2$ in \reef{scalarinpdisp} to a specific value. Since $|g_0|^2$ enters the null condition $a_{1,0}=a_{1,1}$, we can use that to solve for $|g_0|^2$. Start with 
\begin{align}
    0 = a_{1,0} - a_{1,1} = |g_0|^2(v_{0,1}-v_{0,0}) + \<y^{-1}(v_{l,1}-v_{l,0})\>_{\mu_c}
\end{align}
where we used that the high energy average is a linear operation. Using that $v_{l,q} = 0$ for $q>\ell$ and $v_{0,0} = 1$,  we solve for $|g_0|^2$ find 
\begin{align}
  \label{g0rep}
    |g_0|^2 = \<y^{-1}(v_{l,1}-v_{l,0})\>_{\mu_c}.
\end{align}
Using this dispersive representation, we can fixed  $|\bar{g}_0|$ to a specific value $R$ in the numerical bootstrap by making  $|g_0|^2 - R a_{0,0}$ a null constraint in our vector $\vec{V}$. The dispersive representation \reef{g0rep}  determines the corresponding entry in the $\vec{E}_{\ell,y}$ vectors.

To fix more couplings we can use more null conditions.  
For example, if we want to fix both $|\bar{g}_0|^2$ and $|\bar{g}_1|^2$ as discussed in the maintext, we can use $a_{1,0}=a_{1,1}$ along with $0 = a_{3,1} - a_{3,2}$  (that is the lowest-$k$ null relation to which $|g_0|^2$ does not contribute). Then, we find
\begin{align}
    0 &= a_{1,0}-a_{1,1} = |g_0|^2(v_{0,1}-v_{0,0}) + \frac{|g_1|^2}{\mu_2^4}(v_{1,1}-v_{1,0}) + \<y^{-1}(v_{l,1}-v_{l,0})\>_{\mu_c}\\
    0 &= a_{3,1}-a_{3,2} = |g_0|^2(v_{0,1}-v_{0,2}) + \frac{|g_1|^2}{\mu_2^6}(v_{1,1}-v_{1,2}) + \<y^{-3}(v_{l,1}-v_{l,2})\>_{\mu_c} \, .
\end{align}
 Using the fact that $v_{l,q} = 0$ for $q > l$ to simplify these expressions, along with $v_{1,1} = 2$, we solve the linear system 
 that can be solved to give $|g_{0}|^2$ and $|g_{1}|^2$ in terms of their high energy integrals: 
\begin{align}
    |g_{0}|^2 &= \frac{\mu_2^2}{2}\<y^{-3}(v_{l,2}-v_{l,1})\>_{\mu_c}+\<y^{-1}(v_{l,1}-v_{l,0})\>_{\mu_c}\\
    |g_{1}|^2 &= \frac{\mu_2^6}{2}\<y^{-3}(v_{l,2}-v_{l,1})\>_{\mu_c}\, .
\end{align}
In principle, we could use a null constraint of any order to solve for a coefficient. However, to keep track of $k_{\max}$ properly, we restrict ourselves to using only null constraints up to those with $k = k_{\max}$. To have access to the null constraint $0 = a_{3,1} - a_{3,2}$ we need to take $k_{\max} \geq 3$ and the maximal number of couplings we can fix depends on $k_{\max}$. 

When we solve for couplings using null relations, we are no longer strictly enforcing their positivity. Hence, we only use this approach when we either fix or extremize a coupling that is solved for by the null conditions; otherwise, we risk allowing non-unitary theories where these couplings are negative. It is important to note that when we use SDPB to {\em minimize} a coupling $\bar{g}_{\ell,\mu_n}$, it may give a negative value, so we enforce positivity by hand, taking Min$|\bar{g}_{\ell,\mu_n}|^2 = $ Max$(0,$ SDPB minimum$)$.

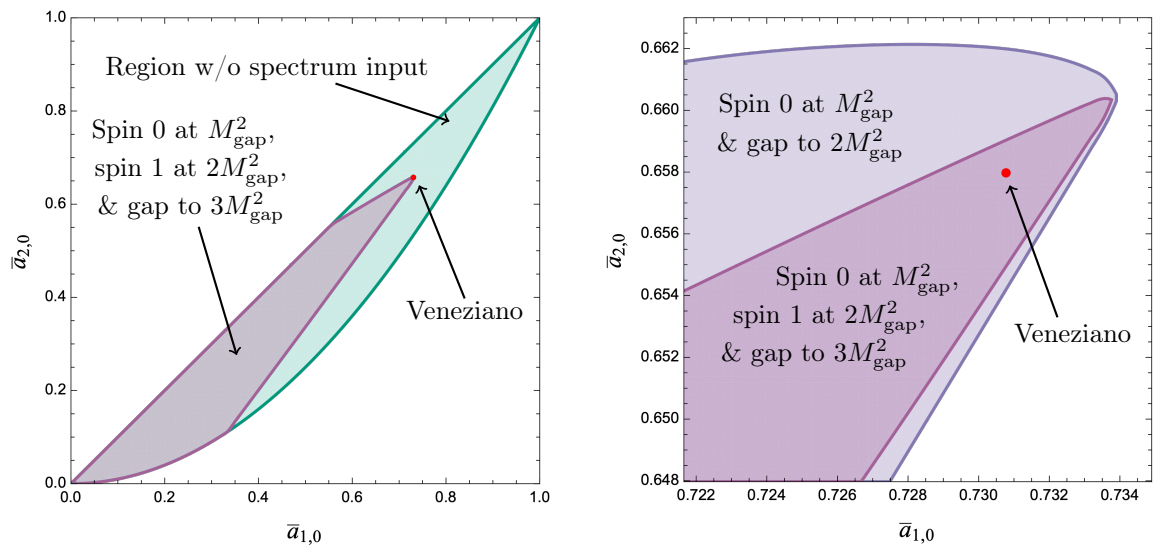
\begin{figure}[t]
\centering
\begin{center}
\begin{tikzpicture}
	\node (image) at (0,0) {\includegraphics[width=\textwidth]{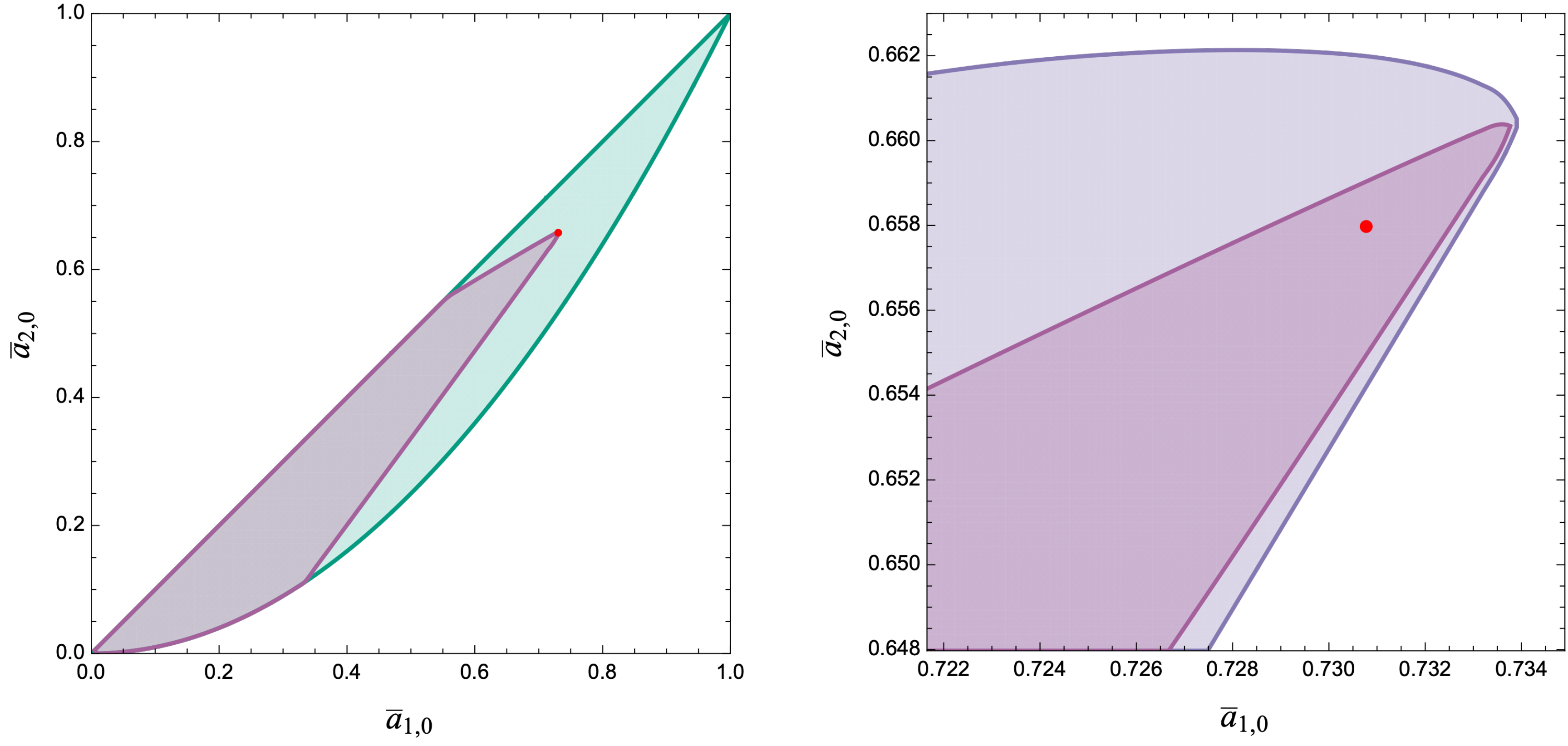}};
        \node at (-4.1,2.8) { {\small Region w/o spectrum input}};
        \node at (-5.2,1.95) 
        {{\small Spin 0 at $M_\text{gap}^2$,}};
        \node at (-5.1,1.45)
        {{\small spin 1 at $2M_\text{gap}^2$,}};
        \node at (-5.1,0.95)
        {{\small\& gap to $3M_\text{gap}^2$}};
        \node at (3.,2.3) {{\small Spin 0 at $M_\text{gap}^2$}};
        \node at (3.05,1.78) {{\small\& gap to $2M_\text{gap}^2$}};
        \node at (3.8,0) {{\small Spin 0 at $M_\text{gap}^2$,}};
        \node at (3.3,-0.5) {{\small spin 1 at $2M_\text{gap}^2$,}};
        \node at (3.1,-1) {{\small \& gap to $3M_\text{gap}^2$}};
        \draw[->,thick] (-3.2,2.6) -- (-1.7,2.1);
        \draw[->,thick] (-5,0.7) -- (-4.5,-1);
        \draw[->,thick] (-1.5,-0.25) -- (-2.1,1.25);  \node at (-1.5,-0.4) {{\small Veneziano}};
        \draw[->,thick] (6.3,-0.5) -- (5.67,1.25);  \node at (6.5,-0.7) {{\small Veneziano}};   \end{tikzpicture}
\end{center}
\caption{
{\bf Left:} The $\bar{a}_{1,0}$ and $\bar{a}_{2,0}$ 
with no spectral assumptions in teal and
when the lowest massive states are assumed to be a scalar at the mass gap $M_\text{gap}^2$, a vector at $2M_\text{gap}^2$, and a further mass-gap to $3M_\text{gap}^2$ in purple. The purple region has a corner very close to the Veneziano amplitude (red dot). 
{\bf Right:} Zoom-in around the corner of the  purple region.
For comparison is also shown in light purple the allowed region when we only input a spin 0 particle at $M_\text{gap}^2$ and a further mass gap to $2M_{\gap}^2$. The bounds for these regions were computed with $k_\text{max}=10$.}
\label{fig:a10a20scavecinp}
\end{figure}

\section{Multi-State Bootstrap of Veneziano}
\label{app:2stateVen}

In this appendix, we compare the corner and island bootstrap of the Veneziano amplitude for two different types of low energy assumptions: (1) the single state input from Section \ref{s:singlestate} with a a scalar at $M_\text{gap}^2$ and no states until the cutoff at $2M_\text{gap}^2$ versus (2) inputting the two lowest mass states: a scalar at $M_\text{gap}^2$,  a vector at $2 M_\text{gap}^2$ and no other states until $3 M_\text{gap}^2$ (which is the ``corner'' value for the cutoff discussed in Section \ref{s:bifurcations}). 

\subsection{Corners}

For the two state input discussed above, we have
\begin{align}\label{inputdispN2string}
    a_{k,q} = |g_{0}|^2v_{0,q}+\frac{|g_{1}|^2}{2^{k+3}}v_{1,q}+\Big\<y^{-k}v_{\ell,q}\Big\>_{3} \, .
\end{align}
With this input, the allowed region of the $(\bar{a}_{1,0},\bar{a}_{2,0})$-plane is shown in Figure \ref{fig:a10a20scavecinp} is somewhat smaller than that of just the single scalar input and still has the string amplitude at the corner. Most importantly, the corner near the string is, as shown on the right of Figure \ref{fig:a10a20scavecinp} quite a bit sharper than with just the single mass input.

\begin{figure}[t]
\centering
\includegraphics[width=\textwidth]{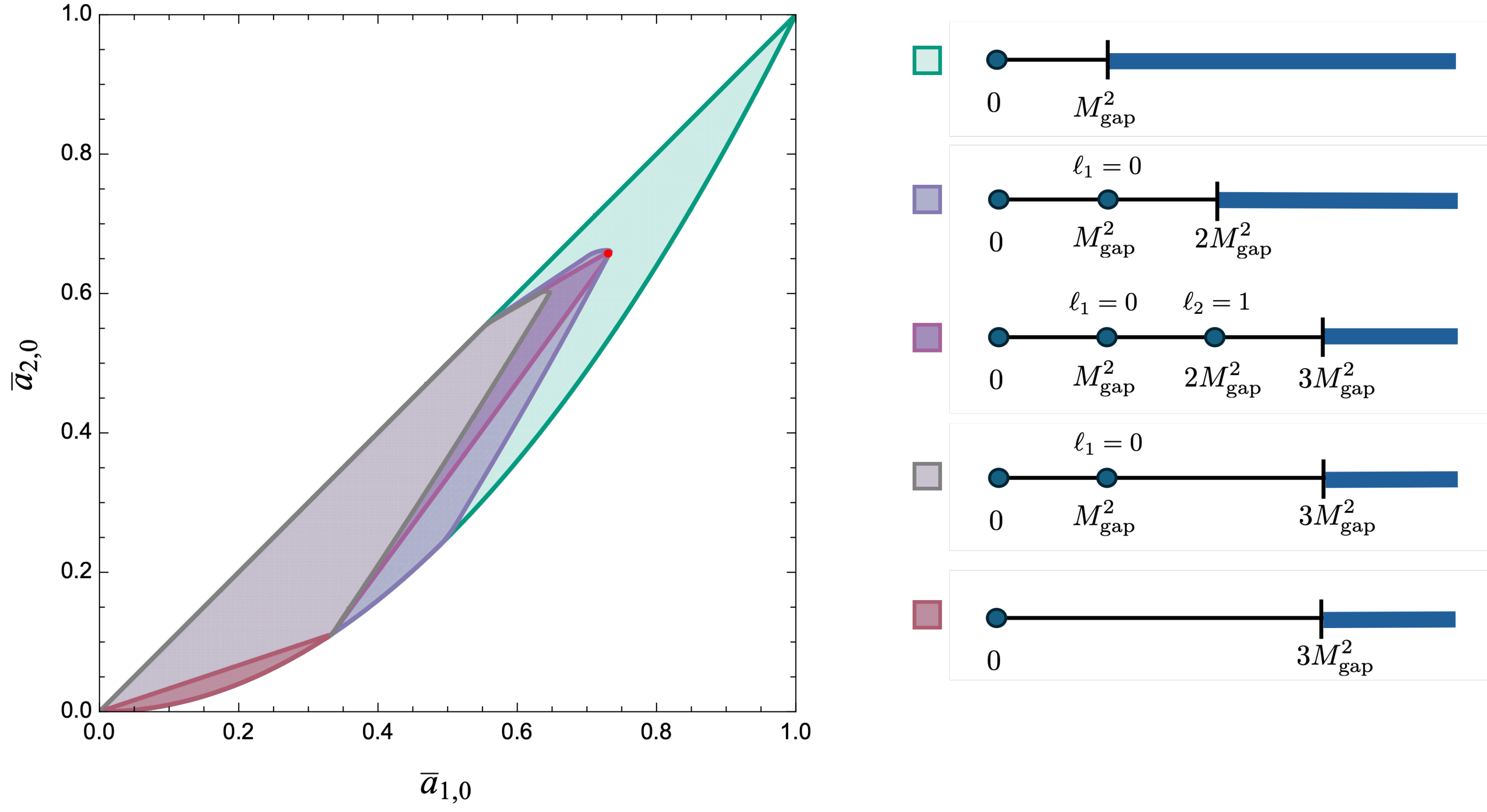}
\caption{Allowed regions for listed spectrum assumptions at $k_{\max} = 10$.}
\label{fig:a10a20build}
\end{figure}

It is interesting to see how the regions are built up and which parts of them are sensitive to the various mass and spin inputs. In Figure \ref{fig:a10a20build}, we show the allowed regions in the $(\bar{a}_{1,0},\bar{a}_{2,0})$-plane for various different spectral assumptions. The red region shows the allowed region with \textit{no} states allowed all the way to the cutoff at $3 M_\text{gap}^2$. A large chunk of parameter space becomes allowed if we turn on the $g_0$ coupling, allowing for a scalar at $M_\text{gap}^2$. If we keep the cutoff at $3 M_\text{gap}^2$, the region, shown in gray excludes the string (red dot), but when the cutoff is reduced to $2 M_\text{gap}^2$, we get the blue-gray region that now includes the string. This is the region studied previously in Section \ref{s:Vcorner} and shown in Figure \ref{fig:a10a20scalarinp}.
Next, we allow for a vector at $2M_\text{gap}$ we get the  region shown in dark purple, for which the Veneziano amplitude is at the sharp corner.

\begin{figure}[t]
\centering
\includegraphics[width=0.6\textwidth]{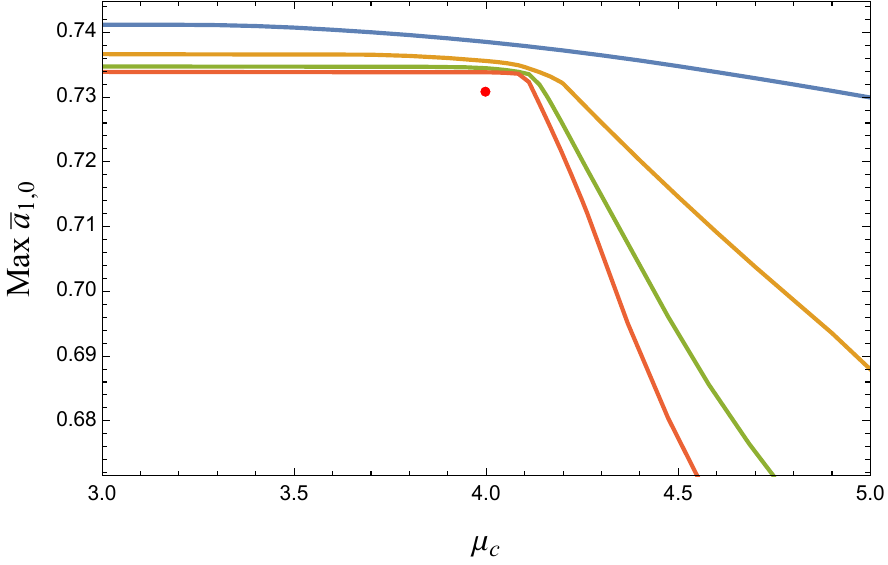}
\caption{The maximal $\bar{a}_{1,0}$ vs. $\mu_c$ at $k_{\max} = 4, 6, 8, 10$ for the spectrum given by \reef{inputdispN3}.}
\label{fig:coupvsmu4}
\end{figure}

\vspace{2mm}
{\bf The Next Corner.} 
The cutoff $\mu_c = 3$ was chosen as the value near the corner in the $\bar{a}_{1,0}$ maximum value. It is interesting to consider what  happens if we continued with further string-y input and take the existence of this spin two state as something ``implied'' by the bootstrap and then use it to find where the next state in the spectrum lives. We then have a dispersion relation of the form
\begin{align}\label{inputdispN3}
    a_{k,q} = |g_{0}|^2v_{0,q}+\frac{|g_{1}|^2}{2^{k+3}}v_{1,q}+\frac{|g_{2}|^2}{3^{k+3}}v_{2,q}+\Big\<y^{-k}v_{\ell,q}\Big\>_{\mu_{c}} \, .
\end{align}
 Figure \ref{fig:coupvsmu4} shows    maximal $\bar{a}_{1,0}$ vs.~the cutoff. The maximal values is almost exactly constant from $\mu_c = 3$ to $\mu_c = 4$ and then has a dramatic falloff, indicating the need for a new state at $ 4 M_\text{gap}^2$. This is corroborated by the SDPB spectrum from Figure \ref{fig:stringspec} which also indicated that there is a spin 2 state at $ 4 M_\text{gap}^2$.

At $ 4 M_\text{gap}^2$, we encounter another feature of the string spectrum: the first daughter trajectory. There is not just a spin three state at $M^2 = 4/\a'$, but also a spin one state exchanged. The fact that a corner appears at (or near) $\mu_4 = 4$ does not tell us what spin the particle there should have, so we can turn to SDPB's spectrum analysis to test for the states that live at the feature. As before, 
the extremal spectrum does not contain any daughter trajectories, so we do not see any purely bootstrap way to proceed along the string spectrum, at least for the values of $k_{\max}$ we can achieve. Additional assumptions, such as the Regge slope help a bit, but the spectrum is still not clean.

\subsection{Islands}\label{s:addislandsinp}
In Section \ref{s:strisl}, we showed that fixing $|\bar{g}_0|^2$, the coupling of the scalar at the mass gap, resulted in 
a maximal allowed cutoff mass $\mu_c M_{\gap}^2$. A unitary theory must have new massive states at or below that maximum value. We found that when  $|\bar{g}_0|^2$ was taken to be its string value, the maximum cutoff mass corresponded precisely to the mass at which the second string state appeared, namely as $2M_{\gap}^2$. Fixing $\mu_c=2$, we found shrinking islands around the Veneziano amplitude Wilson coefficients. 

To get stronger bounds we now impose stronger assumptions on  the spectrum. Specifically, we input information about the state at $2M_{\gap}^2$, as in the previous section.

We first consider a dispersion relation of the form
\begin{align}\label{dispssvcoup}
    \bar{a}_{k,q} = \frac{v_{0,q}}{\zeta_{2}}+\frac{|g_{\ell_2,2}|^2v_{\ell_2,q}}{2^{k+3}\zeta_{2}}+\Big\<y^{-k}v_{\ell,q}\Big\>_{\mu_c} \, ,
\end{align}
where the spin $\ell_2$ of the state at $\mu_2 = 2$ and the cutoff $\mu_c > 2$ are unfixed. We find that, by $k_{\max} = 12$, there are no theories compatible with the bounds \textit{unless} $\ell_2 = 1$, i.e.~there has to be a vector at the state at $2M_{\gap}^2$. With that vector assumed, any state with spin $\ell > 1$ has a coupling that gets suppressed exponentially with $k_{\max}$, similar to what we saw in Section \ref{s:bdspex} for non-scalar states at the mass gap. The maximal coupling to a scalar state at $2M_{\gap}^2$ decreases with $k_{\max}$, but it does not appear to approach zero as quickly as those for $\ell > 1$, so there is not any clear \textit{a priori} reason to rule it out.

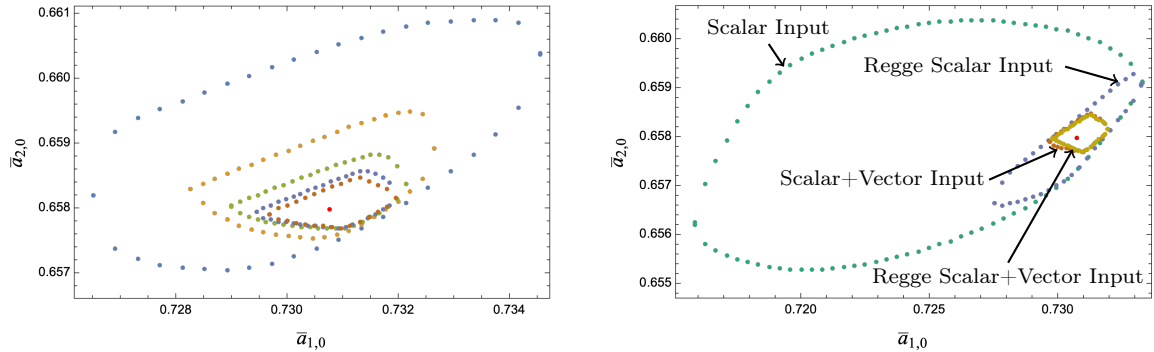
\begin{figure}[t]
\centering
\begin{tikzpicture}
	\node (image) at (0,0) {\includegraphics[width=\textwidth]{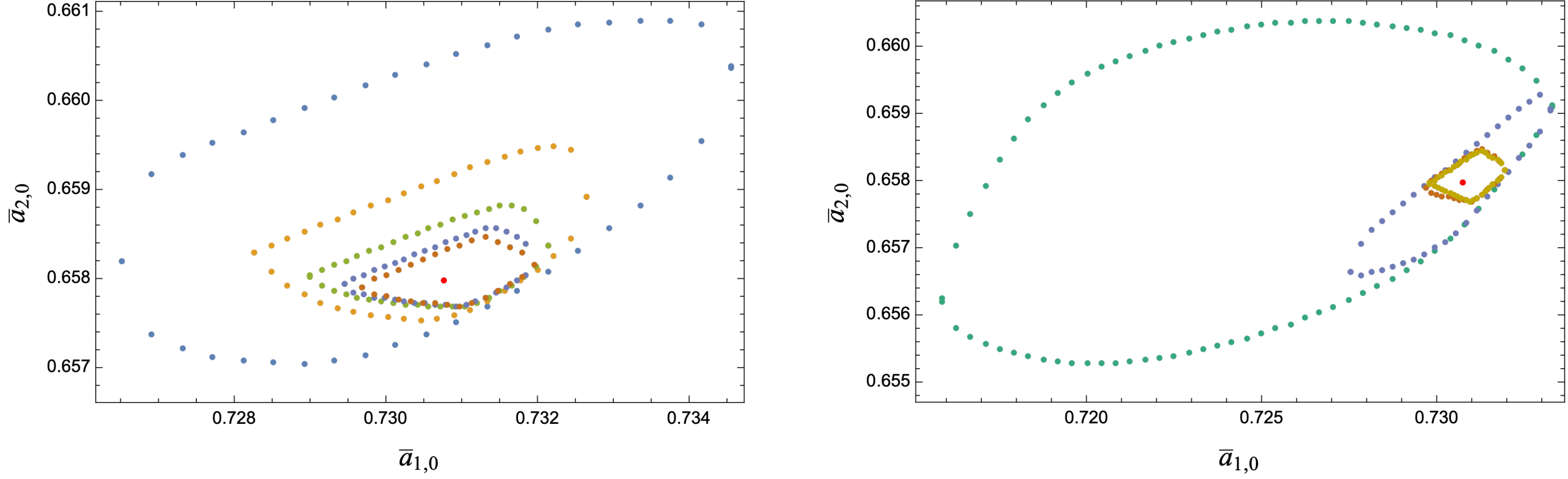}};

        \node at (2.5,2) {\scriptsize{Scalar Input}};
        \node at (5,1.5) {\scriptsize{Regge Scalar Input}};
        \node at (4,0) {\scriptsize{Scalar+Vector Input}};
        \node at (5.65,-1.3) {\scriptsize{Regge Scalar+Vector Input}};
        \draw[->,thick] (5.65,-1.1) -- (6.5,0.4);
        \draw[->,thick] (5.4,0) -- (6.3,0.4);
        \draw[->,thick] (6.4,1.5) -- (7.1,1.3);
        \draw[->,thick] (2.5,1.8) -- (2.7,1.5);
\end{tikzpicture}

\caption{\textbf{Left:} Bounds on the $k_{\max} = 4, 6, 8, 10, 12$ $(\bar{a}_{1,0},\bar{a}_{2,0})$ allowed region with a scalar at the gap and a vector at $\mu_2 = 2$ with their string couplings along with a cutoff $\mu_c = 3$.  \textbf{Right:} Comparing the scalar-only island with smaller scalar with Regge slope, scalar and vector, and scalar and vector with Regge slope island at $k_{\max} = 12$.}
\label{fig:addislands}
\end{figure}

To proceed, we simply make the string-inspired choice to add in just the vector state with its coupling and as well as a gap to $\mu_c = 3$.\footnote{One might hope to dynamically fix this, similar to how we did with just the scalar input, but the results are less clear in this case, so we choose to simply make it an assumption.} The dispersion relation for the Wilson coefficients is then
\begin{align}
    \bar{a}_{k,q} = \frac{v_{0,q}}{\zeta_{2}}+\frac{2v_{1,q}}{2^{k+3}\zeta_{2}}+\Big\<y^{-k}v_{\ell,q}\Big\>_{3} \, ,
\end{align}
using $|\bar{g}_0^{\str}|^2 = 1/\zeta_{2}$  and $|\bar{g}_1^{\str}|^2 = 2/\zeta_{2}$. The bootstrap then gives small shrinking islands around the Veneziano amplitude, as shown on the right of Figure \ref{fig:addislands}. The size of these islands are considerably smaller than the those with only the single state input (e.g.~Figure \ref{fig:SSisland}). A direct comparison of the $k_\text{max} = 12$ islands for the single and double state input is given on the right of Figure \ref{fig:addislands}. For further comparison, we also include in that plot the islands obtained with the same one- and two-state input but with the additional assumption of no states below the leading Regge trajectories (as discussed in Section \ref{s:Regge}.). This assumption has a significant effect on the single-scalar island, but results in hardly any change for the two-scalar island.

\bibliographystyle{JHEP}
\bibliography{SSCorner.bib}

\providecommand{\href}[2]{#2}\begingroup\raggedright\begin{thebibliography}{10}

\bibitem{Caron-Huot:2020cmc}
S.~Caron-Huot and V.~Van~Duong, \emph{{Extremal Effective Field Theories}},
  \href{https://arxiv.org/abs/2011.02957}{{\ttfamily 2011.02957}}.

\bibitem{Arkani-Hamed:2020blm}
N.~Arkani-Hamed, T.-C. Huang and Y.-T. Huang, \emph{{The {EFT}-Hedron}},
  \href{https://doi.org/10.1007/JHEP05(2021)259}{\emph{JHEP} {\bfseries 05}
  (2021) 259} [\href{https://arxiv.org/abs/2012.15849}{{\ttfamily
  2012.15849}}].

\bibitem{Chiang:2021ziz}
L.-Y. Chiang, Y.-t. Huang, W.~Li, L.~Rodina and H.-C. Weng, \emph{{Into the
  EFThedron and UV constraints from IR consistency}},
  \href{https://doi.org/10.1007/JHEP03(2022)063}{\emph{JHEP} {\bfseries 03}
  (2022) 063} [\href{https://arxiv.org/abs/2105.02862}{{\ttfamily
  2105.02862}}].

\bibitem{Albert:2022oes}
J.~Albert and L.~Rastelli, \emph{{Bootstrapping pions at large N}},
  \href{https://doi.org/10.1007/JHEP08(2022)151}{\emph{JHEP} {\bfseries 08}
  (2022) 151} [\href{https://arxiv.org/abs/2203.11950}{{\ttfamily
  2203.11950}}].

\bibitem{Caron-Huot:2022jli}
S.~Caron-Huot, Y.-Z. Li, J.~Parra-Martinez and D.~Simmons-Duffin,
  \emph{{Graviton partial waves and causality in higher dimensions}},
  \href{https://doi.org/10.1103/PhysRevD.108.026007}{\emph{Phys. Rev. D}
  {\bfseries 108} (2023) 026007}
  [\href{https://arxiv.org/abs/2205.01495}{{\ttfamily 2205.01495}}].

\bibitem{Fernandez:2022kzi}
C.~Fernandez, A.~Pomarol, F.~Riva and F.~Sciotti, \emph{{Cornering Large-$N_c$
  QCD with Positivity Bounds}},
  \href{https://arxiv.org/abs/2211.12488}{{\ttfamily 2211.12488}}.

\bibitem{Albert:2023jtd}
J.~Albert and L.~Rastelli, \emph{{Bootstrapping Pions at Large $N$. Part II:
  Background Gauge Fields and the Chiral Anomaly}},
  \href{https://arxiv.org/abs/2307.01246}{{\ttfamily 2307.01246}}.

\bibitem{Alberte:2020bdz}
L.~Alberte, C.~de~Rham, S.~Jaitly and A.~J. Tolley, \emph{{QED positivity
  bounds}}, \href{https://doi.org/10.1103/PhysRevD.103.125020}{\emph{Phys. Rev.
  D} {\bfseries 103} (2021) 125020}
  [\href{https://arxiv.org/abs/2012.05798}{{\ttfamily 2012.05798}}].

\bibitem{Henriksson:2021ymi}
J.~Henriksson, B.~McPeak, F.~Russo and A.~Vichi, \emph{{Rigorous bounds on
  light-by-light scattering}},
  \href{https://doi.org/10.1007/JHEP06(2022)158}{\emph{JHEP} {\bfseries 06}
  (2022) 158} [\href{https://arxiv.org/abs/2107.13009}{{\ttfamily
  2107.13009}}].

\bibitem{Chowdhury:2021ynh}
S.~D. Chowdhury, K.~Ghosh, P.~Haldar, P.~Raman and A.~Sinha, \emph{{Crossing
  Symmetric Spinning S-matrix Bootstrap: {EFT} bounds}},
  \href{https://doi.org/10.21468/SciPostPhys.13.3.051}{\emph{SciPost Phys.}
  {\bfseries 13} (2022) 051}
  [\href{https://arxiv.org/abs/2112.11755}{{\ttfamily 2112.11755}}].

\bibitem{Caron-Huot:2022ugt}
S.~Caron-Huot, Y.-Z. Li, J.~Parra-Martinez and D.~Simmons-Duffin,
  \emph{{Causality constraints on corrections to Einstein gravity}},
  \href{https://arxiv.org/abs/2201.06602}{{\ttfamily 2201.06602}}.

\bibitem{deRham:2022sdl}
C.~de~Rham, L.~Engelbrecht, L.~Heisenberg and A.~L\"uscher, \emph{{Positivity
  bounds in vector theories}},
  \href{https://doi.org/10.1007/JHEP12(2022)086}{\emph{JHEP} {\bfseries 12}
  (2022) 086} [\href{https://arxiv.org/abs/2208.12631}{{\ttfamily
  2208.12631}}].

\bibitem{Ma:2023vgc}
T.~Ma, A.~Pomarol and F.~Sciotti, \emph{{Bootstrapping the chiral anomaly at
  large N$_{c}$}}, \href{https://doi.org/10.1007/JHEP11(2023)176}{\emph{JHEP}
  {\bfseries 11} (2023) 176}
  [\href{https://arxiv.org/abs/2307.04729}{{\ttfamily 2307.04729}}].

\bibitem{CarrilloGonzalez:2023cbf}
M.~Carrillo~Gonz\'alez, C.~de~Rham, S.~Jaitly, V.~Pozsgay and A.~Tokareva,
  \emph{{Positivity-causality competition: a road to ultimate {EFT} consistency
  constraints}},  \href{https://arxiv.org/abs/2307.04784}{{\ttfamily
  2307.04784}}.

\bibitem{Berman:2023jys}
J.~Berman, H.~Elvang and A.~Herderschee, \emph{{Flattening of the {EFT}-hedron:
  supersymmetric positivity bounds and the search for string theory}},
  \href{https://doi.org/10.1007/JHEP03(2024)021}{\emph{JHEP} {\bfseries 03}
  (2024) 021} [\href{https://arxiv.org/abs/2310.10729}{{\ttfamily
  2310.10729}}].

\bibitem{Chiang:2023quf}
L.-Y. Chiang, Y.-t. Huang and H.-C. Weng, \emph{{Bootstrapping string theory
  {EFT}}}, \href{https://doi.org/10.1007/JHEP05(2024)289}{\emph{JHEP}
  {\bfseries 05} (2024) 289}
  [\href{https://arxiv.org/abs/2310.10710}{{\ttfamily 2310.10710}}].

\bibitem{Albert:2023bml}
J.~Albert, J.~Henriksson, L.~Rastelli and A.~Vichi, \emph{Bootstrapping mesons
  at large $n$: Regge trajectory from spin-two maximization},
  \href{https://arxiv.org/abs/2312.15013}{{\ttfamily 2312.15013}}.

\bibitem{Haring:2023zwu}
K.~H\"aring and A.~Zhiboedov, \emph{{The Stringy S-matrix Bootstrap: Maximal
  Spin and Superpolynomial Softness}},
  \href{https://arxiv.org/abs/2311.13631}{{\ttfamily 2311.13631}}.

\bibitem{Veneziano:1968yb}
G.~Veneziano, \emph{{Construction of a crossing - symmetric, Regge behaved
  amplitude for linearly rising trajectories}},
  \href{https://doi.org/10.1007/BF02824451}{\emph{Nuovo Cim. A} {\bfseries 57}
  (1968) 190}.

\bibitem{Plahte:1970wy}
E.~Plahte, \emph{{Symmetry properties of dual tree-graph n-point amplitudes}},
  \href{https://doi.org/10.1007/BF02824716}{\emph{Nuovo Cim. A} {\bfseries 66}
  (1970) 713}.

\bibitem{Stieberger:2009hq}
S.~Stieberger, \emph{{Open \& Closed vs. Pure Open String Disk Amplitudes}},
  \href{https://arxiv.org/abs/0907.2211}{{\ttfamily 0907.2211}}.

\bibitem{Bjerrum-Bohr:2009ulz}
N.~E.~J. Bjerrum-Bohr, P.~H. Damgaard and P.~Vanhove, \emph{{Minimal Basis for
  Gauge Theory Amplitudes}},
  \href{https://doi.org/10.1103/PhysRevLett.103.161602}{\emph{Phys. Rev. Lett.}
  {\bfseries 103} (2009) 161602}
  [\href{https://arxiv.org/abs/0907.1425}{{\ttfamily 0907.1425}}].

\bibitem{Bjerrum-Bohr:2010mia}
N.~E.~J. Bjerrum-Bohr, P.~H. Damgaard, T.~Sondergaard and P.~Vanhove,
  \emph{{Monodromy and Jacobi-like Relations for Color-Ordered Amplitudes}},
  \href{https://doi.org/10.1007/JHEP06(2010)003}{\emph{JHEP} {\bfseries 06}
  (2010) 003} [\href{https://arxiv.org/abs/1003.2403}{{\ttfamily 1003.2403}}].

\bibitem{Bjerrum-Bohr:2010pnr}
N.~E.~J. Bjerrum-Bohr, P.~H. Damgaard, T.~Sondergaard and P.~Vanhove,
  \emph{{The Momentum Kernel of Gauge and Gravity Theories}},
  \href{https://doi.org/10.1007/JHEP01(2011)001}{\emph{JHEP} {\bfseries 01}
  (2011) 001} [\href{https://arxiv.org/abs/1010.3933}{{\ttfamily 1010.3933}}].

\bibitem{Huang:2020nqy}
Y.-t. Huang, J.-Y. Liu, L.~Rodina and Y.~Wang, \emph{{Carving out the Space of
  Open-String S-matrix}},
  \href{https://doi.org/10.1007/JHEP04(2021)195}{\emph{JHEP} {\bfseries 04}
  (2021) 195} [\href{https://arxiv.org/abs/2008.02293}{{\ttfamily
  2008.02293}}].

\bibitem{Simmons-Duffin:2015qma}
D.~Simmons-Duffin, \emph{{A Semidefinite Program Solver for the Conformal
  Bootstrap}}, \href{https://doi.org/10.1007/JHEP06(2015)174}{\emph{JHEP}
  {\bfseries 06} (2015) 174}
  [\href{https://arxiv.org/abs/1502.02033}{{\ttfamily 1502.02033}}].

\bibitem{AKRboot}
J.~Albert, W.~Knop and L.~Rastelli, \emph{{Where is tree-level string theory?
  (to appear)}}, .

\bibitem{Correia:2020xtr}
M.~Correia, A.~Sever and A.~Zhiboedov, \emph{{An analytical toolkit for the
  S-matrix bootstrap}},
  \href{https://doi.org/10.1007/JHEP03(2021)013}{\emph{JHEP} {\bfseries 03}
  (2021) 013} [\href{https://arxiv.org/abs/2006.08221}{{\ttfamily
  2006.08221}}].

\bibitem{Arkani-Hamed:2022gsa}
N.~Arkani-Hamed, L.~Eberhardt, Y.-t. Huang and S.~Mizera, \emph{{On unitarity
  of tree-level string amplitudes}},
  \href{https://doi.org/10.1007/JHEP02(2022)197}{\emph{JHEP} {\bfseries 02}
  (2022) 197} [\href{https://arxiv.org/abs/2201.11575}{{\ttfamily
  2201.11575}}].

\bibitem{Cheung:2022mkw}
C.~Cheung and G.~N. Remmen, \emph{{Veneziano variations: how unique are string
  amplitudes?}}, \href{https://doi.org/10.1007/JHEP01(2023)122}{\emph{JHEP}
  {\bfseries 01} (2023) 122}
  [\href{https://arxiv.org/abs/2210.12163}{{\ttfamily 2210.12163}}].

\bibitem{Geiser:2022exp}
N.~Geiser and L.~W. Lindwasser, \emph{{Generalized Veneziano and Virasoro
  amplitudes}}, \href{https://doi.org/10.1007/JHEP04(2023)031}{\emph{JHEP}
  {\bfseries 04} (2023) 031}
  [\href{https://arxiv.org/abs/2210.14920}{{\ttfamily 2210.14920}}].

\bibitem{Cheung:2023adk}
C.~Cheung and G.~N. Remmen, \emph{{Stringy dynamics from an amplitudes
  bootstrap}}, \href{https://doi.org/10.1103/PhysRevD.108.026011}{\emph{Phys.
  Rev. D} {\bfseries 108} (2023) 026011}
  [\href{https://arxiv.org/abs/2302.12263}{{\ttfamily 2302.12263}}].

\bibitem{Cheung:2023uwn}
C.~Cheung and G.~N. Remmen, \emph{{Bespoke dual resonance}},
  \href{https://doi.org/10.1103/PhysRevD.108.086009}{\emph{Phys. Rev. D}
  {\bfseries 108} (2023) 086009}
  [\href{https://arxiv.org/abs/2308.03833}{{\ttfamily 2308.03833}}].

\bibitem{Lovelace:1968kjy}
C.~Lovelace, \emph{{A novel application of regge trajectories}},
  \href{https://doi.org/10.1016/0370-2693(68)90255-4}{\emph{Phys. Lett. B}
  {\bfseries 28} (1968) 264}.

\bibitem{Shapiro:1969km}
J.~A. Shapiro, \emph{{Narrow-resonance model with regge behavior for pi pi
  scattering}}, \href{https://doi.org/10.1103/PhysRev.179.1345}{\emph{Phys.
  Rev.} {\bfseries 179} (1969) 1345}.

\bibitem{Bianchi:2020cfc}
M.~Bianchi, D.~Consoli and P.~Di~Vecchia, \emph{{On the N-pion extension of the
  Lovelace-Shapiro model}},
  \href{https://doi.org/10.1007/JHEP03(2021)119}{\emph{JHEP} {\bfseries 03}
  (2021) 119} [\href{https://arxiv.org/abs/2002.05419}{{\ttfamily
  2002.05419}}].

\bibitem{Caron-Huot:2021rmr}
S.~Caron-Huot, D.~Mazac, L.~Rastelli and D.~Simmons-Duffin, \emph{{Sharp
  boundaries for the swampland}},
  \href{https://doi.org/10.1007/JHEP07(2021)110}{\emph{JHEP} {\bfseries 07}
  (2021) 110} [\href{https://arxiv.org/abs/2102.08951}{{\ttfamily
  2102.08951}}].

\bibitem{Guerrieri:2020bto}
A.~L. Guerrieri, J.~Penedones and P.~Vieira, \emph{{S-matrix bootstrap for
  effective field theories: massless pions}},
  \href{https://doi.org/10.1007/JHEP06(2021)088}{\emph{JHEP} {\bfseries 06}
  (2021) 088} [\href{https://arxiv.org/abs/2011.02802}{{\ttfamily
  2011.02802}}].

\bibitem{Guerrieri:2021ivu}
A.~Guerrieri, J.~Penedones and P.~Vieira, \emph{{Where Is String Theory in the
  Space of Scattering Amplitudes?}},
  \href{https://doi.org/10.1103/PhysRevLett.127.081601}{\emph{Phys. Rev. Lett.}
  {\bfseries 127} (2021) 081601}
  [\href{https://arxiv.org/abs/2102.02847}{{\ttfamily 2102.02847}}].

\bibitem{Chen:2022nym}
H.~Chen, A.~L. Fitzpatrick and D.~Karateev, \emph{{Nonperturbative bounds on
  scattering of massive scalar particles in d \ensuremath{\geq} 2}},
  \href{https://doi.org/10.1007/JHEP12(2022)092}{\emph{JHEP} {\bfseries 12}
  (2022) 092} [\href{https://arxiv.org/abs/2207.12448}{{\ttfamily
  2207.12448}}].

\bibitem{Haring:2022sdp}
K.~H\"aring, A.~Hebbar, D.~Karateev, M.~Meineri and J.~Penedones, \emph{{Bounds
  on photon scattering}},  \href{https://arxiv.org/abs/2211.05795}{{\ttfamily
  2211.05795}}.

\bibitem{Li:2023qzs}
Y.-Z. Li, \emph{{Effective field theory bootstrap, large-N \ensuremath{\chi}PT
  and holographic QCD}},
  \href{https://doi.org/10.1007/JHEP01(2024)072}{\emph{JHEP} {\bfseries 01}
  (2024) 072} [\href{https://arxiv.org/abs/2310.09698}{{\ttfamily
  2310.09698}}].

\end{thebibliography}\endgroup

\end{document}